\documentclass[traditabstract]{aa}

\usepackage{graphicx}
\usepackage{txfonts}
\usepackage{hyperref}

\usepackage{etoolbox}

\makeatletter

\patchcmd{\NAT@citex}
  {\@citea\NAT@hyper@{%
     \NAT@nmfmt{\NAT@nm}%
     \hyper@natlinkbreak{\NAT@aysep\NAT@spacechar}{\@citeb\@extra@b@citeb}%
     \NAT@date}}
  {\@citea\NAT@nmfmt{\NAT@nm}%
   \NAT@aysep\NAT@spacechar\NAT@hyper@{\NAT@date}}{}{}

\patchcmd{\NAT@citex}
  {\@citea\NAT@hyper@{%
     \NAT@nmfmt{\NAT@nm}%
     \hyper@natlinkbreak{\NAT@spacechar\NAT@@open\if*#1*\else#1\NAT@spacechar\fi}%
       {\@citeb\@extra@b@citeb}%
     \NAT@date}}
  {\@citea\NAT@nmfmt{\NAT@nm}%
   \NAT@spacechar\NAT@@open\if*#1*\else#1\NAT@spacechar\fi\NAT@hyper@{\NAT@date}}
  {}{}

\makeatother

\begin{document}

\title{The {\it Herschel}\thanks{{\it Herschel} is an ESA space observatory with science instruments provided by European-led Principal Investigator consortia and with important participation from NASA.} view of the dominant mode of galaxy growth from $z=4$ to the present day}

\author{C.~Schreiber\inst{1}
    \and M.~Pannella\inst{1,2}
    \and D.~Elbaz\inst{1}
    \and M.~B\'ethermin\inst{1}
    \and H.~Inami\inst{3}
    \and M.~Dickinson\inst{3}
    \and B.~Magnelli\inst{4,5}
    \and T.~Wang\inst{1,6}
    \and H.~Aussel\inst{1}
    \and E.~Daddi\inst{1}
    \and S.~Juneau\inst{1}
    \and X.~Shu\inst{1}
    \and M.~T.~Sargent\inst{1,7}
    \and V.~Buat\inst{8}
    \and S.~M.~Faber\inst{9}
    \and H.~C.~Ferguson\inst{10}
    \and M.~Giavalisco\inst{11}
    \and A.~M.~Koekemoer\inst{10}
    \and G.~Magdis\inst{12,13}
    \and G.~E.~Morrison\inst{14,15}
    \and C.~Papovich\inst{16}
    \and P.~Santini\inst{17}
    \and D.~Scott\inst{18}.
}

\institute{
    Laboratoire AIM-Paris-Saclay, CEA/DSM/Irfu - CNRS - Universit\'e Paris Diderot, CEA-Saclay, pt courrier 131, F-91191 Gif-sur-Yvette, France\\
    \email{corentin.schreiber@cea.fr}
    \and Institut d'Astrophysique de Paris, UMR 7095, CNRS, UPMC Univ. Paris 06, 98bis boulevard Arago, F-75014 Paris, France
    \and National Optical Astronomy Observatory, 950 North Cherry Avenue, Tucson, AZ 85719, USA
    \and Max-Planck-Institut f\"ur Extraterrestrische Physik (MPE), Postfach 1312, D-85741 Garching, Germany
    \and Argelander-Institut f\"ur Astronomie, University of Bonn, auf dem H\"agel 71, D-53121 Bonn, Germany
    \and School of Astronomy and Space Sciences, Nanjing University, Nanjing, 210093, China
    \and Astronomy Center, Dept.~of Physics \& Astronomy, University of Sussex, Brighton BN1 9QH, UK
    \and Aix-Marseille Université, CNRS, LAM (Laboratoire d’Astrophysique de Marseille) UMR7326, 13388, Marseille, France
    \and University of California Observatories / Lick Observatory, University of California, Santa Cruz, CA 95064
    \and Space Telescope Science Institute, Baltimore, MD, USA
    \and Department of Astronomy, University of Massachusetts, Amherst, MA 01003, USA
    \and Department of Physics, University of Oxford, Keble Road, Oxford OX1 3RH
    \and Institute for Astronomy, Astrophysics, Space Applications and Remote Sensing, National Observatory of Athens, GR-15236 Athens, Greece
    \and Institute for Astronomy, University of Hawaii, Honolulu, Hawaii, 96822, USA
    \and Canada-France-Hawaii Telescope, Kamuela, Hawaii, 96743, USA
    \and George P.~and Cynthia W. Mitchell Institute for Fundamental Physics and Astronomy, Department of Physics and Astronomy, Texas A\&M University, College Station, TX 77843, USA
    \and INAF -- Osservatorio Astronomico di Roma, via di Frascati 33, 00040 Monte Porzio Catone, Italy
    \and Department of Physics and Astronomy, University of British Columbia, Vancouver, BC V6T 1Z1, Canada
}

\date{Received 17 September 2014; accepted 23 December 2014}

\abstract {
We present an analysis of the deepest {\it Herschel} images in four major extragalactic fields GOODS--North, GOODS--South, UDS, and COSMOS obtained within the GOODS--{\it Herschel} and CANDELS--{\it Herschel} key programs. The star formation picture provided by a total of $10\ 497$ individual far-infrared detections is supplemented by the stacking analysis of a mass complete sample of $62\ 361$ star-forming galaxies from the {\it Hubble Space Telescope} ({\it HST}) $H$ band-selected catalogs of the CANDELS survey and from two deep ground-based $K_{\rm s}$ band-selected catalogs in the GOODS--{\it North} and the COSMOS-wide field to obtain one of the most accurate and unbiased understanding to date of the stellar mass growth over the cosmic history.

We show, for the first time, that stacking also provides a powerful tool to determine the dispersion of a physical correlation and describe our method called ``scatter stacking", which may be easily generalized to other experiments.

The combination of direct UV and far-infrared UV-reprocessed light provides a complete census on the star formation rates (${\rm SFR}$s), allowing us to demonstrate that galaxies at $z=4$ to $0$ of all stellar masses ($M_\ast$) follow a universal scaling law, the so-called main sequence of star-forming galaxies. We find a universal close-to-linear slope of the $\log_{10}({\rm SFR})$--$\log_{10}(M_\ast)$ relation, with evidence for a flattening of the main sequence at high masses ($\log_{10}(M_\ast/{\rm M}_\odot) > 10.5$) that becomes less prominent with increasing redshift and almost vanishes by $z\simeq2$. This flattening may be due to the parallel stellar growth of quiescent bulges in star-forming galaxies, which mostly happens over the same redshift range. Within the main sequence, we measure a nonvarying ${\rm SFR}$ dispersion of $0.3\,{\rm dex}$: at a fixed redshift and stellar mass, about $68\%$ of star-forming galaxies form stars at a universal rate within a factor $2$. The specific ${\rm SFR}$ (${\rm sSFR}={\rm SFR}/M_\ast$) of star-forming galaxies is found to continuously increase from $z=0$ to $4$.

Finally we discuss the implications of our findings on the cosmic ${\rm SFR}$ history and on the origin of present-day stars: more than two-thirds of present-day stars must have formed in a regime dominated by the ``main sequence'' mode. As a consequence we conclude that, although omnipresent in the distant Universe, galaxy mergers had little impact in shaping the global star formation history over the last $12.5$ billion years.
}


\maketitle

\section{Introduction}

Most extremely star-forming galaxies in the local Universe are heavily dust obscured and show undeniable signs of an ongoing major merger, however such objects are relatively rare \citep{armus1987,sanders1996}. They have been historically classified as Luminous and Ultra Luminous InfraRed Galaxies, LIRGs and ULIRGs, based on their bolometric infrared luminosity over the wavelength range $8$--$1000\,\mu{\rm m}$, by $L_{\rm IR} > 10^{11}\,L_\odot$ and $> 10^{12}\,L_\odot$, respectively. However, they make up for only $2\%$ of the integral of the local IR luminosity function, the remaining fraction mainly produced by more typical isolated galaxies \citep{sanders1996}.

More recently, studies at higher redshift showed that the LIRGs were the dominant population at $z=1$ \citep{chary2001,lefloch2005}, replaced by ULIRGs at $z=2$ \citep{magnelli2013}. This was first interpreted as an increasing contribution of gas-rich galaxy mergers to the global star formation activity of the Universe, in qualitative agreement with the predicted and observed increase of the major merger rate \citep[e.g.,][]{patton1997,lefevre2000,conselice2003}.

The discovery of the correlation between star formation rate (${\rm SFR}$) and stellar mass ($M_\ast$), also called the ``main sequence'' of star-forming galaxies \citep{noeske2007}, at $z\simeq0$ \citep{brinchmann2004}, $z\simeq1$ \citep{noeske2007,elbaz2007}, $z\simeq2$ \citep{daddi2007-a,pannella2009-a,rodighiero2011,whitaker2012-a} $z=3$--$4$ \citep{daddi2009,magdis2010-b,heinis2013,pannella2014} and even up to $z=7$ \citep[e.g.,][]{stark2009,bouwens2012,stark2013,gonzalez2014,salmon2014,steinhardt2014} suggested instead a radically new paradigm. The tightness of this correlation is indeed not consistent with frequent random bursts induced by processes like major mergers of gas-rich galaxies, and favors more stable star formation histories \citep{noeske2007}.

Furthermore, systematic studies of the dust properties of the ``average galaxy'' at different redshifts show that LIRGs at $z=1$ and ULIRGs at $z=2$ bear close resemblance to normal star-forming galaxies at $z=0$. In particular, in spite of having star formation rates (${\rm SFR}$s) higher by orders of magnitude, they appear to share similar star-forming region sizes \citep{rujopakarn2011}, polycyclic aromatic hydrocarbon (PAH) emission lines equivalent widths \citep{pope2008-a,fadda2010,elbaz2011,nordon2012}, $[\ion{C}{II}]$ to far-infrared (FIR) luminosity ($L_{\rm FIR}$) ratios \citep{diaz-santos2013}, and universal FIR spectral energy distributions (SEDs) \citep{elbaz2011}. Only outliers above the ${\rm SFR}$--$M_\ast$ correlation \citep[usually called ``starbursts'', ][]{elbaz2011} show signs of different dust properties: more compact geometry \citep{rujopakarn2011}, excess of ${\rm IR8} \equiv L_{\rm IR}/L_{8\,\mu{\rm m}}$ \citep{elbaz2011}, $[\ion{C}{II}]$ deficit \citep{diaz-santos2013}, increased effective dust temperature \citep{elbaz2011,magnelli2014}, and PAH deficit \citep{nordon2012,murata2014}, indicating that these starburst galaxies are the true analogs of local LIRGs and ULIRGs. In this paradigm, the properties of galaxies are no longer most closely related to their rest-frame bolometric luminosities, but rather to their excess ${\rm SFR}$ compared to that of the main sequence.

This could mean that starburst galaxies are actually triggered by major mergers, but that the precise mechanism that fuels the remaining vast majority of ``normal'' galaxies is not yet understood. Measurements of galactic gas reservoirs yield gas fractions evolving from about $10\%$ in the local Universe \citep{leroy2008} up to $60\%$ at $z\simeq3$ \citep[B\'ethermin et al. 2014, submitted]{tacconi2010,daddi2010,geach2011,magdis2012,saintonge2013,santini2014,genzel2014}. Compared to the observed ${\rm SFR}$, this implies gas-consumption timescales that are much shorter than the typical duty cycle of most galaxies. It is thus necessary to replenish the gas reservoirs of these galaxies in some way. Large volume numerical simulations \citep{dekel2009-a} have shown that streams of cold gas from the intergalactic medium can fulfill this role, allowing galaxies to keep forming stars at these high but steady rates. Since the amount of gas accreted through these ``cold flows'' is directly linked to the matter density of the intergalactic medium, this also provides a qualitative explanation for the gradual decline of the ${\rm SFR}$ from $z=3$ to the present day \citep[e.g.,][]{dave2011}.

This whole picture relies on the existence of the main sequence. However, actual observations of the ${\rm SFR}$--$M_\ast$ correlation at $z>2$ rely mostly on ultraviolet-derived star formation rates, which need to be corrected by large factors to account for dust extinction \citep{calzetti1994,madau1998,meurer1999,steidel1999}. These corrections, performed using the UV continuum slope $\beta$ and assuming an extinction law, are uncertain and still debated. Although dust-corrected ${\rm SFR}$s are able to match more robust estimators on average in the local Universe \citep{calzetti1994,meurer1999} and beyond \citep[e.g.,][]{pannella2009-a,overzier2011,rodighiero2014}, it has been shown for example that these corrections cannot recover the full star formation rate of the most active objects \citep{goldader2002,buat2005,elbaz2007,rodighiero2011,wuyts2011-a,penner2012,oteo2013,rodighiero2014}. More recently, several studies have pointed toward an evolution of the calibration between the UV slope and UV attenuation as a function of redshift, possibly due to changes in the ISM properties \citep[e.g.,][]{pannella2014,castellano2014} or even as a function of environment \citep{koyama2013}. It is therefore possible that using UV-based ${\rm SFR}$ estimates modifies the normalization of the main sequence, and/or its dispersion. In particular, it could be that the tight scatter of the main sequence observed at high redshift \citep[e.g.,][]{bouwens2012,salmon2014} is not real but induced by the use of such ${\rm SFR}$s, thereby questioning the very existence of a main sequence at these epochs. Indeed, a small scatter is a key ingredient without which the main sequence loses its meaning.

Infrared telescopes allow us to measure the bolometric infrared luminosity of a galaxy ($L_{\rm IR}$), a robust star formation tracer \citep{kennicutt1998-a}. Unfortunately, they typically provide observations of substantially poorer quality (both in angular resolution and typical depth) compared to optical surveys. The launch of the {\it Spitzer} space telescope \citep{werner2004} was a huge step forward, as it allowed us to detect for the first time moderately luminous objects at high redshifts ($z<3$) in the mid-infrared (MIR) thanks to the MIPS instrument \citep{rieke2004}. It was soon followed by the {\it Herschel} space telescope \citep{pilbratt2010}, which provided better constraints on the spectrum of the dust emission by observing in the FIR with the PACS \citep{poglitsch2010} and SPIRE instruments \citep{griffin2010}.

Nevertheless only the most luminous star-forming objects can be detected at high redshifts, yielding strongly ${\rm SFR}$ biased samples \citep{elbaz2011}. In particular, most galaxies reliably detected with these instruments at $z\ge3$ are very luminous starbursts, making it difficult to study the properties of ``normal'' galaxies at these epochs. So far only a handful of studies have probed in a relatively complete manner the Universe at $z\gtrsim3$ with IR facilities \citep[e.g.,][]{heinis2014,pannella2014} and most of what we know about normal galaxies at $z>3$ is currently based on UV light alone \citep{daddi2009,stark2009,bouwens2012,stark2013,gonzalez2014,salmon2014}.

Here we take advantage of the deepest data ever taken with {\it Herschel} in the Great Observatories Origins Deep Survey (GOODS, PI: D.~Elbaz), covering the GOODS--North and GOODS--South fields, and the Cosmic Assembly Near-Infrared Deep Extragalactic Legacy Survey (CANDELS, PI: M.~E.~Dickinson) covering a fraction of the Ultra-Deep Survey\footnote{This field is also known as the Subary XMM Deep Survey (SXDS) field.} (UDS) and Cosmic Evolution Survey (COSMOS) fields, to infer stricter constraints on the existence and relevance of the main sequence in the young Universe up to $z=4$. To do so, we first construct a mass-selected sample with known photometric redshifts and stellar masses and then isolate star-forming galaxies within it. We bin this sample in redshift and stellar mass and stack the {\it Herschel} images. This allows us to infer their average $L_{\rm IR}$, and thus their ${\rm SFR}$s. We then present a new technique we call ``scatter stacking'' to measure the dispersion around the average stacked ${\rm SFR}$, taking nondetected galaxies into account. Finally, we cross-match our sample with {\it Herschel} catalogs to study individually detected galaxies.

In the following, we assume a $\Lambda$CDM cosmology with $H_0 = 70\ {\rm km}\,{\rm s}^{-1} {\rm Mpc}^{-1}$, $\Omega_{\rm M} = 0.3$, $\Omega_\Lambda = 0.7$ and a \cite{salpeter1955} initial mass function (IMF), to derive both star formation rates and stellar masses. All magnitudes are quoted in the AB system, such that $M_{\rm AB} = 23.9 - 2.5\log_{10}(S_{\!\nu}\ [\mu{\rm Jy}])$.

\section{Sample and observations}

\begin{table*}[htdp]
\caption{Catalog depths for each field. \label{TAB:depth}}
\begin{center}
\begin{tabular}{lcccccccc}
    \hline
    \hline \\[-2.5mm]
    Field & Area$^{\rm a}$ & NIR ($5\sigma$) & $24\,\mu{\rm m}$ & $100\,\mu{\rm m}$ & $160\,\mu{\rm m}$ & $250\,\mu{\rm m}$& $350\,\mu{\rm m}$ & $500\,\mu{\rm m}$ \\
          &                 &                 & $\mu{\rm Jy}$ ($3\sigma$) & ${\rm mJy}$ ($3\sigma)$ & ${\rm mJy}$ ($3\sigma)$ & ${\rm mJy}$ ($5\sigma)$ & ${\rm mJy}$ ($5\sigma)$ & ${\rm mJy}$ ($5\sigma)$ \\
    \hline \\[-2.5mm]
    GN     & $168\,{\rm arcmin}^2$ & $K_{\rm s}$$ < 24.5$       & $21$       & $1.1$  & $2.7$ & $7.3$  & $7.8$ & $13$ \\
    GS     & $184\,{\rm arcmin}^2$ & $H < 27.4$--$29.7$ & $20$       & $0.8$  & $2.4$ & $7.0$  & $7.5$ & $13$ \\
    UDS    & $202\,{\rm arcmin}^2$ & $H < 27.1$--$27.6$ & $40$       & $1.7$  & $3.9$ & $10$  & $11$ & $13$ \\[1pt]
    COSMOS & & & & & & & & \\
    \hfill{\tiny\it -CANDELS} & $208\,{\rm arcmin}^2$ & $H < 27.4$--$27.8$ & $27$--$40$ & $1.5$  & $3.1$ & $11$  & $14$ & $14$ \\
    \hfill{\tiny\it -UVISTA} & $1.6\,\deg^2$         & $K_{\rm s}$$< 23.4$        & $27$--$40$ & $4.6$  & $9.9$ & ---          & ---         & --- \\[1pt]
    \hline
\end{tabular}
\end{center}
$^{\rm (a)}$ This is the sky coverage of our sample, and may be smaller than the nominal area of the detection image.
\end{table*}

We use the ultra-deep $H$-band catalogs provided by the CANDELS--{\it HST} team \citep{grogin2011,koekemoer2011} in three of the CANDELS fields, namely GOODS--South \citep[GS][]{guo2013}, UDS \citep{galametz2013}, and COSMOS (Nayyeri et al.~in prep.). With the GOODS--North (GN) CANDELS catalog not being finalized at the time of writing, we fall back to a ground-based $K_{\rm s}$-band catalog. To extend our sample to rarer and brighter objects, we also take advantage of the much wider area provided by the $K_{\rm s}$-band imaging in the COSMOS field acquired as part of the UltraVISTA program (UVISTA). In the following, we will refer to this field as ``COSMOS UltraVISTA'', while the deeper but smaller region observed by CANDELS will be called ``COSMOS CANDELS''.

Using either the $H$ or the $K_{\rm s}$ as the selection band will introduce potentially different selection effects. In practice, these two bands are sufficiently close in wavelengths that one does not expect major differences to arise: if anything, the $K_{\rm s}$-band catalogs are potentially more likely to be mass-complete, since this band will probe the rest-frame optical up to higher redshifts. However these catalogs are ground based, and lack both angular resolution and depth when compared to the {\it HST} $H$-band data. It is thus necessary to carefully estimate the mass completeness level of each catalog, and only consider mass-complete regimes in the following analysis.

All these fields were selected for having among the deepest {\it Herschel} observations, which are at the heart of the present study, along with high-quality, multi-wavelength photometry in the UV to NIR. The respective depths of each catalog are listed in Table \ref{TAB:depth}. We next present the details of the photometry and source extraction of each field.

\subsection{GOODS--North}

GOODS--North is one of the fields targeted by the CANDELS--{\it HST} program, and the last to be observed. Consequently, the data reduction was delayed compared to the other fields and there was no available catalog when we started this work. We thus use the ground-based $K_{\rm s}$-band catalog presented in \cite{pannella2014}, which is constructed from the deep CFHT WIRCAM $K_{\rm s}$-band observations of \cite{wang2010}. This catalog contains 20 photometric bands from the NUV to IRAC $8\,\mu{\rm m}$ and was built using {\sc SExtractor} \citep{bertin1996} in dual image mode, with the $K_{\rm s}$-band image as the detection image. Fluxes are measured within a $2\arcsec$ aperture on all images, and the effect of varying point spread function (PSF) and / or seeing is accounted for using PSF-matching corrections. Per-object aperture corrections to total are provided by the ratio of the {\tt \verb|FLUX_AUTO|} as given by {\sc SExtractor} and the aperture $K_{\rm s}$-band flux. This results in a $0.8\arcsec$ angular resolution catalog of $79\ 003$ sources and a $5\sigma$ limiting magnitude of $K_{\rm s}$$ = 24.5$.

The $K_{\rm s}$-band image extends over $0.25\deg^2$, but only the central area is covered by {\it Spitzer} and {\it Herschel}. We therefore only keep the sources that fall inside the coverage of those two instruments, i.e., $15\ 284$ objects in $168\,{\rm arcmin}^2$. We also remove stars identified either from the {\sc SExtractor} flag {\tt \verb|CLASS_STAR|} for bright enough objects ($K_{\rm s}$$ < 20$), or using the $BzK$\xspace color-color diagram \citep{daddi2004-a}. Our final sample consists of $14\ 828$ galaxies, $12\ 317$ of which are brighter than the $5\sigma$ limiting magnitude, with $3\ 775$ spectroscopic redshifts.

The {\it Herschel} images in both PACS and SPIRE were obtained as part of the GOODS--{\it Herschel} program \citep{elbaz2011}. The source catalog of {\it Herschel} and {\it Spitzer} MIPS $24\,\mu{\rm m}$ are taken from the public GOODS--{\it Herschel} DR1. {\it Herschel} PACS and SPIRE $250\,\mu{\rm m}$ flux densities are extracted using PSF fitting at the position of MIPS priors, themselves extracted from IRAC priors. SPIRE $350\,\mu{\rm m}$ and $500\,\mu{\rm m}$ flux densities are obtained by building a reduced prior list out of the $250\,\mu{\rm m}$ detections. This procedure, described in more detail in \cite{elbaz2011}, yields $2\ 681$ MIPS and $1\ 039$ {\it Herschel} detections \citep[$>3\sigma$ in any PACS band or $>5\sigma$ in SPIRE, following][]{elbaz2011} that we could cross-match to the $K_{\rm s}$-band catalog using their IRAC positions.

\subsection{GOODS--South, UDS, \& COSMOS CANDELS}

In GOODS--South, UDS and COSMOS CANDELS we use the official CANDELS catalogs presented, respectively, in \cite{guo2013} (version $121114$), \cite{galametz2013} (version $120720$) and Nayyeri et al. (in prep.) (version $130701$). They are built using {\sc SExtractor} in dual image mode, using the {\it HST} $H$-band image as the detection image to extract the photometry at the other {\it HST} bands. The ground-based and {\it Spitzer} photometry is obtained with TFIT \citep{laidler2007}. The {\it HST} photometry was measured using the {\verb|FLUX_ISO|} from {\sc SExtractor} and corrected to total magnitudes using either the {\tt \verb|FLUX_BEST|} or {\tt \verb|FLUX_AUTO|} measured in the $H$ band, while the ground-based and {\it Spitzer} photometry is already ``total'' by construction. These catalogs gather $16$ photometric bands in GOODS--South, $19$ in UDS, and $27$ in COSMOS, ranging from the $U$ band to IRAC $8\,\mu{\rm m}$, for a total of $34\ 930$ (respectively $35\ 932$ and $38\ 601$) sources, $1\ 767$ (respectively $575$ and $1\ 175$) of which have a spectroscopic redshift. The $H$-band exposure in the fields is quite heterogeneous, the $5\sigma$ limiting magnitude ranging from $27.4$ to $29.7$ in GOODS--South, $27.1$ to $27.6$ in UDS, and $27.4$ to $27.8$ in COSMOS, but it always goes much deeper than the available ground-based photometry. These extreme depths can also become a problem, especially when dealing with sources so faint that they are significantly detected in the {\it HST} images only. The SED of these objects is so poorly constrained that we cannot robustly identify them as galaxies, or compute accurate photometric redshifts. To solve this issue, one would like to only keep sources that have a sufficient wavelength coverage, e.g., imposing a significant detection in at least ten UV to NIR bands, but this would introduce complex selection effects. Here we decide to only keep sources that have an $H$-band magnitude brighter than $26$. This ensures that the median number of UV to NIR bands for each source (along with the $16$th and $84$th percentiles) is $11_{-2}^{+3}$, $16_{-4}^{+3}$ and $21_{-5}^{+5}$, respectively, as compared to $9_{-4}^{+4}$, $13_{-5}^{+5}$ and $18_{-7}^{+7}$ when using the whole catalogs.

As for GOODS--North, we remove stars using a combination of morphology and $BzK$\xspace classification, and end up with $18\ 364$ (respectively $21\ 552$ and $24\ 396$) galaxies with $H < 26$ in $184\,{\rm arcmin}^2$ (respectively $202\,{\rm arcmin}^2$ and $208\,{\rm arcmin}^2$).

In both UDS and COSMOS, the {\it Herschel} PACS and SPIRE images were taken as part of the CANDELS--{\it Herschel} program, and are slightly shallower than those in the two GOODS fields. The MIPS $24\,\mu{\rm m}$ images, however, are clearly shallower, since they reach a noise level of approximately $40\,\mu{\rm Jy}$ ($1\sigma$), as compared to the $20\,\mu{\rm Jy}$ in GOODS. In COSMOS, however, the MIPS map contains a ``deep'' region \citep{sanders2007} that covers roughly half of the COSMOS CANDELS area with a depth of about $30\,\mu{\rm Jy}$.

In those two fields, sources are extracted with the same procedure as in GOODS--North (Inami et al.~in prep). These catalogs provide, respectively, $2\ 461$ and $2\ 585$ MIPS sources as well as $730$ and $1\ 239$ {\it Herschel} detections within the {\it HST} coverage. Since the IRAC priors used in the source extraction come directly from the CANDELS catalog, no cross-matching has to be performed.

The {\it Herschel} images in GOODS--South come from three separate programs. The PACS images are the result of the combined observation of both GOODS--{\it Herschel} and PEP \citep{lutz2011}, while SPIRE images were obtained as part of the HerMES program \citep{oliver2012}. The PACS fluxes are taken from the public PEP DR1 catalog \citep{magnelli2013}, and were extracted using the same procedure as in GOODS--North. For the SPIRE fluxes, we downloaded the individual level-2 data products covering the full ECDFS from the {\it Herschel} ESA archive\footnote{\url{http://www.cosmos.esa.int/web/herschel/science-archive}} and reduced them following the same procedure as the other sets of SPIRE data used in GOODS and CANDELS--{\it Herschel}. This catalog provides $1\ 875$ MIPS and $1\ 058$ {\it Herschel} detections within the {\it HST} coverage, which were cross matched to the CANDELS catalog using their IRAC positions.

\subsection{COSMOS UltraVISTA \label{SEC:sample_uvista}}

Only a small region of the COSMOS field has been observed within the CANDELS program. For the remaining area, we have to rely on ground-based photometry. To this end, we consider two different $K_{\rm s}$-band catalogs, both based on the UltraVISTA DR1 \citep{mccracken2012}.

The first catalog, presented in \cite{muzzin2013-a}, is built using {\sc SExtractor} in dual image mode, with the $K_{\rm s}$-band image as detection image. The photometry in the other bands is extracted using PSF-matched images degraded to a common resolution of \mbox{$\sim1.1\arcsec$} and an aperture of $2.1\arcsec$, except for the {\it Spitzer} bands and {\it GALEX}. Here, an alternative cleaning method is used, where nearby sources are first subtracted using the PSF-convolved $K_{\rm s}$-band profiles ($u^*$ band for {\it GALEX}), then the photometry of the central source is measured inside an aperture of $3\arcsec$. In both cases, aperture fluxes are corrected to total using the ratio of {\tt \verb|FLUX_AUTO|} and aperture $K_{\rm s}$-band flux. In the end, the catalog contains 30 photometric bands ranging from {\it GALEX} FUV to IRAC $8\,\mu{\rm m}$ (we did not use the $24\,\mu{\rm m}$ photometry), for a total of $262\ 615$ objects and a $5\sigma$ limiting magnitude of $K_{\rm s}$$ = 23.4$. As for the CANDELS fields, stars are excluded using a combination of morphological and $BzK$\xspace classification, resulting in a final number of $249\ 823$ galaxies within $1.6\,\deg^2$, $168\ 509$ of which are brighter than the $5\sigma$ limiting magnitude, with $5\ 532$ having spectroscopic redshifts.

The second catalog, presented in \cite{ilbert2013}, is very similar in that, apart from missing {\it GALEX} and {\it Subaru} $g^+$, it uses the same raw images and was also built with {\sc SExtractor}. The difference lies mostly in the extraction of IRAC fluxes. Here, and for IRAC only, {\sc SExtractor} is used in dual image mode, with the {\it Subaru} $i$-band image as the detection image. Since the IRAC photometry was not released along with the rest of the photometry, we could not directly check the consistency of the two catalogs, nor use this photometry to derive accurate galaxy properties. Nevertheless, the photometric catalog comes with a set of photometric redshifts and stellar masses that we can use as a consistency check. These were built using a much more extensive but private set of spectroscopic redshifts, and are thus expected to be of higher quality. A direct comparison of the two photometric redshift estimations shows a constant relative scatter of $4\%$ below $z=2$. At higher redshifts, the scatter increases to $10\%$ because of the ambiguity between the Balmer and Lyman breaks. This ambiguity arises because of the poor wavelength coverage caused by the shallow depths of these surveys, but it takes place in a redshift regime where our results are mostly based on the deeper, and therefore more robust, CANDELS data. We also checked that redoing our analysis with Ilbert et al.'s catalog yielded very similar results in the mass-complete regimes.

Finally, while the {\it Spitzer} MIPS imaging is the same as that in COSMOS CANDELS, the {\it Herschel} PACS images in this wide field were taken as part of the PEP program, at substantially shallower depth \citep{lutz2011}. The {\it Spitzer} MIPS and {\it Herschel} PACS photometry are taken from the public PEP DR1 catalog\footnote{{\url{http://www.mpe.mpg.de/ir/Research/PEP/DR1}}}, itself based on the MIPS catalog of \cite{lefloch2009}, yielding $37\ 544$ MIPS and $9\ 387$ PACS detections successfully cross-matched to the first $K_{\rm s}$ band catalog.

\subsection{Photometric redshifts and stellar masses \label{SEC:zmstar}}

Photometric redshifts (photo-$z$) and stellar masses are derived using the procedure described in \cite{pannella2014}. Briefly, photo-$z$s are computed using EAZY\footnote{\url{http://code.google.com/p/eazy-photoz}.} \citep{brammer2008} in its standard setup. Global photometric zero points are adjusted iteratively by comparing the photo-$z$s to the available spectroscopic redshifts (spec-$z$), and minimizing the difference between the two. We emphasize that, although part of these adjustments are due to photometric calibration issues, they also originate from defects in the adopted SED template library. To estimate the quality of the computed photo-$z$s, we request that the {\tt odds} computed by EAZY, which is the estimated probability that the true redshift lies within $\Delta z = 0.2\times(1 + z_{\rm phot})$ \citep{benitez2000}, be larger than $0.8$. A more stringent set of criteria is adopted in COSMOS CANDELS, because of the lower quality of the photometric catalog. To prevent contamination of our sample from issues in the photometry, we prefer to be more conservative and only keep ${\tt odds} > 0.98$ and impose that the $\chi^2$ of the fit be less than $100$ to remove catastrophic fits. The median $\Delta z \equiv |z_{\rm phot} - z_{\rm spec}|/(1 + z_{\rm spec})$ is respectively $3.0\%$, $3.2\%$, $1.8\%$, $2.0\%$, and $0.8\%$ in GOODS--North, GOODS--South, UDS CANDELS, COSMOS CANDELS, and COSMOS UltraVISTA. We stress however that the representativeness of this accuracy also depends on the spectroscopic sample. In COSMOS UltraVISTA, for example, we only have spec-$z$s for the brightest objects, hence those that have the best photometry. Fainter and more uncertain sources thus do not contribute to the accuracy measurement, which is why the measured value is so low. Lastly, although we use these spec-$z$s to calibrate our photo-$z$s, we do not use them afterwards in this study. The achieved precision of our photo-$z$s is high enough for our purposes, and the selection functions of all spectroscopic surveys we gather here are very different, if not unknown. To avoid introducing any incontrollable systematic, we therefore decide to consistently use photo-$z$s for all our sample.

Stellar masses are derived using FAST\footnote{\url{http://astro.berkeley.edu/\string~mariska/FAST.html}} \citep{kriek2009}, adopting \cite{salpeter1955} IMF\footnote{Using another IMF would systematically shift both our $M_\ast$ and ${\rm SFR}$s by approximately the same amount, and therefore would not affect the shape of the main sequence.}, the \cite{bruzual2003} stellar population synthesis model and assuming that all galaxies follow delayed exponentially declining\footnote{Other star formation histories were considered, in particular with a constant or exponentially declining ${\rm SFR}$. Selecting all galaxies from $z>0.3$ to $z<5$, no systematic offset is found, while the scatter evolves mildly from $0.12\,{\rm dex}$ at $M_\ast = 1\times10^8\,{\rm M}_\odot$ to $0.08$ at $M_\ast = 3\times10^{11}\,{\rm M}_\odot$.} star formation histories (SFHs), parametrized by ${\rm SFR}(t) \propto (t/\tau^2)\exp(-t/\tau)$ with $0.01 < \tau < 10\,{\rm Gyr}$. Dust extinction is accounted for assuming the \cite{calzetti2000} law, with a grid ranging from $A_{V}=0$ to $4$. Metallicity is kept fixed and equal to $Z_\odot$. We assess the quality of the stellar mass estimate with the reduced $\chi^2$ of the fit, only keeping galaxies for which $\chi^2 < 10$.

\subsection{Rest-frame luminosities and star formation rates \label{SEC:sfr}}

Star formation rates are typically computed by measuring the light of young OB stars, which emit the bulk of their light in the UV. However this UV light is most of the time largely absorbed by the interstellar dust, and re-emitted in the IR as thermal radiation. To obtain the total ${\rm SFR}$ of a galaxy, it is therefore necessary to combine the light from both the UV and the IR.

Rest-frame luminosities in the FUV ($1500\,\AA$), $U$, $V$, and $J$ bands are computed with {\it EAZY} by convolving the best-fit SED model from the stellar mass fit with the filter response curves. The FUV luminosity is then converted into ${\rm SFR}$ {\it uncorrected} for dust attenuation using the formula from \cite{daddi2004-a}, i.e.,
\begin{equation}
    {\rm SFR}_{\rm UV} = 2.17\times10^{-10}\,L_{\rm UV}\,[L_\odot]\,.
\end{equation}

The infrared luminosity $L_{\rm IR}$ is computed following the procedure of \cite{elbaz2011}. We fit the {\it Herschel} flux densities with CE01 templates, and compute $L_{\rm IR}$ from the best-fit template. In this procedure, photometric points below $30\,\mu{\rm m}$ rest-frame are not used in the fit since this is a domain that is potentially dominated by active galactic nuclei (AGN) torus emission, and not by star formation \citep[e.g.,][]{mullaney2011}. We come back to this issue in section \ref{SEC:selfinal}. This IR luminosity is, in turn, converted into dust-reprocessed ${\rm SFR}$ using the formula from \cite{kennicutt1998-a}
\begin{equation}
    {\rm SFR}_{\rm IR} = 1.72\times10^{-10}\,L_{\rm IR}\,[L_\odot]\,.
\end{equation}

The total ${\rm SFR}$ is finally computed as the sum of ${\rm SFR}_{\rm UV}$ and ${\rm SFR}_{\rm IR}$. The above two relations are derived assuming a \cite{salpeter1955} IMF and assume that the ${\rm SFR}$ remained constant over the last $100\,{\rm Myr}$.

A substantial number of galaxies in this sample ($50\%$ in the CANDELS fields, $75\%$ in COSMOS UltraVISTA) are detected by {\it Spitzer} MIPS but not by ${\it Herschel}$. Although for these galaxies we only have a single photometric point in the MIR, we can still infer accurate monochromatic ${\rm SFR}$s using the original $L_{\rm IR}$ calibration of the CE01 library. This calibration is valid up to $z<1.5$, as shown in \cite{elbaz2011}, hence we only use MIPS-derived ${\rm SFR}$s for sources not detected by {\it Herschel} over this redshift range. Although there exist other calibrations that are applicable to higher redshifts \citep[e.g.,][]{elbaz2011,wuyts2011-a}, we do not know how they would impact the measurement of the scatter of the main sequence. We therefore prefer not to use them and discard the $24\,\mu{\rm m}$ measurements above $z=1.5$. Galaxies not detected in the MIR ($z<1.5$) or FIR have no individual ${\rm SFR}$ estimates and are only used for stacking. When working with detections alone (section \ref{SEC:indiv}), this obviously leads to an ${\rm SFR}$ selected sample and is taken into account by estimating the ${\rm SFR}$ completeness.

Lastly, there are some biases that can affect our estimates of ${\rm SFR}$ from the IR. In particular, the dust can also be heated by old stars that trace the total stellar mass content rather than the star formation activity \citep[e.g.,][]{salim2009}. Because of the relatively low luminosity of these stars, this will most likely be an issue for massive galaxies with low star formation activity, i.e., typically quiescent galaxies (see, e.g., Appendix \ref{APP:uvj} where we analyze such cases). Since we remove these galaxies from our sample, we should not be affected by this bias. This is also confirmed by the excellent agreement of IR based ${\rm SFR}$ estimates with those obtained from the radio emission \citep[e.g.,][]{pannella2014}, the latter not being affected by the light of old stars.

\subsection{A mass-complete sample of star-forming galaxies \label{SEC:selfinal}}

\begin{table}[htdp]
\caption{Number of object in our sample per field. \label{TAB:sample}}
\begin{center}
\begin{tabular}{lcccccccc}
    \hline
    \hline \\[-2.5mm]
    Field & All galaxies$^{\rm a}$ & SF$^{\rm b}$ & Spec-$z$$^{\rm c}$ & {\it Herschel}$^{\rm d}$ \\
    \hline \\[-2.5mm]
    GN     &  $6\ 973$  &  $5\ 358$  &  $2\ 605$  &  $867$ \\
    GS     &  $5\ 539$  &  $4\ 630$  &  $2\ 275$  &  $947$ \\
    UDS    &  $7\ 455$  &  $6\ 372$  &  $504$  &  $654$ \\
    COSMOS & & & & \\[1pt]
    \hfill{\tiny\it -CANDELS} & $7\ 580$  &  $6\ 599$  &  $811$  &  $976$ \\
    \hfill{\tiny\it -UVISTA} &  $58\ 202$  &  $39\ 375$  &  $3\ 736$  &  $7\ 053$  \\[1pt]
    \hline
\end{tabular}
\end{center}
$^{\rm (a)}$ Number of galaxies in our mass-complete NIR sample, removing stars, spurious sources, and requiring {\it Spitzer} and {\it Herschel} coverage. $^{\rm (b)}$ Final subsample of good quality galaxies classified as star-forming with the $UVJ$\xspace criterion (see section \ref{SEC:selfinal}). $^{\rm (c)}$ Subsample of galaxies with a spectroscopic redshift (various sources, see catalog papers for references). $^{\rm (d)}$ Subsample of galaxies with a detection in any {\it Herschel} band, requiring $>3\sigma$ significance in PACS or $>5\sigma$ in SPIRE \citep[following][]{elbaz2011}.
\end{table}

\begin{figure*}
    \centering
    \includegraphics*[width=18cm]{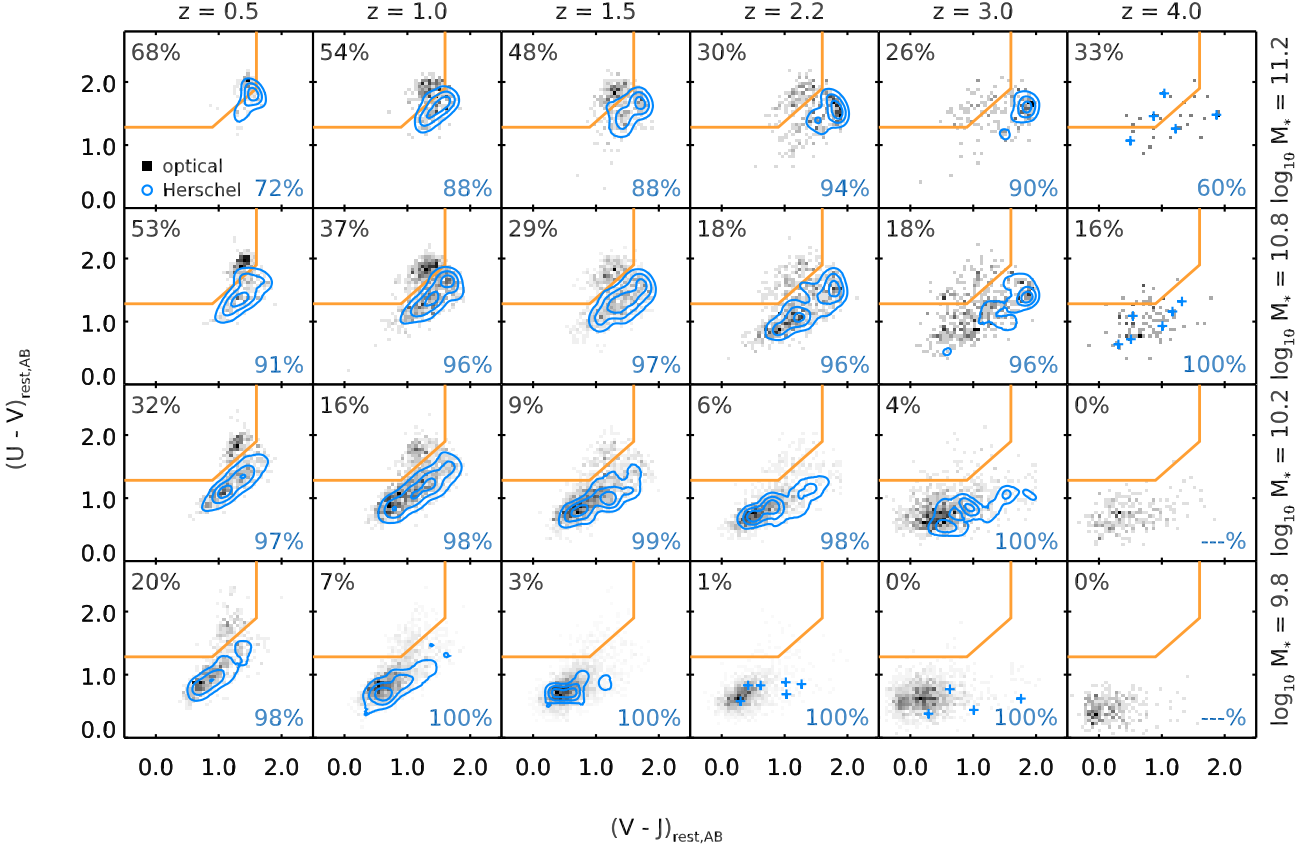}
    \caption{$UVJ$\xspace diagrams in each bin of redshift (horizontally) and mass (vertically) of our CANDELS sample. The central value of the redshift and mass bins are shown at the top and on right-hand side of the figure, respectively. The dividing line between active and passive galaxies is shown as a solid orange line on each plot, with passive galaxies located on the top-left corner. We show in the background the distribution of sources from the $H$-band catalogs in gray scale. We also overplot the position of sources detected with {\it Herschel} as blue contours or, when the source density is too low, as individual blue open circles. On the top-left corner of each plot, we give the fraction of $H$ band-selected galaxies that fall inside the quiescent region, and on the bottom-right corner we show the fraction of {\it Herschel} sources that reside in the star-forming region.}
    \label{FIG:uvj}
\end{figure*}

We finalize our sample by selecting actively star-forming galaxies. Indeed, the observation of a correlation between mass and ${\rm SFR}$ only applies to galaxies that are still forming stars, and not to quiescent galaxies. The latter are not evolving anymore and pile up at high stellar masses with little to no detectable signs of star formation. Nevertheless, they can still show residual IR emission due to the warm inter stellar medium (ISM). This cannot be properly accounted for with the CE01 library, and will be misinterpreted as an ${\rm SFR}$ tracer.

Several methods exist to exclude quiescent galaxies. The most obvious is to select galaxies based on their specific ${\rm SFR}$ (${\rm sSFR} \equiv {\rm SFR}/M_\ast$). Indeed, quiescent galaxies have very low ${\rm SFR}$ by definition, and they are preferentially found at high $M_\ast$. Therefore, they will have very low ${\rm sSFR}$ compared to star-forming galaxies. This obviously relies on the very existence of the correlation between ${\rm SFR}$ and $M_\ast$, and removing galaxies with too low ${\rm sSFR}$ would artificially create the correlation even where it does not exist. On the other hand, selecting galaxies based on their ${\rm SFR}$ alone would destroy the correlation, even where it exists \citep{rodighiero2011,lee2013}. It is therefore crucial that the selection does not apply directly to any combination of ${\rm SFR}$ or $M_\ast$. Furthermore, these methods require that an accurate ${\rm SFR}$ is available for all galaxies, and this is something we do not have since most galaxies are not detected in the mid- or far-IR. We must therefore select star-forming galaxies based on information that is available for all the galaxies in our sample, i.e., involving optical photometry only.

There are several color-magnitude or color-color criteria that are designed to accomplish this. Some, like the $BzK$\xspace approach \citep{daddi2004-a}, are based on the observed photometry and are thus very simple to compute, but they also select a particular redshift range by construction. This is not desirable for our sample, and we thus need to use rest-frame magnitudes. Color-magnitude diagrams \citep[e.g., $U-r$ versus $r$-band magnitude as in][]{baldry2004} tend to wrongly classify some of the red galaxies as passive, while they could also be red because of high dust attenuation. Since high mass galaxies suffer the most from dust extinction \citep{pannella2009-a}, it is thus likely that color-magnitude selections would have a nontrivial effect on our sample. It is therefore important to use another color to disentangle galaxies that are red because of their old stellar populations and those that are red because of dust extinction.

To this end, \cite{williams2009} devised the $UVJ$\xspace selection, based on the corresponding color-color diagram introduced in \cite{wuyts2007}. It uses the $U-V$ color, similar to the $U-r$ from the standard color-magnitude diagram, but combines it to the $V-J$ color to break the age--dust degeneracy. Although the bimodality stands out clearly on this diagram, the locus of the passive cloud has been confirmed by \cite{williams2009} using a sample of massive galaxies in the range $0.8 < z < 1.2$ with little or no [\ion{O}{II}] line emission, while the active cloud falls on the \cite{bruzual2003} evolutionary track for a galaxy with constant ${\rm SFR}$. One can then draw a dividing line that passes between those two clouds to separate one population from the other. We use the following definition, at all redshifts and stellar masses:
\begin{equation}
    {\rm quiescent} = \left\{\begin{array}{rcl}
        U - V &>& 1.3\,, \\
        V - J &<& 1.6\,, \\
        U - V &>& 0.88\times(V - J) + 0.49\,.
    \end{array}\right.
\end{equation}
This definition differs by only $0.1$ magnitude compared to that of \cite{williams2009}. Rest-frame colors can show offsets of similar order from one catalog to another, because of photometric coverage and uncertainties in the zero-point corrections. It is thus common to adopt slightly different definitions to account for these effects \citep[see e.g.,][]{cardamone2010-a,whitaker2011,brammer2011,strazzullo2013,viero2013,muzzin2013-a}. In COSMOS UltraVISTA, we follow the definition given by \cite{muzzin2013-a}.

The corresponding diagram in bins of mass and redshift for the CANDELS fields is shown in Fig.~\ref{FIG:uvj}. Here we also overplot the location of the galaxies detected by {\it Herschel}; because of the detection limit of the surveys, the vast majority of {\it Herschel} detections have high ${\rm SFR}$s. We therefore expect them to fall on the $UVJ$\xspace ``active'' region. This is indeed the case for the vast majority of these galaxies, even when the majority of optical sources are quiescent as is the case at $z=0.5$ and $\log_{10}(M_\ast/{\rm M}_\odot) > 10$. In total, only $5\%$ of the galaxies in our {\it Herschel} sample are classified as passive, and about a third of those have a probability larger than $20\%$ to be misclassified because of uncertainties in their $UVJ$\xspace colors. The statistics in COSMOS UltraVISTA are similar.

The number of galaxies with reliable redshifts and stellar masses (see section \ref{SEC:zmstar}) that are classified with this diagram as actively star-forming are reported in Table \ref{TAB:sample}. These are the galaxies considered in the following analysis. As a check, we also analyze separately the quiescent galaxies in Appendix \ref{APP:uvj}.

Finally, we do not explicitly exclude known AGNs from our sample. We expect AGNs to reside in massive star-forming galaxies \citep{kauffmann2003,mullaney2012,santini2012,juneau2013,rosario2013}. While the most luminous optically unobscured AGNs may greatly perturb the optical photometry, and therefore the measurement of redshift and stellar mass, they will also degrade the quality of the SED fitting because we have no AGN templates in our fitting libraries. This can produce an increased $\chi^2$, hence selecting galaxies with $\chi^2 < 10$ (see section \ref{SEC:zmstar}) helps remove some of these objects. Also, their point-like morphology on the detection image tends to make them look like stars, which are systematically removed from the sample. The more common moderate luminosity AGNs can still be fit properly with galaxy templates \citep{salvato2011}. Therefore, several AGNs do remain in our sample without significantly affecting the optical SED fitting and stellar masses. Still, obscured AGNs will emit some fraction of their light in the IR through the emission of a dusty torus. To prevent pollution of our FIR measurements by the light of such dusty AGNs, we only use the photometry at rest-frame wavelengths larger than $30\,\mu{\rm m}$, where the contribution of the AGN is negligible \citep{mullaney2011}. Indeed, while the most extreme AGNs may affect mid-to-far IR colors, such as $24$-to-$70\,\mu{\rm m}$ color, their far-IR colors are indistinguishable from that of star-forming galaxies \citep{hatziminaoglou2010}. By rejecting the most problematic cases, and mitigating against AGN contribution to the IR, we aim to remove severe contamination while retaining a high sample completeness.

\subsection{Completeness and mass functions \label{SEC:massfunc}}

\begin{figure}
    \centering
    \includegraphics*[width=9cm]{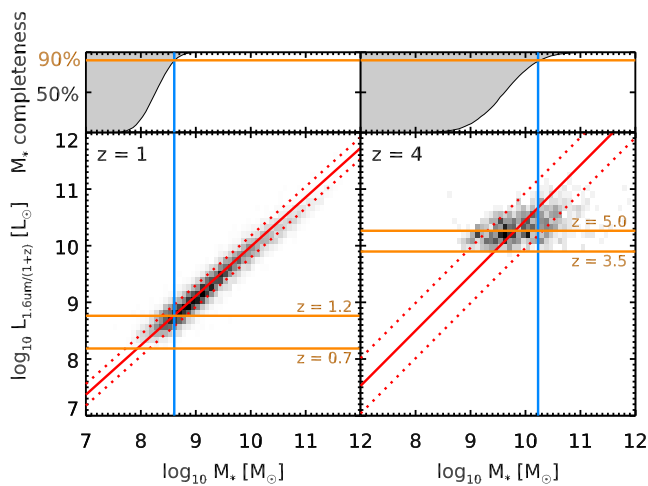}
    \caption{Correlation between the stellar mass and the luminosity in the observed-frame $H$ band at $0.7 < z < 1.2$ (left) and $3.5 < z < 5$ (right) in the three CANDELS fields GOODS--South, UDS, and COSMOS. On the bottom plots, the two horizontal orange lines show the position of the $H=26$ limiting magnitude at $z=z_{\rm min}$ and $z=z_{\rm max}$. The red line is the best-fit relation, and the dotted lines above and below show the $1\sigma$ dispersion ($0.2$ and $0.5\,{\rm dex}$, respectively). The blue vertical line shows the locus of the estimated $90\%$ mass completeness in each redshift bin. The top plots show the evolution of completeness (i.e., the estimated fraction of detected objects) with stellar mass, and the horizontal orange line shows the $90\%$ completeness level.}
    \label{FIG:mtol}
\end{figure}

\begin{figure}
    \centering
    \includegraphics*[width=9cm]{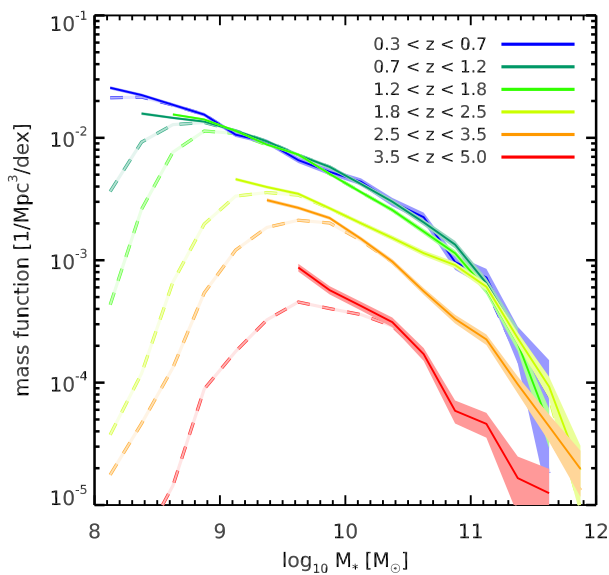}
    \caption{Evolution of the star-forming galaxy stellar mass function with redshift in the three CANDELS fields GOODS--South, UDS, and COSMOS for galaxies brighter than $H=26$. Raw, incomplete counts are shown as dashed lines, while solid lines show the corrected counts. The shaded areas correspond to Poissonian errors.}
    \label{FIG:massfunc}
\end{figure}

\begin{table}[htdp]
\setlength{\tabcolsep}{4pt}
\caption{$\log_{10}(M_\ast/{\rm M}_\odot)$ above which our samples are at least $90\%$ complete, for each catalog. \label{TAB:masscomp}}
\begin{center}
\begin{tabular}{lcccccc}
    \hline
    \hline \\[-2.5mm]
Catalog            & $z=0.5$ & $1.0$ & $1.5$  & $2.2$  & $3.0$  & $4.0$ \\
    \hline \\[-2.5mm]
GN                 & $8.9$   & $9.3$ & $9.8$  & $10.1$ & $10.5$ & $10.7$ \\
CANDELS$^{\rm a}$ & $8.3$   & $8.7$ & $9.0$  & $9.4$  & $9.9$  & $10.3$ \\
COSMOS UVISTA      & $9.1$   & $9.6$ & $10.1$ & $10.6$ & $10.9$ & $11.3$ \\
    \hline
\end{tabular}
\end{center}
$^{\rm (a)}$ These values are valid for GOODS--South, UDS, and COSMOS CANDELS, keeping all sources with $H<26$.
\end{table}

The last step before going through the analysis is to make sure that, in each stellar mass bin we will work with, as few galaxies as possible are missed because of our selection criteria. The fact that we built these samples by starting from an NIR selection makes it much simpler to compute the corresponding mass completeness: the stellar mass of a galaxy at a given redshift is indeed well correlated with the luminosity in the selection band (either $H$ or $K_{\rm s}$), as illustrated in Fig.~\ref{FIG:mtol}, the scatter around the correlation being caused by differences of age, attenuation, and to some extent flux uncertainties and $k$-correction. From our sample, we can actually see by looking at this correlation with various bands ($H$, $K_{\rm s}$, and IRAC channels 1 and 2) that this scatter is minimal ($0.14\,{\rm dex}$) when probing the rest-frame $1.7\,\mu{\rm m}$, but it reaches $0.4$ dex in the rest-frame UV ($3500\,\AA$). While this value is of course model dependent, it stresses the importance of having high-quality NIR photometry, especially the {\it Spitzer} IRAC bands (observed $3$--$5\,\mu{\rm m}$).

To estimate the mass completeness, we decided to use an empirical approach, where we do not assume any functional form for the true mass function. Instead, we directly compute the completeness assuming that, at a given redshift, the stellar mass is well estimated by a power law of the luminosity (measured either from the observed $H$ or $K_{\rm s}$ band), i.e., ~$M_\ast = C\,L^\alpha$, plus a Gaussian scatter in log space. We fit this power law and estimate the amplitude of the scatter using the detected galaxies, as shown in Fig.~\ref{FIG:mtol}. Using this model (red solid and dotted lines) and knowing the limiting luminosity in the selection band (orange horizontal lines), we can estimate how many galaxies we miss at a given stellar mass, using, e.g., a Monte Carlo simulation. At a given stellar mass, we generate a mock population of galaxies with uniform redshift distribution within the bin and estimate what would be their luminosity in the selection band by using the above relation and adding a Gaussian scatter to the logarithm of the luminosity. The completeness is then computed as the fraction of galaxies that have a luminosity greater than the limiting luminosity at the considered redshift. We consider our catalogs as ``complete'' when the completeness reaches at least $90\%$.

The same procedure is used on COSMOS UltraVISTA and GOODS--North separately, and the estimated completeness levels are all reported in Table \ref{TAB:masscomp}. We compared the values obtained in GOODS--North with those reported in \cite{pannella2014}, where the completeness is estimated following \cite{rodighiero2010} using a stellar population model. The parameters of the model chosen in \cite{pannella2014} are quite conservative, and their method consistently yields mass limits that are on average $0.3\,{\rm dex}$ higher than ours. In COSMOS UltraVISTA, we obtain values similar to that of \cite{muzzin2013}.

Finally, we build stellar mass functions by simply counting the number of galaxies in bins of redshift and stellar masses in the three CANDELS fields that are $H$-band selected, and normalize the counts by the volume that is probed. These raw mass functions are presented in Fig.~\ref{FIG:massfunc} as dashed lines. Assuming that the counts follow a Schechter-like shape, i.e., rising with a power law toward low stellar mass, the incompleteness of our sample is clearly visible. We then use the estimated completeness (top panel in Fig.~\ref{FIG:mtol}) to correct the stellar mass functions. Here, we limit ourselves to reasonable corrections of at most a factor two in order not to introduce too much uncertainty in the extrapolation. The resulting mass functions are shown as solid lines in Fig.~\ref{FIG:massfunc}, with shaded areas showing the Poisson noise. The obtained mass functions are in good agreement with those already published in the literature \citep[e.g.,][]{ilbert2013}.

\section{Deriving statistical properties of star-forming galaxies}

Because of the limitations of the {\it Herschel} surveys (the result of photometric or confusion noise), we cannot derive robust individual ${\rm SFR}$s for all the sources in our sample (see section \ref{SEC:sfr}). Indeed, the fraction of star-forming galaxies detected in the FIR ranges from 80\% at $M_\ast > 3\times10^{10}\,{\rm M}_\odot$ and $z < 1$, to almost 0\% for $M_\ast < 10^{10}\,{\rm M}_\odot$ and $z > 1$. Above $z=1$, the completeness in FIR detections reaches better than $60\%$ only above $M_\ast = 10^{11} {\rm M}_\odot$ and up to $z=2.5$. Below this mass and above that redshift, the FIR completeness is lower than $20$--$30\%$.

We overcome these limitations by stacking the {\it Herschel} images. Stacking is a powerful and routinely used technique that combines the signal of multiple sources at various positions on the images, known from deeper surveys \citep[see, e.g.,][where it was first applied to FIR images]{dole2006}. This effectively increases the signal to noise ratio of the measurement, allowing us to probe fainter fluxes than can be reached by the usual source extraction. The price to pay is that we lose information about each individual source, and only recover statistical properties of the considered sample. Commonly, this method is used to determine the average flux density of a selected population of objects. We will show in the following that it can also be used to obtain information on the flux \emph{distribution} of the sample, i.e., not only its average flux, but also how much the stacked sources scatter around this average value.

This scatter is crucial information. If we measure an average correlation between ${\rm SFR}$ and $M_\ast$, as has been measured in several other studies at different redshifts, this correlation cannot be called a ``sequence'' if the sources show a large dispersion around it.

Several studies have already measured this quantity. \cite{noeske2007} and \cite{elbaz2007} at $z=1$ reported a $1\sigma$ dispersion in $\log_{10}({\rm SFR})$ of around $0.3\,{\rm dex}$ from {\it Spitzer} MIPS observations of a flux-limited sample. At $z=2$, \cite{rodighiero2011} reported $0.24\,{\rm dex}$, using mostly UV-derived ${\rm SFR}$s, while \cite{whitaker2012-a} reported $0.34\,{\rm dex}$ from {\it Spitzer} MIPS observations. These two studies tested the consistency of their ${\rm SFR}$ estimator on average, but we do not know how they impact the measure of the dispersion. The variation found in these two studies suggests that this is indeed an issue \citep[see for example the discussion in][]{speagle2014}. On the one hand, UV ${\rm SFR}$s have to be corrected for dust extinction. If one assumes a single extinction law for the whole sample, one might artificially reduce the dispersion. On the other hand, MIPS $24\,\mu{\rm m}$ at $z=2$ probes the rest-frame $8\,\mu{\rm m}$. While \cite{elbaz2011} have shown that it correlates well with $L_{\rm IR}$, this same study also demonstrates that it misses a fraction of $L_{\rm IR}$ that is proportional to the distance from the main sequence. This can also have an impact on the measured dispersion.

Here we measure for the first time the ${\rm SFR}$--$M_\ast$ main sequence and its dispersion with a robust ${\rm SFR}$ tracer down to the very limits of the deepest {\it Herschel} surveys to constrain its existence and relevance at higher redshifts and lower stellar masses.

\subsection{Simulated images \label{SEC:simu}}

All the methods described in this section have been extensively tested to make sure that they are not affected by systematic biases or, if they are, to implement the necessary corrections. We conduct these tests on simulated {\it Herschel} images that we set up to be as close as possible to the real images, in a statistical sense. In other words, we reproduce the number counts, the photometric noise, the confusion noise, and the source clustering. The algorithms, the methodology, and the detailed results are described fully in Appendix \ref{APP:simu}.

\subsection{The stacking procedure \label{SEC:stackmethod}}

\begin{figure}
    \centering
    \includegraphics*[width=9cm]{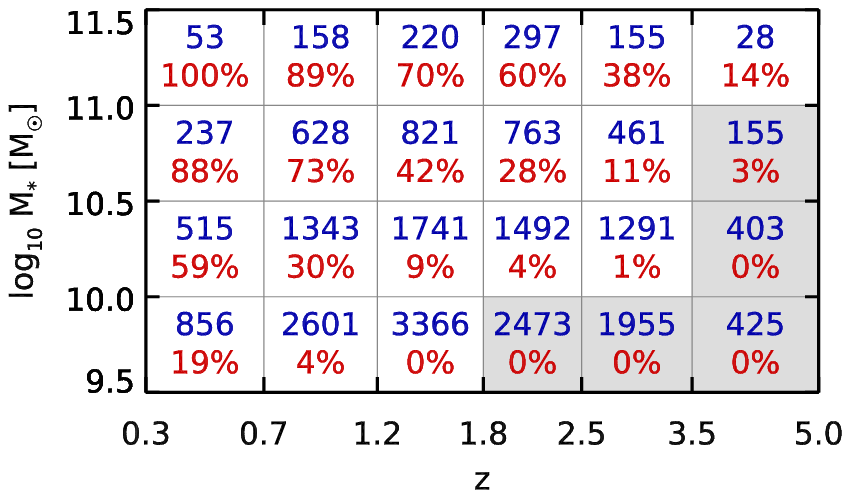}
    \caption{Redshift and stellar mass bins chosen for stacking. We display in each bin (from top to bottom) the total number of star-forming $H$ or $K_{\rm s}$-band galaxies that are stacked in the CANDELS fields, and the fraction of galaxies individually detected with {\it Herschel}. The bins where we do not detect any stacked signal are shown with a gray background.}
    \label{FIG:stack_bins}
\end{figure}

\begin{figure}
    \centering
    \includegraphics*[width=9cm]{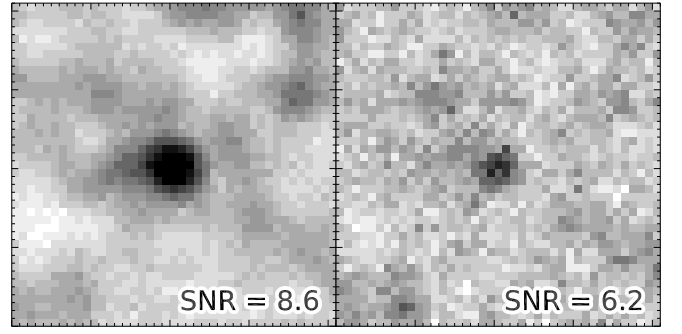}
    \caption{Stack of 155 galaxies at $z=3$ and $\log_{10}(M_\ast/{\rm M}_\odot)=11.3$ in the SPIRE $250\,\mu{\rm m}$ images. {\bf Left:} mean flux image, {\bf Right:} ${\rm MAD}$ dispersion image. Measuring the dispersion is more difficult than measuring the flux, since the signal is always fainter. $38\%$ of these galaxies are individually detected by {\it Herschel}, and only $25\%$ are detected in the SPIRE $250\,\mu{\rm m}$ channel.}
    \label{FIG:stack}
\end{figure}

\begin{figure*}
    \centering
    \includegraphics*[width=18cm]{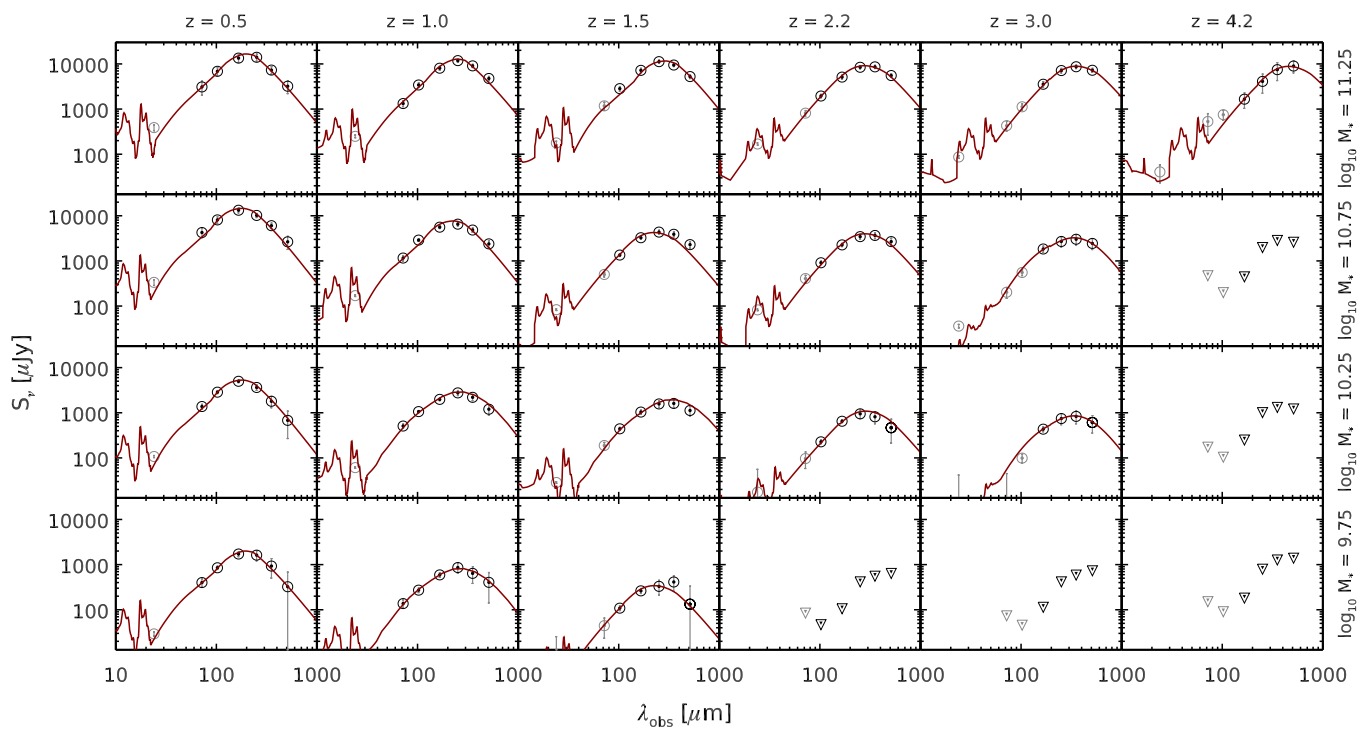}
    \caption{Stacked SEDs of our star-forming mass-selected samples in bins of redshift (horizontally) and stellar mass (vertically). Stacked points are shown as empty circles, and the best-fit CE01 template is shown as the solid red curve. Gray data points were not used in the fit because they are probing rest-frame wavelengths below $30\,\mu{\rm m}$. The data points have been corrected for the contribution of galaxy clustering (see Table \ref{TAB:clustering}). In the bins where the signal is too low (typically $<5\sigma$), we plot $3\sigma$ upper limits as downward triangles.}
    \label{FIG:seds}
\end{figure*}

We divide our star-forming galaxy sample into logarithmic bins of stellar mass and redshift, as shown in Fig.~\ref{FIG:stack_bins}, to have a reasonable number of sources in each bin. We then go to the original {\it Herschel} images of each field and extract $N \times N$ pixel cutouts around each source in the bin, thus building a pixel cube. We choose $N=41$ for all {\it Herschel} bands, which is equivalent to $8$ times the full-width at half maximum (FWHM) of the PSF, and $N=61$ for {\it Spitzer} MIPS ($13\times{\rm FWHM}$), as a substantial fraction of the {\it Spitzer} flux is located in the first Airy ring. Since the maps were reduced in a consistent way across all the CANDELS fields, we can safely merge together all the sources in a given bin, allowing us to go deeper while mitigating the effects of cosmic variance.

In parallel, we also stack the sources of the COSMOS UltraVISTA catalog in the wider but shallower FIR images. These stacked values are mostly used as consistency checks, since they do not offer any advantage over those obtained in the CANDELS fields: the shallow {\it Herschel} exposure is roughly compensated by the large area, but the mass completeness is much lower.

In the literature, a commonly used method consists of stacking only the undetected sources on the \emph{residual} maps, after extracting sources brighter than a given flux threshold. This removes most of the contamination from bright neighbors, and thus lowers the confusion noise for the faint sources, while potentially introducing a bias that has to be corrected. Detected and stacked sources are then combined using a weighted average \citep[as in, e.g.,][]{magnelli2009}. We prefer here to treat both detected and undetected sources homogeneously in order not to introduce any systematic error tied to either the adopted flux threshold or the details of the source extraction procedure. Although simpler, this procedure nevertheless gives accurate results when applied to our simulated images. Indeed, the contribution of bright neighbors is a random process: although it is clear that each source suffers from a varying level of contamination, statistically they are all affected in the same way. In other words, when a sufficient number of sources are stacked, the contribution of neighbors tends to average out to the same value $\mu_{\rm gal}$ on all pixels, which is the contribution of galaxies to the Cosmic InfraRed Background (CIRB). But this is only true in the absence of galaxy clustering \citep{bethermin2010-a}. When galaxies are clustered, there is an increased probability of finding a neighbor close to each stacked galaxy \citep{chary2010}, so that $\mu_{\rm gal}$ will be larger toward the center of the stacked image. \cite{kurczynski2010} proposed an alternative stacking technique \citep[implemented by][in the {\tiny SIMSTACK} code]{viero2013} that should get rid of most of this bias, and that consists of simultaneously fitting for the flux of all sources within a given volume (i.e., in a given redshift bin). It is however less versatile, and in particular it is not capable of measuring flux dispersions. B\'ethermin et al. (2014, submitted) also show that is can suffer from biases coming from the incompleteness of the input catalog.

The next step is to reduce each cube into a single image by combining the pixels together. There are several ways to do this, the two most common being to compute the mean or the median flux of all the cutouts in a given pixel. The advantage of the mean stacking is that it is a linear operation, thus one can exactly understand and quantify its biases \citep[e.g.,][]{bethermin2010-a}. More specifically, it can be shown that the mean stacked value corresponds to the covariance between the input source catalog and the map \citep{marsden2009}. Median stacking, on the other hand, has the nice property of naturally filtering out bright neighbors and catastrophic outliers and thus produces cleaner flux measurements. On the down side, we show in Appendix \ref{APP:medmean} that this measurement is systematically biased in a nontrivial way \citep[see also][]{white2007}. Correcting for this bias requires some assumptions about the stacked flux distribution, e.g., the dispersion. Since this is a quantity we want to measure, we prefer to use mean over median stacking. An example of a mean stacked cutout from the SPIRE $250\,\mu{\rm m}$ images is shown in Fig.~\ref{FIG:stack} (left). However, in two bins at low masses and high redshifts ($z=1.5$ and $\log_{10}(M_\ast/{\rm M}_\odot) = 9.75$, as well as $z=3.0$ and $\log_{10}(M_\ast/{\rm M}_\odot) = 10.25$), the mean stacked fluxes have signal to noise ratios that are too low and thus cannot be used, while the median stacked fluxes are still robustly measured. To extend our measurement of the main sequence ${\rm SFR}$, we allow ourselves to use the median stacked fluxes in these particular bins only. This is actually a regime where we expect the median stacking to most closely measure the mean flux (see Appendix \ref{APP:medmean}), hence this should not introduce significant biases. Lastly, we are interested in the \emph{mode} of the main sequence, which is not strictly speaking the mean ${\rm SFR}$ we measure. We calibrated the difference between those two quantities with our simulations, and in all the following we refer to the ${\rm SFR}$ of the main sequence as the mode of the distribution. For example, for a log-normal distribution of $\sigma = 0.3\,{\rm dex}$, this difference is about $0.1\,{\rm dex}$.

To measure the stacked flux, we choose to use PSF fitting in all the stacked bands. In all fields, we use the same PSFs as those used to extract the photometry of individual objects, and apply the corresponding aperture corrections. This method assumes that the stacked image is a linear combination of: 1) a uniform background; and 2) the PSF of the instrument, since none of our sources is spatially resolved. The measured flux is then obtained as the best-fit normalization factor applied to the PSF that minimizes the residuals. In practice, we simultaneously fit both the flux and the background within a fixed aperture whose radius is $0.9$ times the FWHM of the PSF. The advantage of this choice is that although we use less information in the fit, the background computed this way is more local, and the flux measurement is more robust against source clustering. Indeed, the amplitude of the clustering is a continuous function of angular distance: although a fraction of clustered sources will fall within a radius that is much smaller than the FWHM of the PSF and will bias our measurements no matter what, the rest will generate signal over a scale that is larger than the PSF itself, such that it will be resolved. Estimating the background within a small aperture will therefore remove the contribution of clustering coming from the largest scales.

We quantify the expected amount of flux boosting due to source physical clustering using our simulated maps. We show in Appendix \ref{APP:clustering} that it is mostly a function of beam size, i.e., there is no effect in the PACS bands but it can boost the SPIRE fluxes by up to $25\%$ at $500\,\mu{\rm m}$. We also compare our flux extraction method to other standard approaches and show that it does reduces the clustering bias by a factor of $1.5$ to $2.5$, while also producing less noisy flux measurements. The value of $0.9\times{\rm FWHM}$ was chosen to get the lowest clustering amplitudes and flux uncertainties.

To obtain an estimate of the error on this measure, we also compute the standard deviation $\sigma_{\rm RES}$ of the residual image (i.e., the stacked image minus the fitted source) and multiply it by the PSF error scaling factor
\begin{equation}
    \sigma_{\rm IMG} = \sigma_{\rm RES} \times \left(|P^2| - \frac{|P|^2}{N_{\rm pix}}\right)^{-1/2}\,, \label{EQ:psferror}
\end{equation}
where $N_{\rm pix}$ is the number of pixels that are used in the fit, $|P|$ is the sum of all the pixels of the PSF model within the chosen aperture, and $|P^2|$ the sum of the squares of these pixels. This is the formal error on the linear fit performed to extract the flux (i.e., the square root of the diagonal element corresponding to the PSF in the covariance matrix), assuming that all pixels are affected by a similar uncorrelated Gaussian error of amplitude $\sigma_{\rm RES}$. In practice, since the PSFs that we use are all sampled by roughly the same number of pixels (approximately two times the Nyquist sampling), this factor is always close to $0.5$ divided by the value of the central pixel of the PSF. Intuitively, this comes from the fact that the error on the measured flux is the combination of the error on all the pixels that enter in the fit, weighted by the amplitude of the PSF. It is thus naturally lower than the error on one single pixel. In other words, using PSF fitting on these stacks allows for  measuring fluxes that are twice as faint as those obtained when using only the central pixel of the image. Simple aperture photometry yields $\sigma_{\rm APER} = \sigma_{\rm RES} \times \left(\!\sqrt{N_{\rm pix} + {N_{\rm pix}}^2/N_{\rm bg}}\right)/|P|$, where $N_{\rm bg}$ is the number of pixels used to estimate the background (e.g., within an annulus around the source). If $N_{\rm bg}$ is sufficiently large ($\gtrsim N_{\rm pix}$), this error is lower than that obtain with our PSF fitting technique because the background is estimated independently of the flux. The price to pay is that this background is not local, hence the aperture flux will be most sensitive to clustering. Finally, if there is no clustering, PSF fitting will give the lowest errors of all methods, provided the full PSF is used in the fit. The optimal strategy is therefore always to use PSF fitting, varying the aperture within which the fit is performed depending on the presence of clustering.

To be conservative, we compute an alternative error estimate using bootstrapping: we randomly discard half of the sources, stack the remaining ones, measure the stacked flux, and repeat this procedure $100$ times. The error $\sigma_{\rm BS}$ is then computed as the standard deviation of the measured flux in these $100$ realizations, divided by $\sqrt{2}$, since we only work with half of the parent sample. Using our simulated images, we show in Appendix \ref{APP:errors} that accurate error estimates are obtained by keeping the maximum error between $\sigma_{\rm IMG}$ and $\sigma_{\rm BS}$. For the SPIRE bands, however, the same simulations show that both error estimates are systematically underestimated and need to be corrected by a factor of $\sim1.7$. We demonstrate in Appendix \ref{APP:errors} that this comes from the fact that the error budged in the SPIRE bands is mostly generated by the random contribution of nearby sources rather than instrumental or shot noise. In this case, the error on each pixel is largely correlated with that of its neighbors, and the above assumptions do not hold.

We apply the above procedure to all the redshift and stellar mass bins of Fig.~\ref{FIG:stack_bins} and stack all the MIR to FIR images, from MIPS $24\,\mu{\rm m}$ to SPIRE $500\,\mu{\rm m}$. Using the measured mean fluxes, we build effective SEDs\footnote{These SEDs are effective in the sense that they are not necessarily the SED of the average galaxy in the sample: they are potentially broadened by the range of redshifts and dust temperatures of the galaxies in the stacked samples. In practice, we checked that the broadening due to the redshift distribution is negligible, and the photometry is well fitted by standard galaxy templates, as can be seen in Fig.~\ref{FIG:seds}.} in each bin, shown in Fig.~\ref{FIG:seds}. We fit the {\it Herschel} photometry with CE01 templates, leaving the normalization of each template free and keeping only the best-fit, and obtain the mean $L_{\rm IR}$. As for the individual detections, we do not use the photometry probing rest-frame wavelengths below $30\,\mu{\rm m}$ (see section \ref{SEC:sfr}). The MIPS $24\,\mu{\rm m}$ photometry is used as a check only. Converting the measured $L_{\rm IR}$ to ${\rm SFR}_{\rm IR}$ with the \cite{kennicutt1998-a} relation and adding the mean observed ${\rm SFR}_{\rm UV}$ (non-dust-corrected contribution), we obtain the mean total ${\rm SFR}$ in each bin.

\subsection{Measuring flux dispersion with \emph{scatter stacking}}

To measure the flux dispersion, we introduce a new method. The idea is to come back to the pixel cube and build a \emph{dispersion} image by measuring the scatter of each pixel around its average value. Stacked pixels away from the center measure the background fluctuations (the combination of photometric noise and random contribution from nearby sources), while pixels in the central region show enhanced dispersion due to flux heterogeneities in the stacked population, as in Fig.~\ref{FIG:stack}. In particular, if all the stacked sources had the same flux, the dispersion map would be flat.

Again, this can be achieved in different ways. Computing the standard deviation of pixels is the most straightforward approach, but it suffers from similar issues as mean stacking with respect to bright neighbor contamination, in a more amplified manner because pixels are combined in quadrature. Our simulations also show that this method is not able to reliably measure high dispersion values. We thus use the median absolute deviation (MAD), which is more effective in filtering out outliers while providing the same information.

\begin{figure}
    \centering
    \includegraphics*[width=8cm]{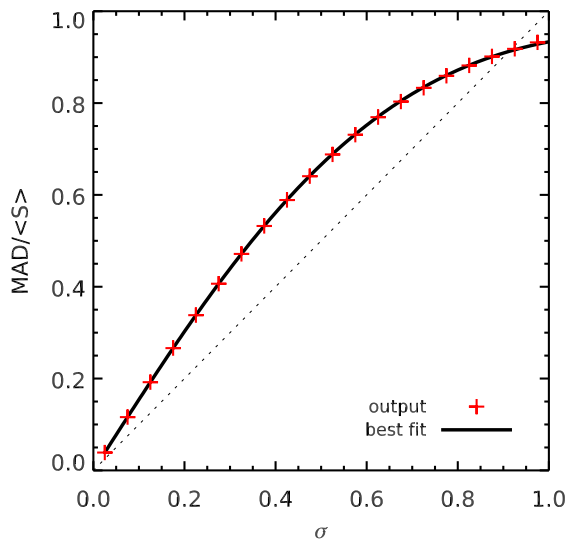}
    \caption{Median absolute deviation (${\rm MAD}$) computed by solving Eq.~\ref{EQ:MAD_prob} numerically for a log-normal distribution of $\left<S\right> = 1$ as a function of the chosen $\sigma$. The solid line is the best-fit of Eq.~\ref{EQ:MAD_log} to the numerical solutions, and the dashed line is the one-to-one correlation.}
    \label{FIG:sigma_mad}
\end{figure}

The ${\rm MAD}$ is formally defined as the half-width of the range that is centered on the median flux $\left<S\right>$ and contains $50\%$ of the whole sample. In other words
\begin{equation}
    \phi\left(\left<S\right> + {\rm MAD}\right) - \phi\left(\left<S\right> - {\rm MAD}\right) = \frac{1}{2}\,,
    \label{EQ:MAD_prob}
\end{equation}
where $\phi$ is the cumulative probability distribution function of the flux.

To interpret this value in terms of more common dispersion indicators, we will convert the ${\rm MAD}$ to a $\log$-dispersion $\sigma$ assuming that fluxes follow a Gaussian distribution in $\log_{10}(S)$, i.e., a log-normal distribution in $S$. There are two reasons that justify this choice: 1) it allows for direct comparison of our measured dispersions to the data from literature that quote standard deviations of $\log_{10}({\rm SFR})$; and 2) log-normal distribution are good models for describing ${\rm sSFR}$ distributions in the regimes where we can actually detect individual sources \citep[see, e.g.,][and also section \ref{SEC:indiv}]{rodighiero2011,sargent2012,gladders2013,guo2013}. For this family of distributions,
\begin{equation}
    \phi(S) = \frac{1}{2}\,{\rm erfc}\left(-\frac{\log_{10}\left(\frac{S}{\left<S\right>}\right)}{\,\sqrt{2}\,\sigma}\right)\,,
    \label{EQ:lognormal_cumprob}
\end{equation}
where ${\rm erfc}$ is the complementary error function. In this case there is no analytical solution to Eq.~\ref{EQ:MAD_prob}, but it can be solved numerically. It turns out that one can relate the ${\rm MAD}$ and $\left<S\right>$ directly to $\sigma$ (see Fig.~\ref{FIG:sigma_mad}) {\it via} the following equation, which was fit on the output of the numerical analysis\footnote{This analysis was performed with \emph{Mathematica}.} (for $\sigma \in [0.05, 1.0]\,{\rm dex}$):
\begin{equation}
    \frac{{\rm MAD}}{\left<S\right>} \simeq \frac{1.552\,\sigma}{1 + 0.663\,{\sigma}^2}\,,
    \label{EQ:MAD_log}
\end{equation}
with a maximum absolute error of less than $0.01$. This relation can, in turn, be inverted to obtain $\sigma$. Defining the ``normalized'' median absolute deviation ${\rm NMAD} \equiv {\rm MAD}/\left<S\right>$, and only keeping the positive solution of Eq.~\ref{EQ:MAD_log}, we obtain
\begin{equation}
    \sigma \simeq \frac{1.171}{{\rm NMAD}}\,\left(1 - \,\sqrt{1 - \left(\frac{{\rm NMAD}}{0.953}\right)^2}\right)\,.
    \label{EQ:log_MAD}
\end{equation}

Therefore, measuring the ${\rm MAD}$ allows us to obtain the intrinsic log-normal flux dispersion $\sigma$ of the stacked sample. To do so, we perform PSF fitting on the squared images (since the dispersion combines quadratically with background noise) and fit a constant background noise plus the square of the PSF on all the pixels within a fixed radius of $0.6\times{\rm FWHM}$. Here we do not use the same $0.9\,\times{\rm FWHM}$ cut as for the flux extraction, since the MAD does not fully preserve the shape of the PSF when its pixels are low in signal to noise (see below). We thus restrain ourselves to a more central region to prevent being dominated by these faint pixels. Again, this value was chosen using the simulated maps in order to produce the least biased and least uncertain measurements.

\begin{figure}
    \centering
    \includegraphics*[width=9cm]{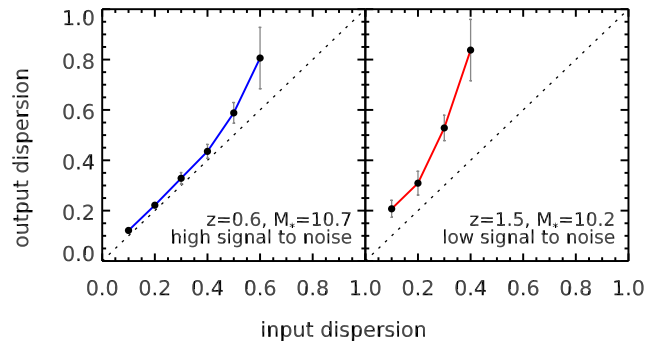}
    \caption{Correction procedure for the measured dispersion. Each point is a simulated dispersion measurement with a different input value. Error bars show the scatter observed among the $20$ realizations. The dashed line shows the one-to-one relation. The plots display two examples of simulated dispersions for the PACS $100\,\mu{\rm m}$ band, at $z=0.6$ for $M_\ast = 3\times10^{10}$ (left panel), and at $z=1.5$ for $M_\ast = 2\times10^{10}\,{\rm M}_\odot$ (right panel). These bins were chosen to illustrate the two regimes of high and low signal to noise, respectively.}
    \label{FIG:dispcor}
\end{figure}

Even then, the dispersion measured with this method is slightly biased toward higher values, but this bias can be quantified and corrected in a self-consistent way with no prior information using Monte Carlo simulations. For each source in the stack, we extract another cutout at a random position in the map. We then place a fake source at the center of each random cutout, whose flux follows a log-normal distribution of width $\sigma_{\rm MC}$, and with a mean flux equal to that measured for the real sources. We apply our scatter stacking technique to measure the dispersion on the resulting mock flux cube, and compare it to $\sigma_{\rm MC}$. We repeat this procedure for different values of $\sigma_{\rm MC}$ (from $0.1$ to $0.7\,{\rm dex}$), and derive the relation between the intrinsic and measured dispersion. Examples are shown in Fig.~\ref{FIG:dispcor}. To average out the measurement error, we repeat this procedure $20$ times for each value of $\sigma_{\rm MC}$. In practice, this correction is mostly negligible, except for the lowest measured mass bins at any redshift where it reaches up to $0.1\,{\rm dex}$.

\subsection{${\rm SFR}$ dispersion from \emph{scatter stacking}}

\begin{figure}
    \centering
    \includegraphics*[width=9cm]{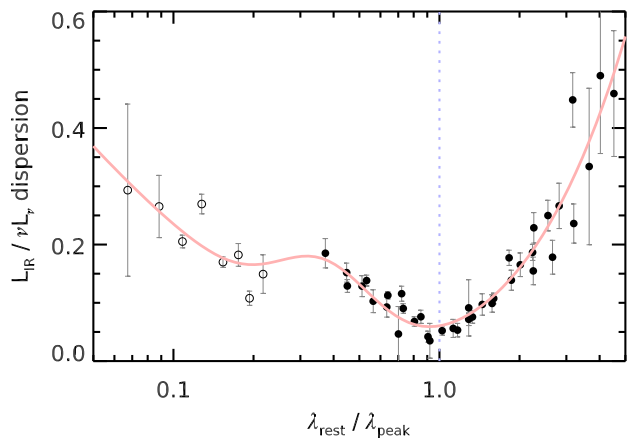}
    \caption{Dispersion of the ratio $L_{\rm IR} / \nu L_\nu$ as a function of wavelength in bins of redshift and for the five {\it Herschel} bands in the four CANDELS fields. The wavelength is normalized here to the ``peak'' wavelength, where the FIR SED in $\nu L_\nu$ reaches its maximum (calibrated from our stacked SEDs, Fig.~\ref{FIG:seds}). The $L_{\rm IR}$ is computed by fitting all the available {\it Herschel} bands (we require a minimum of three) together with CE01 templates, while $\nu L_\nu$ is the flux in a single {\it Herschel} band converted to rest-frame luminosity. Open symbols denote measurements where $\nu L_\nu$ comes from MIPS $24\,\mu{\rm m}$. Error bars come from simple bootstrapping. The contribution of photometric errors was statistically removed. The red line shows a fit to the data points to guide the eye.}
    \label{FIG:lirdisp}
\end{figure}

The procedure described in the previous section allows us to measure the log-normal \emph{flux} dispersion, while we are interested in the dispersion in ${\rm SFR}$.

The first step is to obtain the $\log_{10}(L_{\rm IR})$ dispersion $\sigma_{\rm IR}$. Using detected sources, we observe that the dispersion in $L_{\rm IR}$ of a population of galaxies having the same flux at a given redshift depends on the rest-frame wavelength probed, as illustrated in Fig.~\ref{FIG:lirdisp}. The data points in this figure are produced by looking at multiple bins of redshift, and measuring the scatter of the correlation between $L_{\rm IR}$, measured by fitting all available FIR bands, and the flux in each {\it Herschel} band converted to rest-frame luminosity ($\nu\,L_\nu$). By spanning a range of redshift, the five {\it Herschel} bands will probe a varying range of rest-frame wavelengths, allowing us to observe the behavior of the $L_{\rm IR}$ scatter with rest-frame wavelength. The smaller dispersions are found at wavelengths close to the peak of the SED, in which case the dispersion drops as low as $0.05\,{\rm dex}$. This is due to galaxies showing a variety of effective dust emissivities and temperatures that both influence the shape of the FIR SED, respectively longward and shortward of the peak.

Therefore, to obtain $\sigma_{\rm IR}$, we simply measure the flux dispersion of the {\it Herschel} band that is the closest to the peak. We thus first measure the peak wavelength $\lambda_{\rm peak}$ from the stacked SEDs (Fig.~\ref{FIG:seds}), and interpolate the measured log-normal flux dispersions at $\lambda_{\rm peak}$. By construction, this also tends to select {\it Herschel} measurements with the highest signal to noise ratio.

One then has to combine the dispersion in $L_{\rm IR}$ with that in $L_{\rm UV}$, since we combine both tracers to derive the total ${\rm SFR}$. This is not straightforward, as the two quantities are not independent (i.e., at fixed ${\rm SFR}$, more attenuated objects will have higher $L_{\rm IR}$ and lower $L_{\rm UV}$). In particular, we see on individual detections that the dispersion of ${\rm SFR} = {\rm SFR}_{\rm IR} + {\rm SFR}_{\rm UV}$ is actually \emph{lower} than that of ${\rm SFR}_{\rm IR}$ alone.

To address this issue, we choose to work directly on ``SFR stacks''. First, we use our observed FIR SEDs to derive $L_{\rm IR}$ monochromatic conversion factors for all bands in each of our redshift and stellar mass bins. Second, in each stacked bin, we convert all cutouts to ${\rm SFR}_{\rm IR}$ units, using the aforementioned conversion factor and the \cite{kennicutt1998-a} relation. Third, we add to each individual cutout an additional amount of ${\rm SFR}$ equal to the non-dust-corrected ${\rm SFR}_{\rm UV}$, as a centered PSF. Finally, to correct for the smearing due to the width of the redshift and mass bins, we also use our observed relation between mass, redshift, and ${\rm SFR}$ (given below in Eq.~\ref{EQ:sfrms}) and normalize each cutout to the reference mass and redshift of the sample by adding ${\rm SFR}_{\rm MS}(z_{\rm ref}, M_{\ast, \rm ref}) - {\rm SFR}_{\rm MS}(z, M_\ast)$. This last step is a small correction: it reduces the measured dispersion by only $0.02$ to $0.03\,{\rm dex}$.

We stack these cutouts and again run the dispersion measurement procedure, including the bias correction. Interpolating the measured dispersions in the five {\it Herschel} bands at $\lambda_{\rm peak}$ as described earlier, we obtain $\sigma_{{\rm SFR}}$. As expected, the difference between the flux dispersion at the peak of the SED and the ${\rm SFR}$ dispersion is marginal, except for the lowest mass bins where it can reach $0.05\,{\rm dex}$. This is mainly caused by the increasing contribution of the escaping UV light to the total ${\rm SFR}$, as ${\rm SFR}_{\rm IR}/{\rm SFR}_{\rm UV}$ approaches unity in these bins.

A remaining bias that we do not account for in this study is the impact of errors on the photo-$z$s and stellar masses. As pointed out in section \ref{SEC:zmstar}, the measured few percent accuracy on the photo-$z$s only applies to the bright sources, and we do not know the reliability of the fainter sources. We measure statistical uncertainties on both these quantities, but this does not take systematic errors coming from the library or gaps in the photometry into account. Intuitively, one can expect these errors to increase the dispersion, but this would be true only if the true error was purely random. It could be that our SED fitting technique is too simplistic in assuming a universal IMF, metallicity, and SFH functional form for all galaxies, and as such erases part of the diversity of the population. This could in turn \emph{decrease} the measured dispersion \citep[see discussion in][]{reddy2012-a}. It is therefore important to keep in mind that our measurement is tied to the adopted modeling of stellar mass.

\section{Results}

\subsection{The ${\rm SFR}$ of main-sequence galaxies}

\begin{figure}
    \centering
    \includegraphics*[width=9cm]{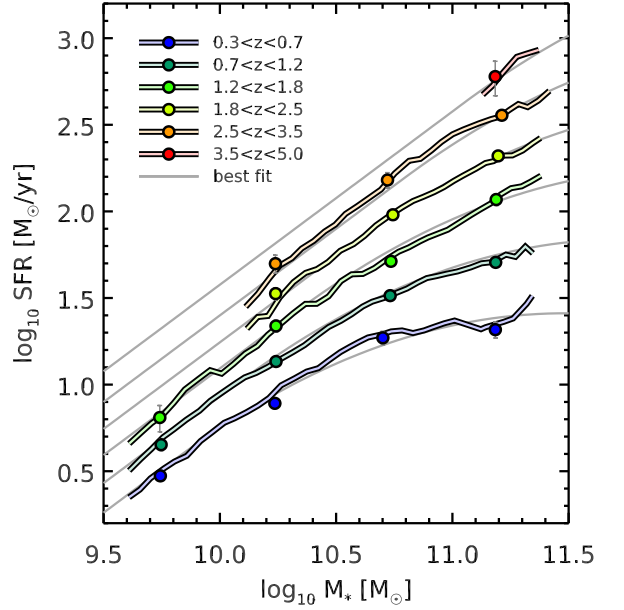}
    \caption{Evolution of the average ${\rm SFR}$ of star-forming galaxies with mass and redshift. Our results from stacking are shown as colored filled circles, the colors corresponding to the different redshifts as indicated in the legend. We complement these measurements by stacking sliding bins of mass (see text) for visualization purposes only to better grasp the mass dependence of the ${\rm SFR}$. In the background, we show as light gray curves our best-fit relation for the main sequence (Eq.~\ref{EQ:sfrms}).}
    \label{FIG:sfrms_cmp}
\end{figure}

\begin{figure*}
    \centering
    \includegraphics*[width=9cm]{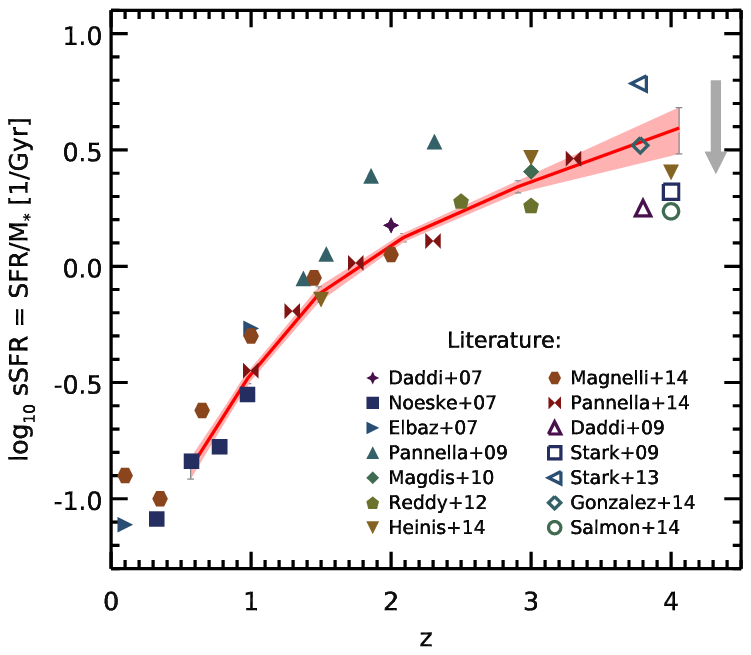}
    \includegraphics*[width=9cm]{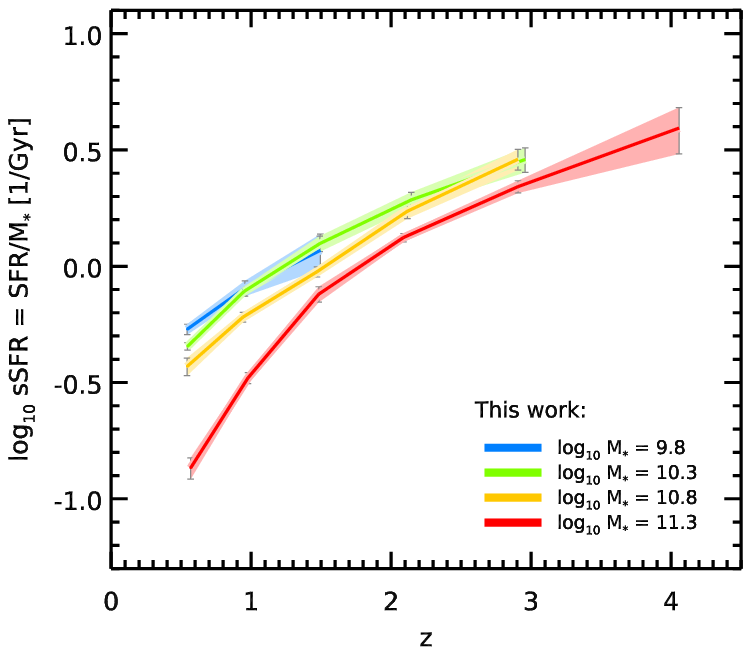}
    \caption{Evolution of the average ${\rm sSFR}$ of star-forming galaxies with redshift. {\bf Left:} comparison of our results at $M_\ast = 2\times10^{11}\,{\rm M}_\odot$ (red curve) to published values in the literature (filled and open symbols). Filled symbols compile various results that were derived from mass-complete samples with ${\rm SFR}$s computed either from the IR \citep{daddi2007-a,noeske2007,elbaz2007,magdis2010-b,reddy2012-a,heinis2014,magnelli2014,pannella2014} or the radio \citep{pannella2009-a,pannella2014}. When possible, these were rescaled to a common stellar mass of $2\times10^{11}\,{\rm M}_\odot$ using the corresponding published ${\rm SFR}$--$M_\ast$ relations. Results from stacking have been corrected by $-0.1\,{\rm dex}$ to reach the mode of the main sequence (see discussion in section \ref{SEC:stackmethod}). Open symbols show results from the literature that make use of the Lyman break selection technique (LBGs) and where the ${\rm SFR}$s are obtained from the UV light alone \citep{daddi2009,stark2009,stark2013,gonzalez2014,salmon2014}. These samples are mostly composed of galaxies of much lower stellar mass, typically $3\times10^{9}\,{\rm M}_\odot$, so the extrapolation to $10^{11}\,{\rm M}_\odot$ is more uncertain. We therefore simply quote the published values. The gray arrow shows how the open symbols would move if we were to apply a mass correction assuming the $z=4$ main sequence slope of \cite{bouwens2012}. When necessary, data from the literature have been converted to a Salpeter IMF. {\bf Right:} same figure showing our other stacked mass bins with different colors.}
    \label{FIG:ssfrz}
\end{figure*}

The first results we present concern the evolution of the main sequence with redshift, as well as its dependence on stellar mass. In section \ref{SEC:ssfrz} we start by describing the redshift evolution of the ${\rm sSFR} \equiv {\rm SFR}/M_\ast$, and we then address the mass dependence of the slope of the main sequence in section \ref{SEC:ssfrm}.

These results are summarized in Fig.~\ref{FIG:sfrms_cmp} where, for the sake of visualization, we also run our full stacking procedure on sliding bins of mass, i.e., defining a fine grid of $M_\ast$ and selecting galaxies within mass bins of constant logarithmic width of $0.3\,{\rm dex}$. The data points are not independent anymore, since a single galaxy is included in the stacked sample of multiple neighboring points, but this allows us to better grasp the evolution of the main sequence with mass. These ``sliding averages'' of the ${\rm SFR}$ are displayed as solid colored lines, while the points obtained with regular mass bins are shown as filled circles.

By fitting these points (filled circles only), we parametrize the ${\rm SFR}$ of main-sequence galaxies with the following formula, defining $r \equiv \log_{10}(1+z)$ and $m \equiv \log_{10}(M_\ast / 10^{9}\,{\rm M}_\odot)$:
\begin{eqnarray}
\log_{10}({\rm SFR}_{\rm MS} [{\rm M}_\odot / {\rm yr}]) =  m - m_0 + a_0\,r \hspace{2.5cm} \nonumber \\
\hspace{0.9cm} - a_1 \, \big[{\rm max}(0, m - m_1 - a_2\,r)\big]^2\,,
\label{EQ:sfrms}
\end{eqnarray}
with $m_0 = 0.5 \pm 0.07$, $a_0 = 1.5 \pm 0.15$, $a_1 = 0.3 \pm 0.08$, $m_1 = 0.36 \pm 0.3$ and $a_2 = 2.5 \pm 0.6$. The choice of this parametrization is physically motivated: we want to explicitly describe the two regimes seen in Fig.~\ref{FIG:sfrms_cmp} and explored in more detail in section \ref{SEC:ssfrm}, namely a sequence of slope unity whose normalization increases with redshift (first terms), and a ``bending'' that vanishes both at low masses and high redshifts (last term). The precise functional form however is arbitrary, and was chosen as the simplest expression that accurately reproduces the bending behavior. This ${\rm SFR}$ will be used in the following as a reference for the locus of the main sequence.

\subsection{Redshift evolution of the ${\rm sSFR}$: the importance of sample selection and dust correction \label{SEC:ssfrz}}

We show in Fig.~\ref{FIG:ssfrz} the evolution of ${\rm sSFR}$ ($\equiv {\rm SFR}/M_\ast$) as a function of both redshift and stellar mass. Our results at $z\leq3$ are in good agreement with previous estimates from the literature, showing the dramatic increase of the ${\rm sSFR}$ with redshift. At $z=4$, we still measure a rising ${\rm sSFR}$, reaching $5\,{\rm Gyr}^{-1}$, i.e., a mass doubling timescale of only $200\,{\rm Myr}$.

At this redshift, however, our measurement is substantially higher than UV-based estimates \citep{daddi2009,stark2009}. More recent results \citep{bouwens2012,stark2013,gonzalez2014} seem to be in better agreement, but it is important to keep in mind that these studies mostly focus on relatively low mass galaxies, i.e., typically $3\times10^9\,{\rm M}_\odot$. Therefore the quoted ${\rm sSFR}$ values only formally apply to galaxies in this range, i.e., to galaxies a factor of $10$ to $100$ times less massive than those in our sample. Extrapolating their measurements to match the mass range we are working with requires that we know the slope of the ${\rm sSFR}$--$M_\ast$ relation. In their study, \cite{bouwens2012} measured this slope from $M_\ast = 10^8$ to $10^{10}\,{\rm M}_\odot$ at $z=4$ and found it to be around $-0.27$. Assuming that this holds for all masses, this means that we should reduce the ${\rm sSFR}$ by about $0.4\,{\rm dex}$ to be able to compare it directly to our result. This is illustrated by the gray arrow in Fig.~\ref{FIG:ssfrz}.

Previous observations of the ${\rm sSFR}$ ``plateau'' \citep{daddi2009} could be the consequence of two key issues. First, selection effects: these studies are based either on Lyman break galaxies (LBGs) or rest-frame FUV-selected samples that, while less prone to lower redshift contaminants, are likely to miss highly attenuated and thus highly star-forming galaxies. Our sample is mass-complete, so we do not suffer from such biases. Second, failure of dust extinction correction: UV-based ${\rm SFR}$ estimates are plagued by uncertainties in dust attenuation. Most studies rely on observed correlations between UV SED features and dust attenuation that are calibrated in the local Universe, such as the ${\rm IRX}$--$\beta$ relation \citep{meurer1999}. Recent studies tend to show that these correlations are not universal and evolve with redshift, possibly due to subsolar metallicity \citep{castellano2014}, ISM conditions, or dust geometry \citep[][]{oteo2013,pannella2014}.

\subsection{Mass evolution of the ${\rm SFR}$ and varying slope of the main sequence \label{SEC:ssfrm}}

It is also worth noting the dependence of the ${\rm SFR}$ on stellar mass from Fig.~\ref{FIG:sfrms_cmp}. Low mass bins ($M_\ast < 3\times10^{10}\,{\rm M}_\odot$) are well fit with a slope of unity. Many studies have reported different values of this slope, ranging from $0.4$ to unity \citep{brinchmann2004,noeske2007,elbaz2007,daddi2007-a,santini2009,pannella2009-a,rodighiero2011}. A slope of unity can be interpreted as a signature of the universality of the star formation process, since it implies a constant star formation timescale $\tau \equiv 1/{\rm sSFR}$ at all stellar masses, with $M_\ast(t) \sim \exp(t/\tau)$. As suggested by \cite{peng2010}, it is also a necessary ingredient for explaining the observed shape invariance of the stellar mass function of star-forming galaxies.

We find however that the ${\rm SFR}$ of the highest mass bin ($M_\ast \sim 2\times10^{11}\,{\rm M}_\odot$) falls systematically below the value expected for a linear relation, effectively lowering the high mass slope of the ${\rm SFR}$--$M_\ast$ relation to $0.8$ at high redshift, down to an almost flat relation at $z=0.5$. Other studies obtain similar ``broken'' shapes for the ${\rm SFR}$--$M_\ast$ sequence \citep{rodighiero2010,whitaker2012-a,magnelli2014}. Our results are also in very good agreement with \cite{whitaker2014}, who used a very similar approach, albeit only using MIPS $24\,\mu{\rm m}$ for stacking.

The reason for this bending of the slope is still unknown. \cite{abramson2014} showed that the relation between the disk mass $M_{\rm disk}$ and ${\rm SFR}$ has a slope close to one with no sign of bending at $z\simeq0$, suggesting that the bulge plays little to no role in star formation. We will investigate if this explanation holds at higher redshifts in a forthcoming paper.

\subsection{Mass evolution of the ${\rm SFR}$ dispersion around the main sequence \label{SEC:sfrdisp}}

\begin{figure}
    \centering
    \includegraphics*[width=9cm]{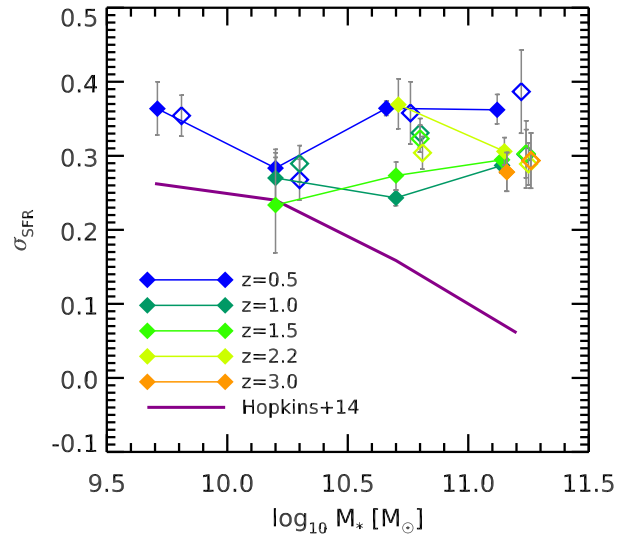}
    \caption{Evolution of the $\log_{10}({\rm SFR})$ dispersion as a function of both redshift and stellar mass. Each color is showing a different redshift bin. Filled symbols show the result of scatter stacking, while open symbols show the dispersion estimated from individual {\it Herschel} detections above the main sequence (see text). The open symbols have been shifted up by $0.1\,{\rm dex}$ in mass for clarity. Errors are from bootstrapping in all cases. We compare these to the typical scatter of the SFHs in the numerical simulation of \cite{hopkins2014} shown as a solid purple line.}
    \label{FIG:msdisp}
\end{figure}

We present in Fig.~\ref{FIG:msdisp} the evolution of the measured ${\rm SFR}$ dispersion $\sigma_{{\rm SFR}}$ as a function of both redshift and stellar mass. We show our measurements only from stacking {\it Herschel} bands. {\it Spitzer} MIPS is more sensitive and thus allows measurements down to lower stellar masses, but it is less robust as an ${\rm SFR}$ indicator. This is mostly an issue at $z \simeq 2$, where the $24\,\mu{\rm m}$ is probing the rest-frame $8\,\mu{\rm m}$. \cite{elbaz2011} have shown that the $8\,\mu{\rm m}$ luminosity $L_8$ correlates very well with $L_{\rm IR}$ ($0.2\,{\rm dex}$ scatter), except for starburst galaxies. Inferring ${\rm SFR}$ from $8\,\mu{\rm m}$ thus has the tendency to erase part of the starburst population, effectively reducing the observed ${\rm SFR}$ dispersion. We checked that our results are nevertheless in good agreement between MIPS and {\it Herschel}, with MIPS derived dispersions being smaller on average by only $0.03 \pm 0.02\,{\rm dex}$.

As a sanity check, we also show an estimation of $\sigma_{{\rm SFR}}$ from individual {\it Herschel} detections. We select all galaxies in our {\it Herschel} sample that fall in a given bin of redshift and mass, and compute their offset from the main sequence $R_{\rm SB} \equiv {\rm SFR}/{\rm SFR}_{\rm MS}$, where ${\rm SFR}_{\rm MS}$ is the average ${\rm SFR}$ of ``main sequence'' galaxies given in Eq.~\ref{EQ:sfrms}. Following \cite{elbaz2011}, we call this quantity the ``starburstiness''. Because of the sensitivity of {\it Herschel}, this sample is almost never complete, and is biased toward high values of $R_{\rm SB}$: since this sample is ${\rm SFR}$ selected, all the galaxies at low mass are starbursts. To avoid completeness issues, we remove the galaxies that have $R_{\rm SB} < 1$, i.e., galaxies that are below the main sequence, and compute the $68$th percentile of the resulting $R_{\rm SB}$ distribution. By construction, this value does not need to be corrected for the width of the redshift and mass bins. However, it is only probing the \emph{upper} part of the ${\rm SFR}$--$M_\ast$ correlation, while the stacked measurements also take undetected sources below the sequence into account. In spite of this difference, the values obtained are in very good agreement with the stacked values. There is a tendency for these to be slightly higher by $0.03\,{\rm dex}$ on average, and this could be due to uncertainties in the individual ${\rm SFR}$ measurements. We conclude that the ${\rm SFR}$ distributions must be quite symmetric. This however does not rule out a ``starburst'' tail, i.e., a subpopulation of galaxies with an excess of star formation. Indeed, simulating a log-normal distribution of $R_{\rm SB}$ with a dispersion of $0.3\,{\rm dex}$ and adding $3\%$ more sources with an excess ${\rm SFR}$ of $0.6\,{\rm dex}$ \citep[following][]{sargent2012} gives a global dispersion measured with ${\rm MAD}$ of $0.309\,{\rm dex}$, while the $68$th percentile of the $R_{\rm SB} > 1$ tail is $0.319\,{\rm dex}$, a difference of only $0.01\,{\rm dex}$, which is well within the uncertainties.

\subsubsection{Implications for the existence of the main sequence}

Probably the most striking feature of Fig.~\ref{FIG:msdisp} is that $\sigma_{{\rm SFR}}$ remains fairly constant over a large fraction of the parameter space we explore, only increasing for the lowest redshift bin and at high stellar masses. This increase is most likely caused by the same phenomenon that bends the sequence at high stellar mass (see section \ref{SEC:ssfrz}, e.g., a substantial population of bulge-dominated objects that blur the correlation). On average, {\it Herschel} stacking thus gives $\sigma_{\rm SFR} = 0.30^{+0.06}_{-0.06}\,{\rm dex}$, with a random error of $0.01\,{\rm dex}$, and can be considered almost constant. Doing the same analysis in COSMOS UltraVISTA consistently yields $\sigma_{\rm SFR} = 0.33^{+0.03}_{-0.03}\,{\rm dex}$, with a random error of $0.01\,{\rm dex}$, showing that this result is not tied to specifics of our input $H$-band catalogs.

More importantly, this value of $0.3\,{\rm dex}$ means that, at a given stellar mass, $68\%$ of actively star-forming galaxies have the same ${\rm SFR}$ within a factor of two. This confirms the existence of the main sequence of star-forming galaxies for all of the stellar mass range probed here and up to $z=3$, i.e., over more than $80\%$ of the history of the universe. A more illustrative picture is shown later in Fig.~\ref{FIG:sfrms}, and we discuss the implication of this finding in section \ref{SEC:disc_feedback}.

\subsection{Contribution of the main sequence to the cosmic ${\rm SFR}$ density}

\begin{figure}
    \centering
    \includegraphics*[width=9cm]{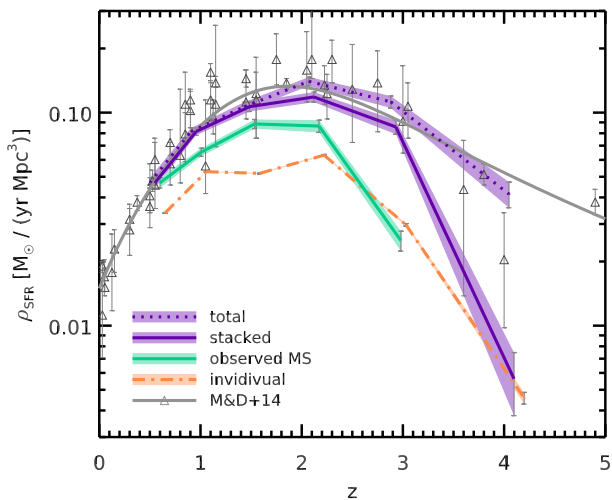}
    \caption{Evolution of the cosmic star formation rate density $\rho_{{\rm SFR}}$ with redshift. The orange dash-dotted line traces the ${\rm SFR}$ density inferred from individual {\it Spitzer} MIPS (for $z < 1.5$) and {\it Herschel} detections alone. The solid purple line represents the contribution of stacked sources with significant signal ($> 5\sigma$), and the dotted line is the extrapolation of the stacked ${\rm SFR}$ down to $M_\ast = 3 \times 10^9\,{\rm M}_\odot$ assuming constant ${\rm sSFR}$ and using the mass functions of Fig.~\ref{FIG:massfunc}. The green line shows the fraction of $\rho_{{\rm SFR}}$ in regimes where we have probed the existence of the main sequence. The lines are slightly offset in redshift for clarity. Light shaded regions in the background show the corresponding $1\sigma$ statistical errors. We compare these to the literature compilation of \cite{madau2014}, shown as open triangles, with their best-fit plotted as a solid gray line.}
    \label{FIG:sfrd}
\end{figure}

\begin{figure}
    \centering
    \includegraphics*[width=9cm]{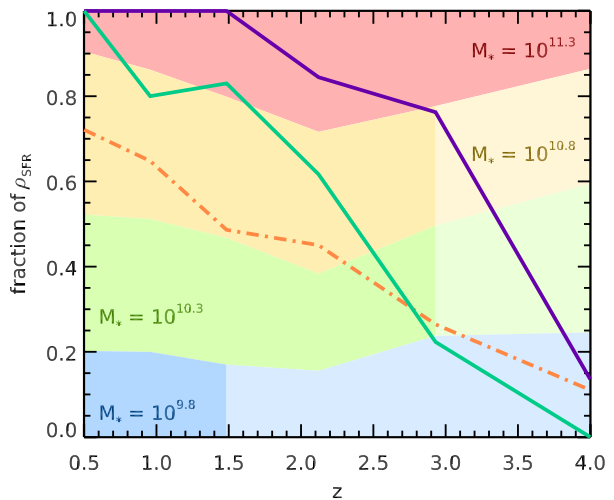}
    \caption{Contribution to the total $\rho_{{\rm SFR}}$ (purple dotted line in Fig.~\ref{FIG:sfrd}) as a function of redshift for the various sub-samples of Fig.~\ref{FIG:sfrd}. Background colors represent how galaxies of different stellar masses contribute to the total $\rho_{{\rm SFR}}$ (from top to bottom: $\log_{10}(M_\ast/{\rm M}_\odot) = 11.2$, $10.8$, $10.2$ and $9.8$), lighter colors indicating regions where $\rho_{{\rm SFR}}$ is extrapolated. The colored lines are defined as in Fig.~\ref{FIG:sfrd}: the solid purple line shows the contribution of stacked sources with significant signal, the green line shows the contribution of galaxies in the regimes where we have probed the existence of the main sequence, and the orange line is the contribution of individually detected FIR sources.}
    \label{FIG:sfrd_frac}
\end{figure}

\begin{figure}
    \centering
    \includegraphics*[width=9cm]{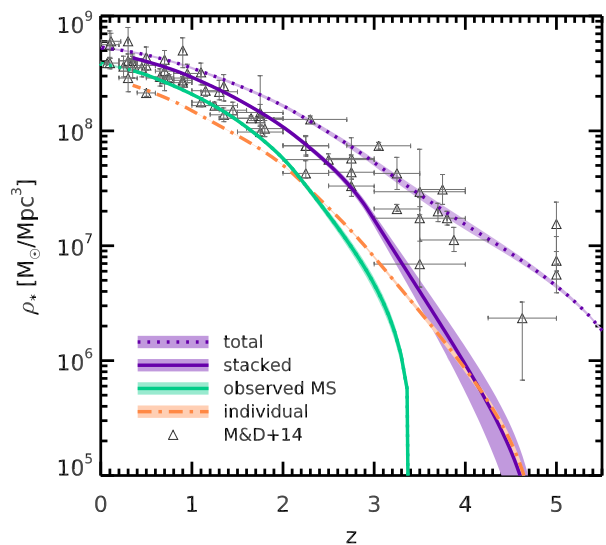}
    \caption{Predicted evolution of the cosmic stellar mass density $\rho_{\ast}$ with redshift. The lines show the inferred mass density by extrapolating our stacked ${\rm SFR}$s down to $M_\ast = 3\times10^9\,{\rm M}_\odot$ and out to $z=6$ using the trend from \cite{madau2014} and integrating as a function of time. Stellar lifetimes are accounted for, and the mass of stellar remnants is included in $\rho_{\ast}$ (see text). Colors are the same as in Fig.~\ref{FIG:sfrd}: the solid purple line shows the contribution of stacked sources with significant signal, the green line shows the contribution of galaxies in the regimes where we have probed the existence of the main sequence, and the orange line is the contribution of individually detected FIR sources. Shaded regions in the background show the corresponding $1\sigma$ statistical errors. We compare these results to the literature compilation of \cite{madau2014} shown as open triangles.}
    \label{FIG:rhostar}
\end{figure}

Using our stacked ${\rm SFR}$s, we can infer the contribution of each of our stacked bins to the cosmic star formation rate density $\rho_{{\rm SFR}}$ \citep{lilly1996,madau1996}. To this end, we use the stellar mass functions described in section \ref{SEC:massfunc} and extrapolate our results to obtain a prediction for the total $\rho_{{\rm SFR}}$, assuming a main-sequence slope of unity for low mass galaxies, and integrating the mass functions down to $M_\ast = 3 \times 10^9\,{\rm M}_\odot$ (i.e., $\sim0.03\,M^\star$). The results of this analysis are presented in Figs.~\ref{FIG:sfrd} and \ref{FIG:sfrd_frac}, and compared to the literature compilation of \cite{madau2014} (where luminosity functions are integrated down to $0.03\,L^\star$, and should thus match our measurements to first order).

We also infer the total stellar mass density $\rho_{\ast}$ by integrating $\rho_{{\rm SFR}}$ as a function of time. At each time step, we create a new population of stars whose total mass is given by $\rho_{{\rm SFR}}$, and let it evolve with time. We account for stellar mass loss using the \cite{salpeter1955} IMF to model the population, allowing stars to evolve and die assuming the stellar lifetimes of \cite{bressan1993} for solar metallicity. As stars die, some of the matter is left in the form of stellar remnants that are traditionally also included in $\rho_{\ast}$, i.e., neutron stars and white dwarfs. We parametrize the masses of these remnants following \cite{prantzos1998}. The contribution of these remnants continuously rises with time to reach about $12\%$ at $z=0$. The result is presented in Fig.~\ref{FIG:rhostar}.

One can see from these figures that individual {\it Herschel} detections in the ultra-deep GOODS and CANDELS surveys (orange dash-dotted line) unveil about $50\%$ of the star formation budget below $z=2$, but less than $10\%$ at $z=4$. In total, and over the redshift range probed here, these galaxies have built $49\%$ of the mass of present day stars, and are thus to be considered as major actors in the stellar mass build up in the Universe. Stacking (purple line) allows us to go much deeper, since we reach almost $100\%$ of the total $\rho_{{\rm SFR}}$ at $z<2$, and accounts for $83\%$ of the mass of present day stars. Extrapolating our observations to lower stellar masses using the mass functions and to $z=0$ using the best-fit $\rho_{{\rm SFR}}$ of \cite{madau2014}, we obtain an estimate of the total amount of star formation in the Universe (purple dotted line). Integrating it to $z=0$ gives $\rho_{\ast}(z=0) = (5.3 \pm 0.1)\times10^8\,{\rm M}_\odot\,{\rm Mpc}^{-3}$, consistent with the value reported by \cite{cole2001} and \cite{bell2003} (our error estimate being purely statistical).

Although the range in redshift and stellar mass over which we are able to probe the existence of the main sequence is limited, it nevertheless accounts for $66\%$ of the mass of present day stars. This number climbs up to $73\%$ if we take other studies that have observed a tight correlation down to $z=0$ \citep{brinchmann2004} into account. We show in the next section that starburst galaxies make up about $15\%$ of the ${\rm SFR}$ budget in all the redshift and mass bins that we probe with individual detections, and that the remaining fraction is accounted for by a single population of ``main sequence" galaxies. Subtracting these $15\%$ from the above $73\%$, we can say that at least $62\%$ of the mass of present day stars was formed by galaxies belonging to the main sequence. In other words, whatever physical phenomenon shapes the main sequence is the dominant mode of star formation in galaxies.

\subsection{Quantification of the role of starburst galaxies and the surprising absence of evolution of the population \label{SEC:indiv}}

\subsubsection{An overview of the main sequence}

\begin{figure*}
    \centering
    \includegraphics*[width=17cm]{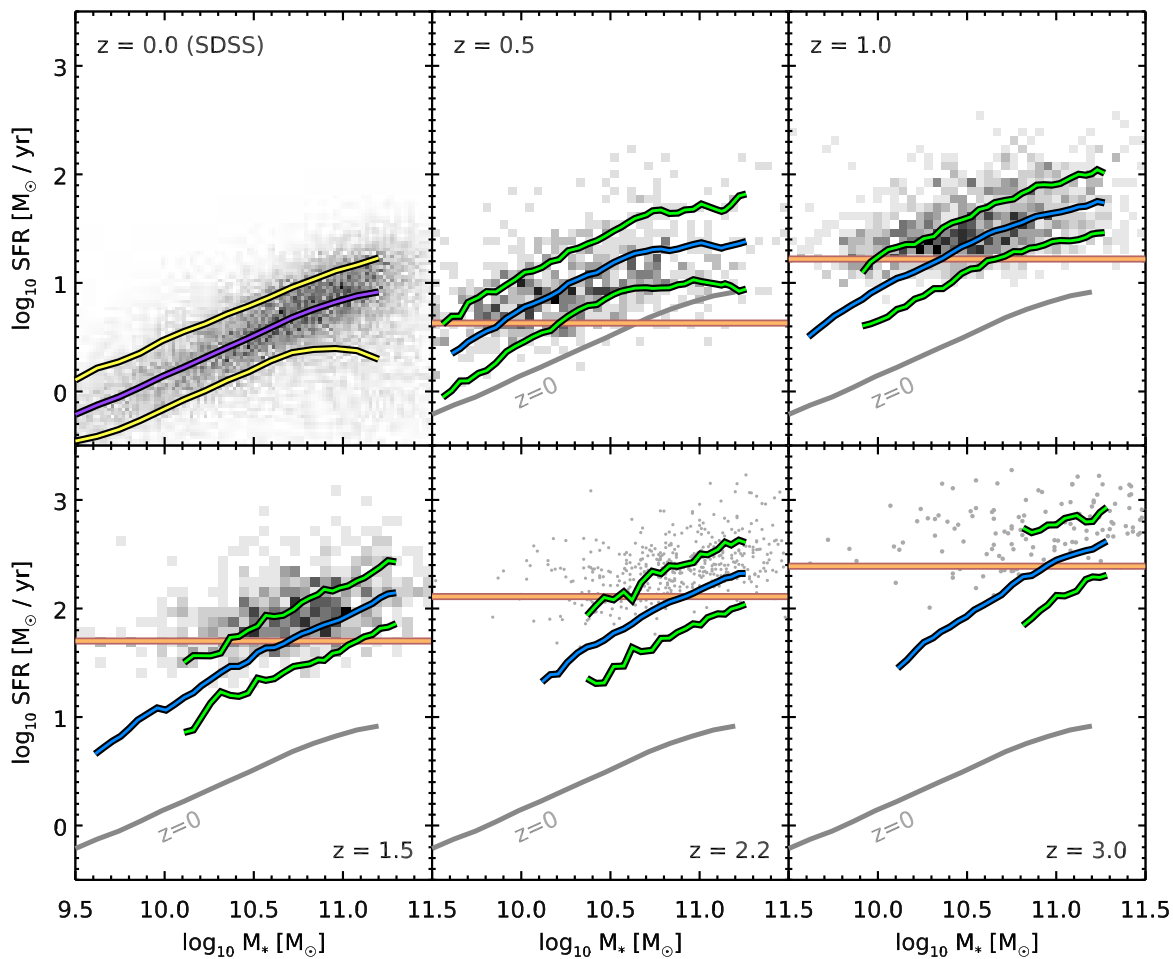}
    \caption{Compilation of both detections and stacking results on the ${\rm SFR}$--$M_\ast$ plane for the CANDELS fields. The top left panel shows the results obtained with the Sloan Digital Sky Survey (SDSS) in the local Universe, as presented in \cite{elbaz2007}, while each subsequent panel displays our result for increasing redshifts. The blue line shows the average stacked ${\rm SFR}$ (section \ref{SEC:ssfrz}), and the green lines above and below show the $1\sigma$ dispersion obtained with scatter stacking (section \ref{SEC:sfrdisp}). Both of these were performed on sliding bins of mass for the sake of visualization, and for this figure only. The ${\rm SFR}$ detection limit of each sample is indicated with a solid orange line. We also show the sliding median and percentiles of the SDSS distribution with purple and yellow lines, respectively, to emphasize that both the ${\rm SFR}$ tracer and the sample selection are different (see text). This correlation, observed in the local Universe, is reproduced as a gray line on each panel. The density of individual detections is shown in gray scale in the background, except for the two highest redshift bins where we show the individual galaxies as gray filled circles.}
    \label{FIG:sfrms}
\end{figure*}

We summarize the previous results in Fig.~\ref{FIG:sfrms}. Here we show the distribution of individually detected galaxies on the ${\rm SFR}$--$M_\ast$ plane at various redshifts. The locus of our stacked ${\rm SFR}$s (solid blue lines) may not appear to coincide with the average of the detections because of the ${\rm SFR}$ detection limit, symbolized by the horizontal dashed line. We discuss later on (in Fig.~\ref{FIG:rsbhist}) the distribution of these detected sources and confirm that the stacks and the detections are in perfect agreement.

We also show for reference the $z=0$ sample taken from the Sloan Digital Sky Survey \citep[SDSS DR4, ][]{brinchmann2004} as presented in \cite{elbaz2007}. In this data set, actively star-forming galaxies are selected according to their rest-frame $U-V$ colors only (i.e., what is usually referred to as the ``blue cloud''), and ${\rm SFR}$s are estimated from the dust-corrected ${\rm H}_\alpha$ line. These differences of observables and sample selection are likely to affect the shape of the main sequence. In particular, it is clear that the bending at high mass is less pronounced in the SDSS sample, and this is likely due to the selection. Therefore, the comparison of this $z=0$ data set with our own sample should be done with caution. This nevertheless resembles our own results quite closely and allows us to paint a consistent picture from $z=0$ to $z=3$.

\subsubsection{``Starburstiness'' distributions}

Although the depth of the {\it Herschel} surveys is limited, there is still a lot to be learned from the individually detected sources, in particular for the bright starburst galaxies. Now that we have a good definition of the main sequence, we can study these galaxies in more detail. \cite{rodighiero2011} have used similar data in COSMOS and found that the distribution of star-forming galaxies on and off the main sequence is bimodal: a population of normal star-forming galaxies shapes the main sequence with a log-normal distribution of ${\rm sSFR}$ at a given mass, while another smaller population of ``starbursts'' boosts the high ${\rm sSFR}$ counts. Their work was restricted to $z=2$ because of the $BzK$\xspace selection, so we want to extend it here to a mass-complete sample over wider range of redshifts to see what we can learn about the starburst population.

\begin{figure*}
    \centering
    \includegraphics*[width=18cm]{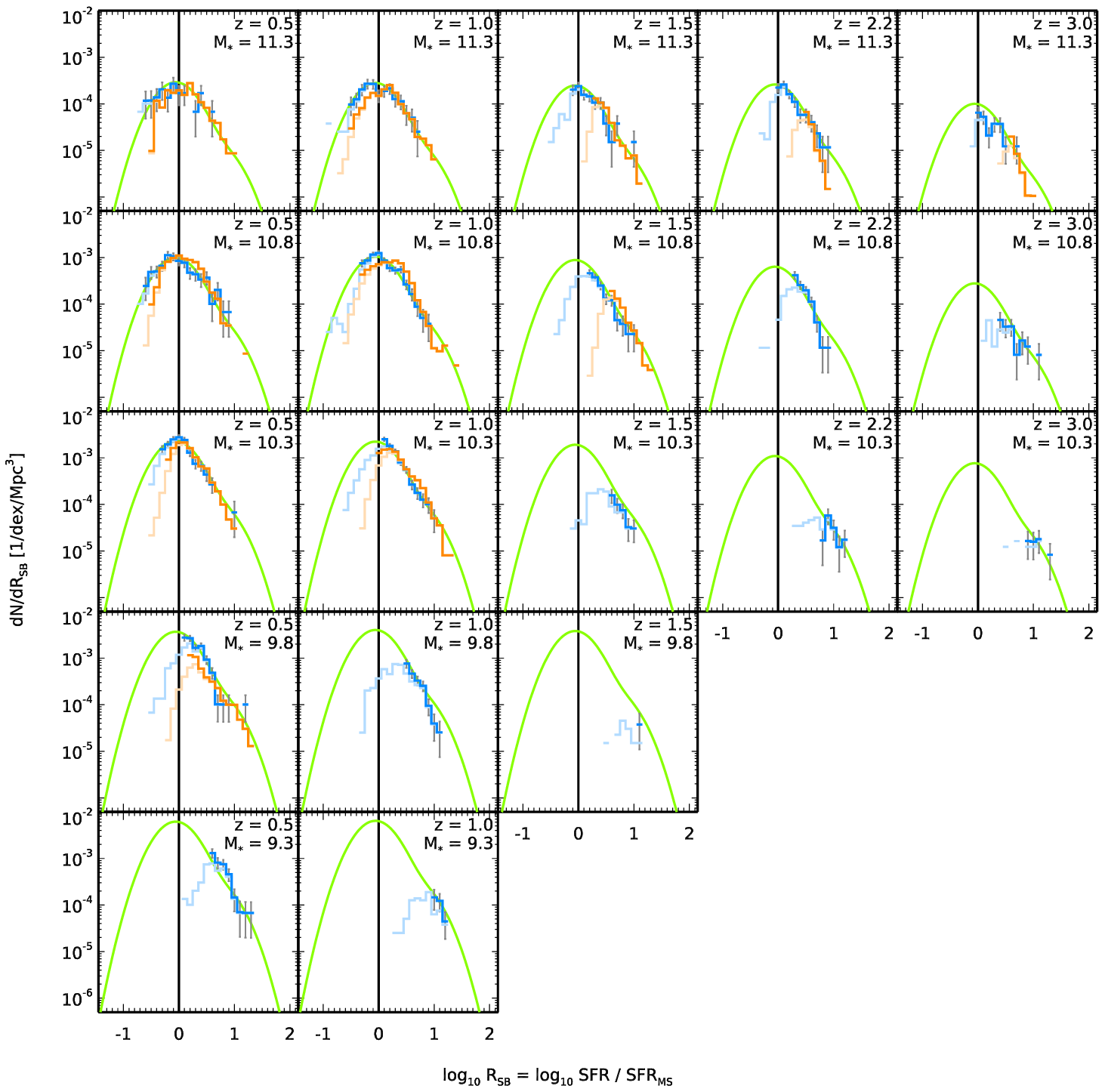}
    \caption{Starburstiness ($R_{\rm SB} \equiv {\rm SFR}/{\rm SFR}_{\rm MS}$) histograms of individual {\it Herschel} and {\it Spitzer} MIPS (for $z < 1.2$) detections in each of our redshift and stellar mass bins. The blue and orange lines correspond to the counts in the CANDELS and COSMOS $2\,\deg^2$ fields, respectively. We also show the incomplete counts in light colors in the background. The green curve shows our best-fit to the combined data set, and is the same for all bins except for the normalization, which is set by the mass function. The black vertical line shows the locus of the main sequence. Error bars indicate Poissonian noise.}
    \label{FIG:rsbhist}
\end{figure*}

In Fig.~\ref{FIG:rsbhist} we show the distributions of ``starburstiness'' $R_{\rm SB}$, defined as the ratio between the actual ${\rm SFR}$ of each galaxy and ${\rm SFR}_{\rm MS}$, the ${\rm SFR}$ they would have if they were exactly following the main sequence defined in Eq.~\ref{EQ:sfrms}. We analyze these distributions in the same bins that were used for stacking, to make the comparison simpler. Since the CANDELS fields have a relatively similar depth, we group them together into a single distribution (blue curve), and following \cite{rodighiero2011} we keep the COSMOS UltraVISTA sources apart (orange curve) where the catalog is mass-complete.

As was the case for the stellar mass functions discussed in section \ref{SEC:massfunc}, these distributions are affected by completeness issues. To correct this, we use a procedure very similar to that used for the mass functions. We assume that the total $L_{\rm IR}$ of a galaxy at a given redshift is well modeled from the rest-frame monochromatic luminosity in each {\it Herschel} band by a power law plus a Gaussian scatter in logarithmic space. In each bin of redshift and stellar mass, we select galaxies that are detected in at least three {\it Herschel} bands, fit this power law and measure the dispersion as in Fig.~\ref{FIG:mtol}. In this case, this dispersion is mainly due to differences in dust temperature, and is found to be minimal at the peak of the FIR emission (see Fig.~\ref{FIG:lirdisp}). Then, for each {\it Herschel} band, in each redshift and mass bin, we then generate a mock population of $10\ 000$ galaxies with uniform redshift and mass distribution within the bin and attribute a starburstiness with uniform probability to each mock galaxy. We multiply this starburstiness by the ${\rm SFR}_{\rm MS}$ of the galaxy computed from its redshift and mass, subtract the average observed ${\rm SFR}_{\rm UV}$ in this bin (we assume no scatter in ${\rm SFR}_{\rm UV}$ for simplicity), convert the remaining ${\rm SFR}_{\rm IR}$ into $L_{\rm IR}$, and finally the $L_{\rm IR}$ into monochromatic luminosity in the considered {\it Herschel} band, adding a random logarithmic scatter whose amplitude is given by the dispersion measured earlier. The completeness is then given as the fraction of mock galaxies with simulated monochromatic luminosity larger than the limiting luminosity at the corresponding redshift.

Since we include in our sample all sources provided that they are detected in at least one {\it Herschel} band, we then take the maximum completeness among all bands. In Fig.~\ref{FIG:rsbhist}, raw incomplete counts are shown as light curves in the background, and corrected counts are shown as darker lines. Error bars indicate Poisson noise and for clarity are only shown for the CANDELS counts.

In all fields, the low $R_{\rm SB}$ counts at $z<1.2$ come from MIPS derived ${\rm SFR}$s. Since the MIPS imaging in COSMOS UltraVISTA is only half as deep as the deepest CANDELS fields (see section \ref{SEC:sample_uvista} and Table \ref{TAB:depth}), the two curves probe almost similar ranges of $R_{\rm SB}$. At $z\ge1.2$ (i.e., starting from the bin at $z=1.5$) MIPS is not used any more, and the difference in depth of the {\it Herschel} surveys becomes quite obvious. Reassuringly, we see very good agreement between the two data sets where they overlap.

\subsubsection{Evolution of the fraction of starbursts}

\begin{figure}
    \centering
    \includegraphics*[width=9cm]{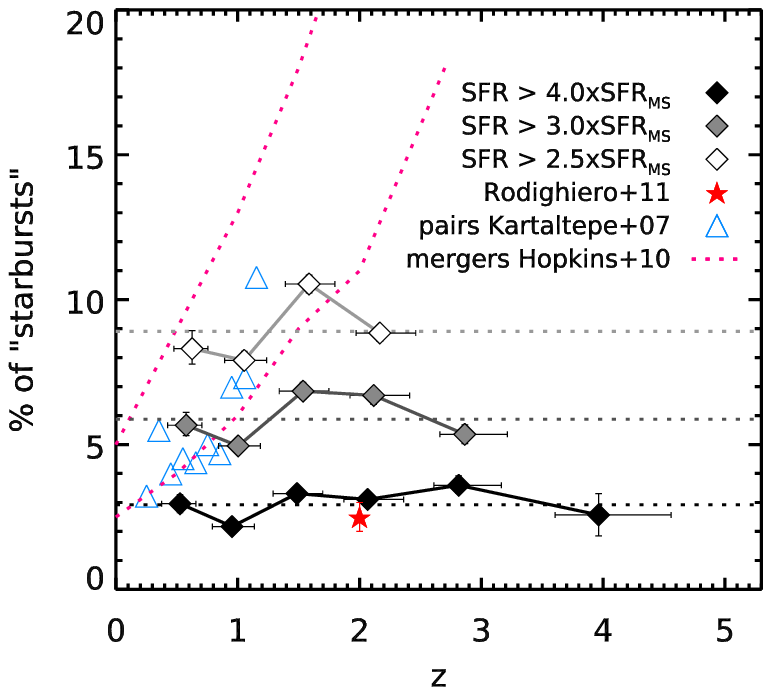}
    \caption{Evolution of the observed ``starburst'' fraction with redshift, where starbursts galaxies are defined as having an ${\rm SFR}$ enhanced by at least a factor $X_{\rm SB}$ compared to the ${\rm SFR}$ on the main sequence. Our results are shown for $X_{\rm SB}=4$, $3$ and $2.5$ as diamonds (black, gray, and white, respectively), slightly offset in redshift for clarity. Only points where the starburst sample is complete are shown, and error bars are estimated using bootstrapping. We also show the value observed by \cite{rodighiero2011} at $z=2$ as a filled red star, which was obtained with $X_{\rm SB}=4$. These figures are compared qualitatively to the observed pair fraction reported by \cite{kartaltepe2007} as open blue triangles, and the range of major merger fractions predicted by \cite{hopkins2010-a} is shown with dashed purple lines. It is clear that, both in observations and simulations, the merger fraction evolves significantly faster than the observed starburst fraction, the latter remaining almost constant regardless of the precise definition of what is a ``starburst''.}
    \label{FIG:sbfrac}
\end{figure}

From these distributions, we can derive interesting statistical properties of our star-forming galaxy sample. In particular, \cite{rodighiero2011} reported that only $2$ to $3\%$ of the galaxies in their $z=2$ sample were in a ``starburst'' mode, with an ${\rm SFR}$ increased by more than a factor $4$ (or $0.6\,{\rm dex}$) compared to the main sequence (i.e., $R_{\rm SB} > 4$). Using our data set, we are able to measure this fraction at different redshifts and look for an evolution of this population. To do so, we select in each redshift bin all star-forming galaxies more massive than $5\times10^{10}\,{\rm M}_\odot$ (this mass threshold is chosen to avoid ${\rm SFR}$ completeness issues), and compute the fraction of objects for which the observed ${\rm SFR}$ is at least a factor $X_{\rm SB}$ above the main sequence. Following \cite{rodighiero2011}, we choose $X_{\rm SB} = 4$. However, to make sure that our results are not affected by this somewhat abritrary choice, we also do this analysis with $X_{\rm SB} = 3$ and $2.5$. By lowering this threshold, the number of objects increases and the statistics become more robust, at the price of having a higher number of nonstarburst contaminants scattering from the main sequence. We could have overcome this problem by fitting the observed counts, decomposing the total ${\rm SFR}$ distribution as coming from two populations: a main-sequence component and a starburst component, as was done in \cite{sargent2012}. While such a deconvolution provides a more physical definition of a ``starburst'', it is also dependent on the model one choses to describe the starburst population. Also, except in a few low redshift bins, our data do not probe a wide enough range to be able to robustly perform this decomposition. We therefore choose this simpler approach of a fixed $R_{\rm SB}$ threshold for now, and will come back to the decomposition later. The results are presented in Fig.~\ref{FIG:sbfrac}. Between $z=0.5$ and $z=4$ and for $X_{\rm SB} = 4$, we measure a roughly constant value ranging between $2$ and $4\%$, and no clear trend with redshift emerges. We discuss the implication of this fact in section \ref{SEC:disc_mergers}.

\subsubsection{Quantifying the contribution of starbursts to the total ${\rm SFR}$ budget}

\begin{figure}
    \centering
    \includegraphics*[width=9cm]{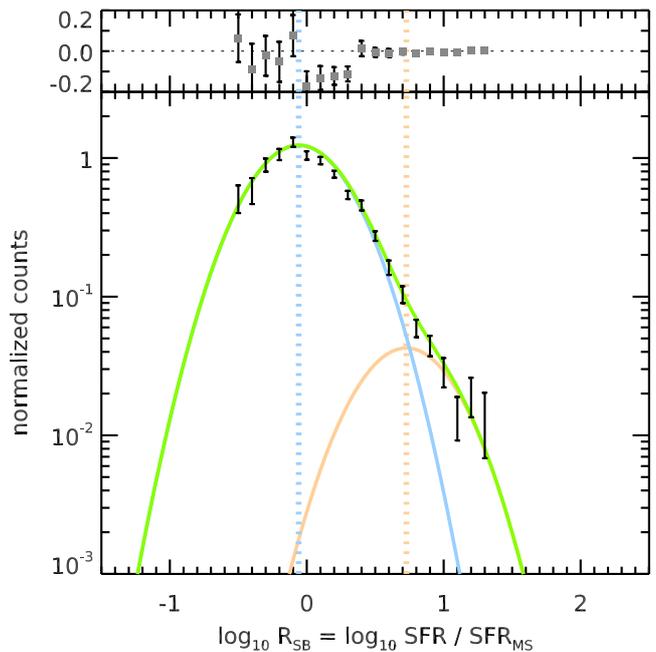}
    \caption{Combined starburstiness ($R_{\rm SB}$) distributions from Fig.~\ref{FIG:rsbhist} normalized to the total number of star-forming galaxies in each bin. The green line shows our best-fit model from Eq.~\ref{EQ:ms_model}, and the blue and orange lines show the contributions of main sequence and starburst galaxies, respectively. The residuals of the fit are shown at the top of the figure.}
    \label{FIG:ms_fit}
\end{figure}

We now normalize the counts by the integral of the stellar mass function in all bins and, supported by our findings on the constant width of the main sequence (Fig.~\ref{FIG:msdisp}) and on a constant starburst fraction (Fig.~\ref{FIG:sbfrac}), we assume that the $R_{\rm SB}$ distribution does not vary. With this same assumption of an unvarying distribution, \cite{sargent2012} managed to reconstruct the IR luminosity function at various redshifts. With the increased statistics, we are now able to perform a two-component decomposition of the whole distribution. We thus fit all the counts simultaneously with a double log-normal distribution following \cite{sargent2012}. The chosen parametrization for the fit is
\begin{eqnarray}
\phi_{R_{\rm SB}}(x) = \frac{1 - f_{\rm SB} - f_{\rm miss}}{\sqrt{2\,\pi}\,\sigma_{\rm MS}}\,\exp\left[-\frac{\log_{10}(x/x_0)^2}{2\,{\sigma_{\rm MS}}^2}\right] \nonumber \\
+ \frac{f_{\rm SB}}{\sqrt{2\,\pi}\,\sigma_{\rm SB}}\,\exp\left[-\frac{\log_{10}(x/B_{\rm SB})^2}{2\,{\sigma_{\rm SB}}^2}\right]\,, \label{EQ:ms_model}
\end{eqnarray}
where $\sigma_{\rm MS}$ and $\sigma_{\rm SB}$ are the widths of the main sequence and starburst distributions, respectively, $f_{\rm SB}$ is the fraction of starbursts, and $B_{\rm SB}$ is the median multiplicative ${\rm SFR}$ boost of starburst galaxies. We also introduce $f_{\rm miss}$ as the fraction of star-forming galaxies that are neither ``main sequence'' nor ``starburst'' galaxies (e.g., ``green valley'' galaxies), and $x_0$ the median $R_{\rm SB}$ of main-sequence galaxies. By construction, the latter two parameters should be close to $0$ and $1$, respectively, but we allow them to vary to check for the consistency between the detections and the stacks.

The result is shown in Fig.~\ref{FIG:ms_fit}. Leaving all parameters free, the fit of the starburst population is highly uncertain, so we decided to fix $\sigma_{\rm SB} = \sigma_{\rm MS}$, and fit the logarithm of the counts. We obtain $\sigma_{\rm MS} = 0.31\pm0.02\,{\rm dex}$, $f_{\rm SB} = 3.3\%\pm1.5\%$, $B_{\rm SB} = 5.3\pm 0.4$, $f_{\rm miss} = 0\%\pm2\%$, and $x_0 = 0.87\pm0.04$.

These numbers depend heavily on the chosen parametrization of the starburst population. For example, not imposing $\sigma_{\rm SB} = \sigma_{\rm MS}$ would change the values of $B_{\rm SB}$ considerably, hence the measured values should be used with caution. The integrated contribution of the starburst population is however well constrained \citep{sargent2012}. Taking these numbers at face value, we reach a similar conclusion as \cite{rodighiero2011} and \cite{sargent2012}, i.e., that starbursts are rare and happen in only about $3\%$ of galaxies at a given instant. However, they form stars on average $\sim 5$ times faster than their main-sequence counterparts, and thus contribute to $\sim 15\%$ of the ${\rm SFR}$ budget. It is worth noting that the bimodality, if any, is not clearly apparent in our data, and the high $R_{\rm SB}$ counts can also be fit with a single power law (with a slope close to $-2$). While our goal is not to demonstrate the validity of this bimodal decomposition, we want to stress that the absence of a ``gap'' in the distribution between the peaks of the two components does not rule out the bimodal hypothesis.

The main-sequence distribution, on the other hand, is very well constrained and both its average and the measured $\sigma_{\rm MS}$ are in agreement with the stacked value. The fact that $f_{\rm miss}$ is close to zero means that we are able to recover essentially \emph{all} the star-forming galaxies with this model. More precisely, if there is another population of star-forming galaxies, we can say with $70\%$ probability that it can only make up for less than $2\%$ of the counts.

Last but not least, the accuracy of the fit in all the bins (as shown in Fig.~\ref{FIG:rsbhist}) confirms the validity of our hypothesis of a universal $R_{\rm SB}$ distribution.

\section{Discussion}

\subsection{Connection of the main-sequence dispersion with feedback processes \label{SEC:disc_feedback}}

The nonevolution of the main-sequence dispersion, as described in section \ref{SEC:sfrdisp}, is intriguing. Indeed, this dispersion can originate from several completely different processes. On the one hand, the scatter within the star formation history (SFH) of individual galaxies, i.e., bursts of star formation due to minor or major merging and feedback from AGNs or supernova winds, will naturally broaden the distribution of ${\rm SFR}$. On the other hand, the scatter may also be due to one or more missing variables, such as age, metallicity, geometry, or environment. For example, \cite{salmi2012} found, using $24\,\mu{\rm m}$ based ${\rm SFR}$s at $z\simeq1$, that the dispersion of the main sequence could artificially be reduced to about $0.15\,{\rm dex}$ by introducing the rest-frame $U-V$ color as well as $z$-band clumpiness as extra variables. This also shows that most of the observed scatter of the main sequence is physical and not due to measurement errors.

\cite{hopkins2014} have computed the expected scatter of SFH from a set of numerical simulations, and found it to be a strong function of halo mass, and thus of stellar mass. Performing abundance matching using their $M_\ast$--$M_{\rm halo}$ relation, one finds that they predict a variation of the ${\rm SFR}$ (averaged over $200\,{\rm Myr}$, hence comparable to the timescale of our FIR ${\rm SFR}$ tracer) of about $0.1\,{\rm dex}$ at $M_\ast > 10^{11}\,{\rm M}_\odot$, rising up to $0.4\,{\rm dex}$ as stellar mass decreases down to $10^8\,{\rm M}_\odot$. They also find that this evolution is coming predominantly from the rising importance of stellar feedback, and not from merging or global gravitational instabilities. Intuitively, the smaller the galaxy, the more sensitive it is to the impact of stellar winds and super novae, since the characteristic length scale over which these phenomena tend to heat and blow away the gas is more or less constant. Since there are other components that add up to the total scatter in ${\rm SFR}$ (age, environment, metallicity, etc.), this prediction should be considered as a lower limit.

The predicted values of \cite{hopkins2014} are shown as the purple line in Fig.~\ref{FIG:msdisp}. The dependence of their prediction on stellar mass is clear, yet we seem to measure a constant value. Even though there are other sources of scatter at play, it would be a strange conspiracy for them to exactly counterbalance the evolution of the scatter within the SFH to maintain a constant main-sequence scatter \citep[see however][]{sparre2014}. Our interpretation is thus the following.

Stellar feedback is a necessary ingredient in numerical simulations. Without it, galaxies would consume their gas too efficiently, and with the amount of infalling gas they receive from the inter-galactic medium, they would end up today with extremely high stellar masses that are not observed. The \emph{real} strength of the stellar feedback is poorly constrained, so it is usually considered as a free parameter and fine-tuned to reproduce the local stellar mass density. However, our observations show that it cannot be arbitrarily high. Other processes can be considered to either decrease the star formation efficiency of galaxies, or reduce the amount of infalling gas they receive \citep[e.g.,][]{gabor2014}.

\subsection{Connection between starbursts and mergers \label{SEC:disc_mergers}}

We have shown in section \ref{SEC:indiv} that the starburst population is not evolving, both in relative numbers and ${\rm SFR}$ excess with respect to the main sequence. This is intriguing in many aspects. Both observations \citep[and references therein]{lefevre2000,kartaltepe2007,lotz2011} and numerical simulations \citep[e.g.,][]{somerville2008,hopkins2010-a} predict an increase of the major merger rate with increasing redshift, typically proportional to $(1+z)^m$. Although the slope $m$ of the evolution of the merger fraction is quite uncertain \citep[see discussion in][]{kampczyk2007}, it is always found to be positive, ranging from $m \simeq 0$ up to $m \simeq 6$. For example, \cite{kartaltepe2007} analyzed the fraction of close pairs from $z=0$ to $z=1.2$, and found $m = 3.1 \pm 0.1$. Their $z=0$ value of $0.7\% \pm 0.1\%$ is comparable to our observed starburst fraction with $X_{\rm SB} = 4$, however extrapolating this relation to $z=2$ would predict a pair fraction of about $50\%$ \citep[$20\%$ if we consider instead the numerical simulation of][]{hopkins2010-a}. If all or a constant fraction of those pairs do lead to gas-rich major mergers, this would have a huge impact on the number of starburst, at odds with our observations.

On the other hand, \cite{perret2014} ran several numerical simulations of mergers of $z=2$ clumpy galaxies, and found little to no impact of the merger on star formation when compared to isolated galaxies. Their point is that by $z=2$ star formation is already fairly active in isolated galaxies and actually close to a saturation point due to feedback processes. When the merger happens, it therefore cannot increase the total ${\rm SFR}$ by a large amount because star formation is already at its maximum. So even if mergers were more frequent in the past, they were also less efficient at triggering bursts of star formation, and this could explain why we are not seeing a huge increase in the number of starburst galaxies. This goes in the same direction as the results of \cite{hopkins2010} who found in their simulations that merger-driven bursts contribute to the same fraction ($5$--$10\%$) of the IR luminosity function at all redshifts, but it does not explain why the fraction of such bursts remains constant over time.

Although the most extreme starburst events are unambiguously associated with major mergers in the local Universe \citep[e.g.,][]{armus1987}, another interpretation of our results is that the situation may be different at earlier epochs, and that some other phenomena may be responsible for such bursts of star formation, such as large scale dynamical instabilities \citep[e.g.,][]{dekel2009}.

\section{Conclusions}

We have put together a catalog of star-forming galaxies that is mass-complete above $2\times10^{10}\,{\rm M}_\odot$ and extends up to $z=4$, using the deep UV to NIR observations in the CANDELS fields. By stacking the {\it Herschel} images at the positions of these galaxies, using bins of mass and redshift, we measured their average star formation rates in a dust-unbiased way. We then derived a new technique called ``scatter stacking'' to measure the scatter in ${\rm SFR}$ around the average stacked value. We also analyzed sources individually detected on the {\it Herschel} images to study the ${\rm SFR}$ distribution in more detail over a more limited range of redshift and stellar mass.

We observe a continuously rising ${\rm sSFR} \equiv {\rm SFR}/M_\ast$ up to $z=4$, with no clear sign of a saturation or plateau at the highest redshifts. Previous observations of this type of saturation are mostly based on LBG samples that lack observations in the FIR to reliably constrain the dust extinction. Earlier results are likely due to a combination of selection effects and biases in the dust extinction correction. It is therefore mandatory to have mass-complete samples and rest-frame MIR or FIR data to provide reliable constraints on the star formation activity of actively star-forming galaxies.

We find that the slope of the ${\rm SFR}$--$M_\ast$ relation is close to unity, except for high mass galaxies ($M_\ast \gtrsim 10^{10.5}\,{\rm M}_\odot$), where the slope is shallower. Furthermore, the high mass slope is evolving from $\sim 0.8$ at high redshifts down to almost $0$ at $z\sim0.5$. One possible explanation is the increasing contribution of the bulge to the stellar mass of these galaxies, while the star formation rates come mostly from the disk \citep{abramson2014}.

At fixed mass and redshift, the scatter around the average ${\rm SFR}$ appears to be constant and close to $0.3\,{\rm dex}$ from $M_\ast = 3\times10^9\,{\rm M}_\odot$ to $2\times10^{11}\,{\rm M}_\odot$, with no clear redshift dependence. We therefore confirm the existence of the ``main sequence'' of star-forming galaxies over a large range of mass and redshift with a robust star formation rate tracer. We show that at least $66\%$ of present day stars were formed in main-sequence galaxies. Consequently, whatever physical process produces the main sequence is the dominant mode of stellar growth in galaxies.

The nonevolution of the ${\rm SFR}$ scatter with mass can be connected to the expected strength of stellar feedback. State-of-the-art numerical simulations indeed predict that stellar feedback generates additional scatter in the star formation histories of galaxies, a scatter whose amplitude is strongly anticorrelated with halo mass and thus galaxy mass. Our observations provide useful constraints for numerical simulations where stellar feedback is often used as an efficient star formation regulator. We show here that it cannot be arbitrarily high.

Refining the above analysis with individual {\it Herschel} detections, we look for starburst galaxies whose ${\rm SFR}$s are systematically larger than those of main-sequence galaxies. In agreement with \cite{sargent2012} and extending their analysis to higher redshifts and more complete samples, we find that the fraction of these starburst galaxies does not evolve with time. This questions the usual interpretation of starburst as the consequence of triggering by major mergers. Several studies, both of simulations and observations, indeed show that the fraction of mergers was substantially higher in the past. An alternative explanation is that mergers may be less efficient at creating bursts of star formation within high redshift galaxies.

We have pushed {\it Herschel} as far as possible to study the main sequence of star-forming galaxies, but it is still necessary to dig deeper than that, i.e., probing higher redshifts or lower stellar masses. Most of what we know at present about the high redshift Universe ($z > 4$) comes from rest-frame UV-based studies, and we have shown here that dust extinction plays an important role even at these redshifts. Therefore it will be necessary to explore these epochs of the Universe with an independent and more robust ${\rm SFR}$ tracer to confirm the pioneering results obtained with the UV light alone. Probing lower stellar masses will also be an important challenge since, owing to their small sizes, low mass ($M_\ast < 3\times10^{9}\,{\rm M}_\odot$) galaxies are probably most sensitive to smaller scale physics, e.g., stellar or AGN feedback.

Valuable insights already come from the study of lensed galaxies. This technique allows us to observe galaxies about an order of magnitude fainter than the nominal instrument depths, either by chance in blank fields \citep[e.g., the {\it Herschel} ATLAS,][]{eales2010}, or by explicitly targeting large galaxy clusters \citep[e.g., the Herschel Lensing Survey,][]{egami2010}. Studying these regimes on statistically relevant samples and with a dust-unbiased ${\rm SFR}$ tracer will only be possible with a new generation of instruments. The most promising candidate available today for the high redshift Universe is certainly the Atacama Large Millimeter/submillimeter Array (ALMA), and interesting science is already on its way. In particular, we are now waiting for the completion of Cycle 2 observations targeting a mass-complete sample of $z=4$ star-forming galaxies down to $\log_{10}(M_\ast/{\rm M}_\odot) = 10.7$. With only a few minutes of on-source integration, these data will allow us to probe ${\rm SFR}$s about five times lower than those available with the deepest {\it Herschel} surveys. As for the low mass galaxies, substantial progress is likely to happen in a few years thanks to the exceptional MIR capabilities of the \emph{James Webb Space Telescope} (\emph{JWST}).

\begin{acknowledgements}

The authors want to thank the anonymous referee for his/her comments that clearly improved the consistency and overall quality of this paper.

CS and DE are grateful to F.~Bournaud for the enlightening discussions that motivated certain aspects of this paper. SJ acknowledges support from the EU through grant ERC-StG-257720.

Most of the numerical analysis conducted in this work have been performed using {\tt phy++}, a free and open source C++ library for fast and robust numerical astrophysics (\url{http://cschreib.github.io/phypp/}).

This work is based on observations taken by the CANDELS Multi-Cycle Treasury Program with the NASA/ESA {\it HST}, which is operated by the Association of Universities for Research in Astronomy, Inc., under NASA contract NAS5-26555.

This research was supported by the French Agence Nationale de la Recherche (ANR) project ANR-09-BLAN-0224 and by the European Commission through the FP7 SPACE project ASTRODEEP (Ref.No: 312725).
\end{acknowledgements}

\bibliographystyle{aa}
\bibliography{full}

\appendix

\section{The $UVJ$\xspace selection \label{APP:uvj}}

\begin{figure*}
    \centering
    \includegraphics*[width=18cm]{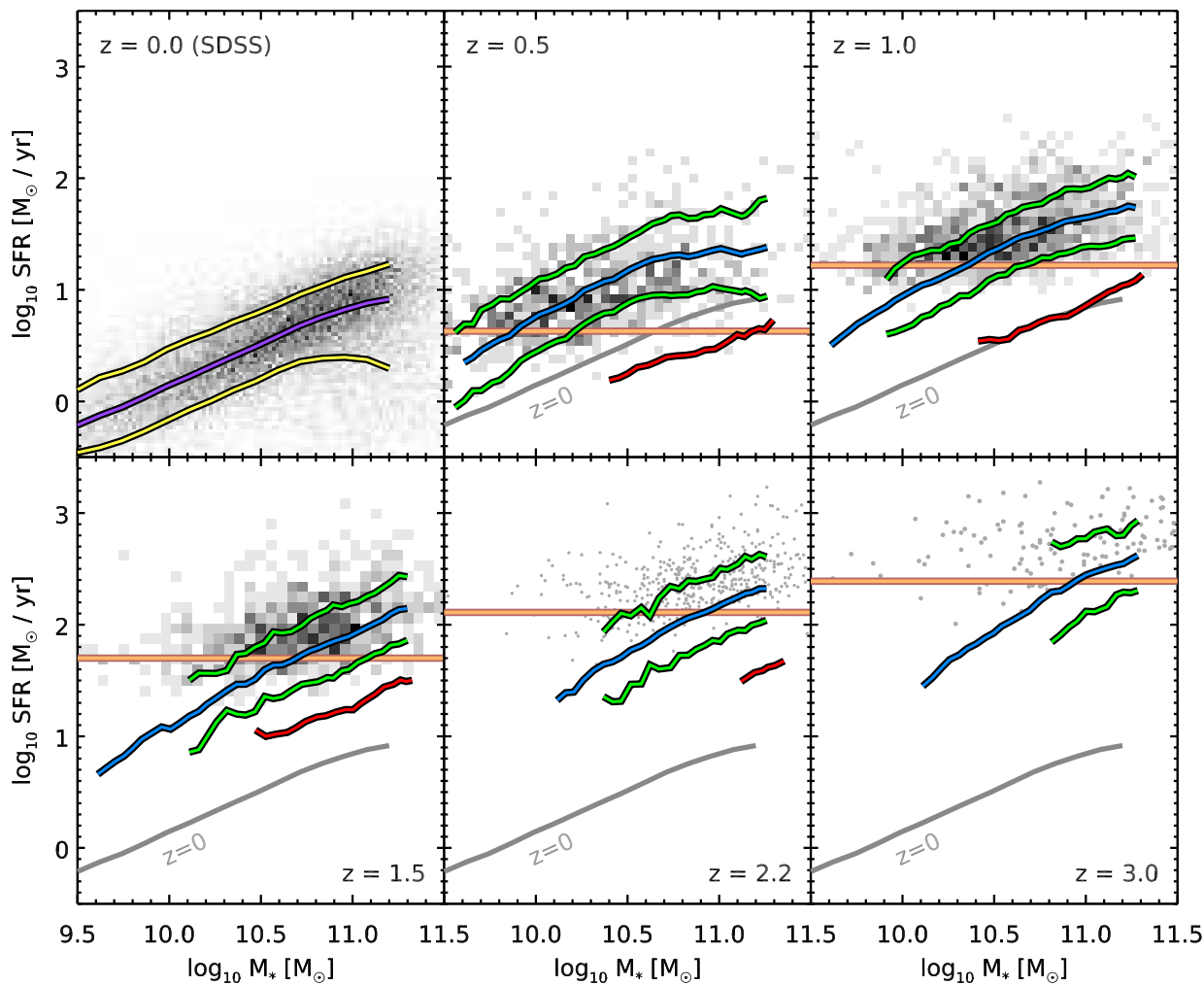}
    \caption{Same as Fig.~\ref{FIG:sfrms}, this time also showing the location of $UVJ$\xspace passive galaxies. In each panel, the blue line shows the average stacked ${\rm SFR}$ (section \ref{SEC:ssfrz}), and the green lines above and below show the $1\sigma$ dispersion obtained with scatter stacking. The orange horizontal line shows the detection limit of {\it Herschel} in ${\rm SFR}$. The red line shows the stacked ${\rm SFR}$ of $UVJ$\xspace passive galaxies, naively assuming that all the IR light comes from star formation. This is a conservative upper limit, since in these galaxies dust is predominantly heated by old stars, and the effective dust temperature inferred from the FIR SED is much colder than for actively star-forming galaxies of comparable mass.}
    \label{FIG:sfrms_passive}
\end{figure*}

To further test the reliability of the $UVJ$\xspace selection technique, we have separately stacked the galaxies classified as quiescent. The result is presented in Fig.~\ref{FIG:sfrms_passive}. On this plot we show what the location of the quiescent galaxies would be on the ${\rm SFR}$--$M_\ast$ plane assuming that \emph{all} their IR luminosity is coming from star formation. This is certainly wrong because in these massive galaxies dust is mostly heated by old stars, so the ${\rm SFR}$ we derive is actually an upper limit on the true star formation activity of these galaxies. However, even with this naive assumption, the derived ${\rm SFR}$s are an order of magnitude lower than that of the star-forming sample. We also observe that the effective dust temperature, inferred from the wavelength at which the FIR emission peaks, is lower and this is expected if dust is indeed mainly heated by less massive stars.

\section{Tests of our methods on simulated images \label{APP:simu}}

\begin{figure*}
    \centering
    \includegraphics*[width=18cm]{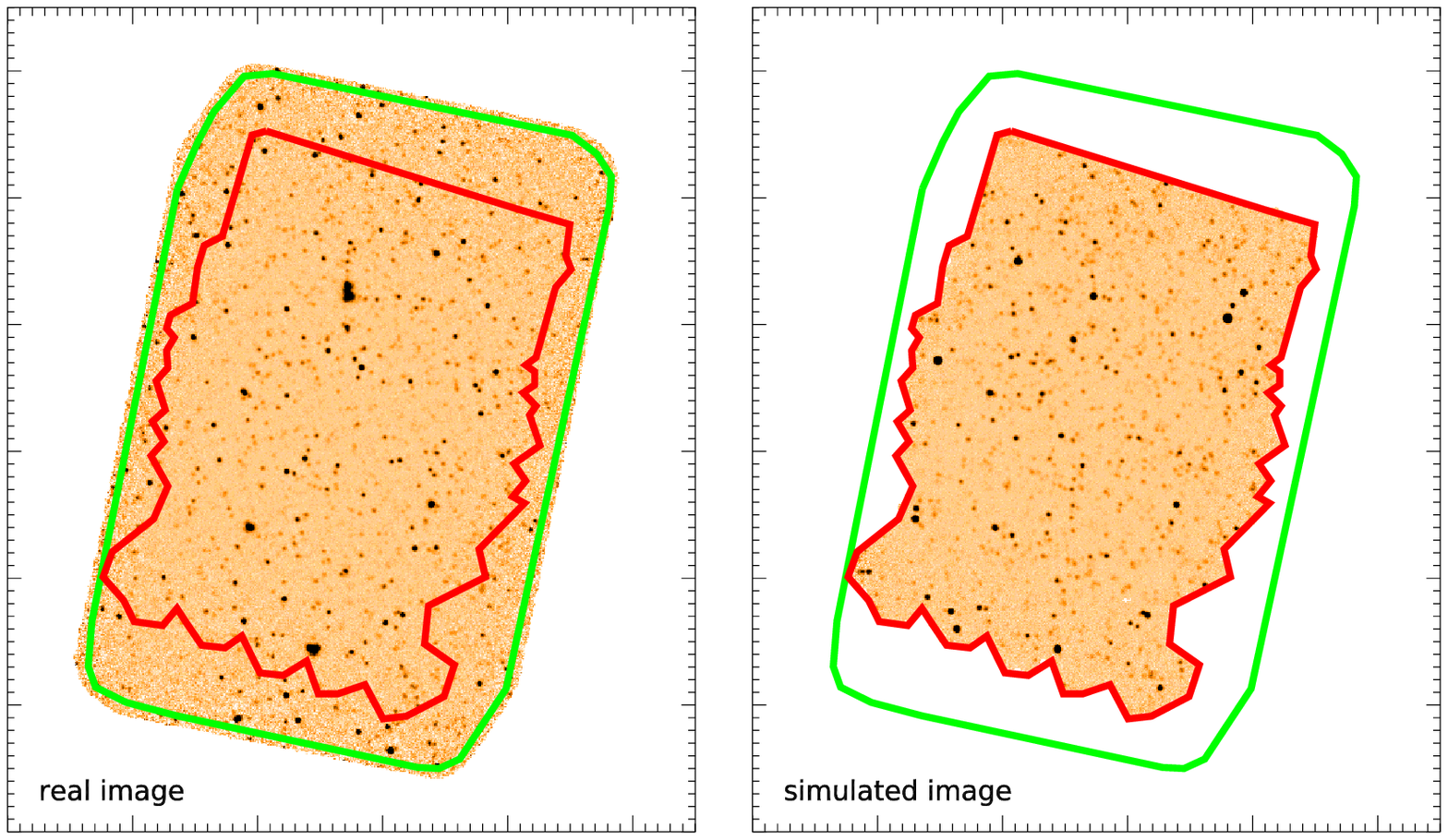}
    \caption{Real {\it Herschel} PACS $100\,\mu{\rm m}$ image (left) and one of our simulations (right). The green region shows the extent of the PACS coverage, while the red region shows the {\it Hubble} ACS coverage, i.e., the extent of our input catalog. The two images are shown here with the same color bar.}
    \label{FIG:simmap}
\end{figure*}

To test all of these procedures, we build a set of simulated images. We design these to be as close as possible to the real images in a \emph{statistical} sense, i.e., the same photometric and confusion noise, and the same number counts.

To do so, we start from our observed $H$-band catalogs, knowing redshifts and stellar masses for all the galaxies. Using our results from stacking {\it Herschel} images, we can attribute an ${\rm SFR}$ to each of these galaxies. We then add a random amount of star formation, following a log-normal distribution of dispersion $0.3\,{\rm dex}$. We also put $2\%$ of our sources in \emph{starburst} mode, where their ${\rm SFR}$ is increased by $0.6\,{\rm dex}$. Next, we assign an FIR SED to each galaxy following the observed trends with redshift (no mass dependence) and excess ${\rm SFR}$ \citep{magnelli2014}. Starburst galaxies are also given warmer SEDs.

From these simulated source catalogs, we generate a list of fluxes in all {\it Herschel} bands. Given noise maps (either modeled from RMS maps assuming Gaussian noise, or constructed from the difference between observing blocks), we build simulated images by placing each source as a PSF centered on its sky position, with a Gaussian uncertainty of $0.45\arcsec$ and a maximum offset of $0.9\arcsec$. We randomly reposition the sources inside the fields using uniform distributions in right ascension and declination, to probe multiple realizations of confusion. These simulated images have pixel distribution, or $P(D)$ plots, very close to the observed images, and are thus good tools to study our methods. An example is shown in Fig.~\ref{FIG:simmap} for the GOODS--South field at $100\,\mu{\rm m}$.

We produce $400$ sets of simulated catalogs and images, each with a different realization of photometric noise, confusion noise and ${\rm SFR}$. We then run our full stacking procedure on each, using the same setup as for the real images (i.e., using the same redshift and mass bins), to test the reliability of our flux extraction and the accuracy of the reported errors.

\subsection{Mean and median stacked fluxes \label{APP:medmean}}

\begin{figure*}
    \centering
    \includegraphics*[width=9cm]{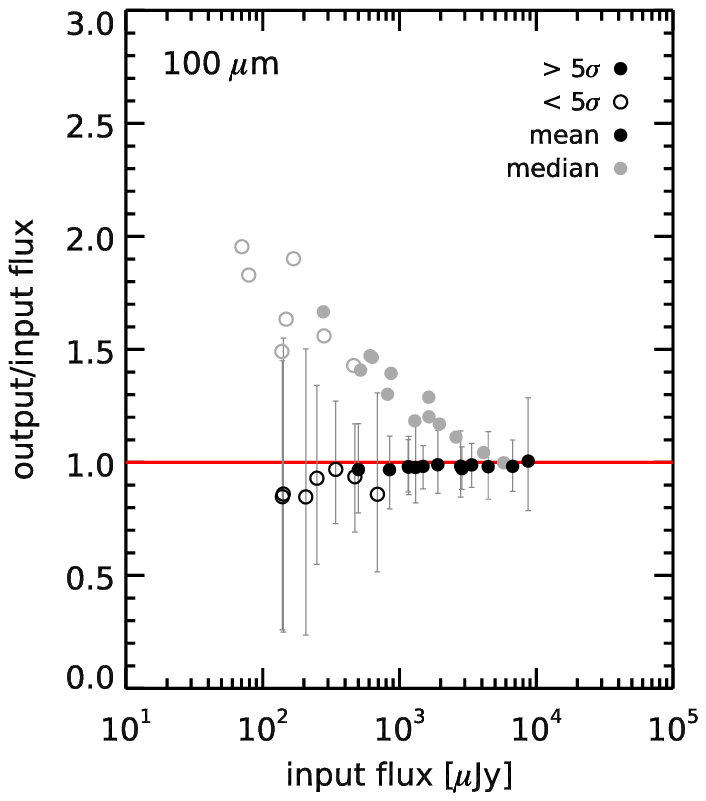}
    \includegraphics*[width=9cm]{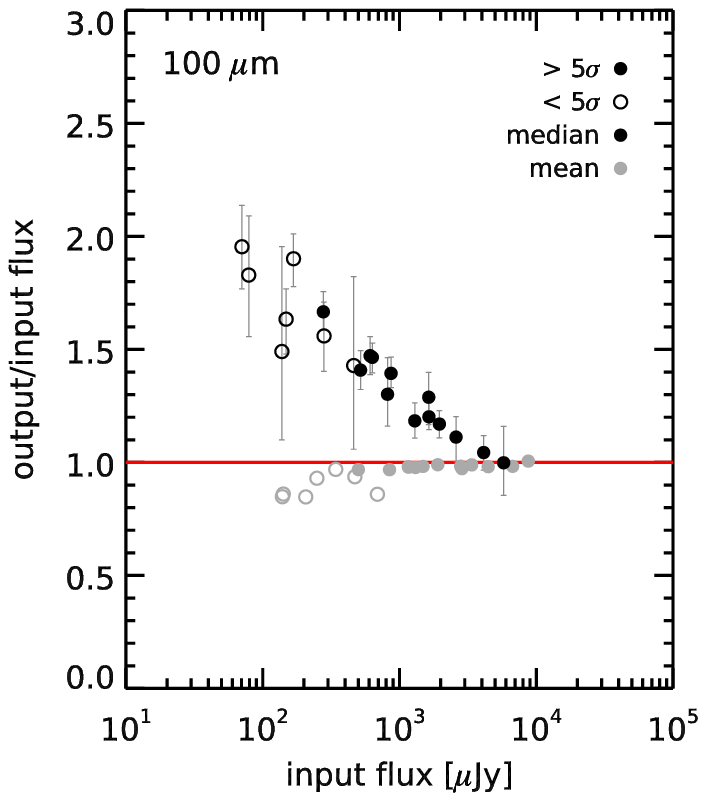}
    \caption{Comparison of measured stacked flux densities from the simulated images with the real flux densities that were put into the $100\,\mu{\rm m}$ map (the other wavelengths behave the same). The stacked sources were binned in redshift and mass using exactly the same bins as those that were used to analyze the real images. {\bf Left:} mean stacked flux densities, {\bf right:} median stacked flux densities. Each point shows the median $S_{\rm output}/S_{\rm input}$ among all the $400$ realizations, while error bars show the $16$th and $84$th percentiles of the distribution. Filled circles indicate measurements that are individually significant at $>5\sigma$ on average, i.e., those we would actually use, while open circles indicate measurements at $<5\sigma$ to illustrate the trend. On each plot, gray circles show the values obtained with the other method (i.e., median and mean) for the sake of direct comparison. It is clear that mean fluxes are more noisy, while median fluxes exhibit a systematic bias.}
    \label{FIG:medmean}
\end{figure*}

\begin{figure}
    \centering
    \includegraphics*[width=8cm]{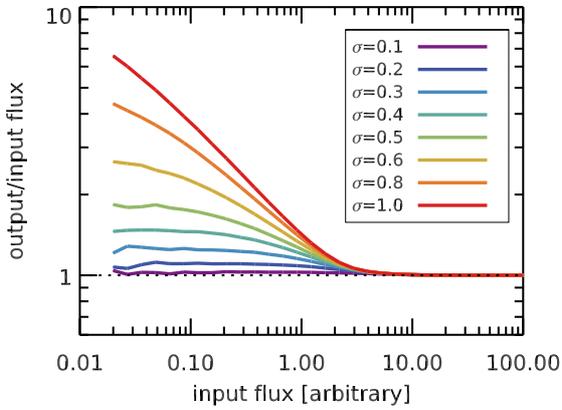}
    \caption{Monte Carlo analysis showing evidence for a systematic bias in median stacking. These values have been obtained by computing medians of log-normally distributed values in the presence of Gaussian noise of fixed amplitude ($\sigma_{\rm noise} = 1$ in these arbitrary flux units, so that the input flux is also the $S/N$).}
    \label{FIG:medbias_th}
\end{figure}

For each of the $400$ realizations we compare the measured flux densities using both mean and median stacking to the expected mean and median flux densities, respectively. The results are shown in Fig.~\ref{FIG:medmean} for the PACS $100\,\mu{\rm m}$ band. The other bands show similar behavior.

Although less noisy, median fluxes are biased toward higher values (at most by a factor $2$ here). This is because the median is not a linear operation, so it is not true in general that $\left<a + b\right> = \left<a\right> + \left<b\right>$, where $\left<.\right>$ denotes the median. In particular, this means that if we compute the median of our noisy stacked image and subtract the median value of the noise, we do not exactly recover the median flux density. We will call this effect the \emph{noise bias} in what follows. \cite{white2007} show that this bias arises when: 1) the signal to noise ratio of stacked sources is low; and 2) the distribution of flux is skewed toward either faint or bright sources. The latter is indeed true in our simulations, since we used a log-normal distribution for the ${\rm SFR}$. Correcting for this effect is not trivial, as it requires knowledge of the real flux distribution. Indeed, Fig.~\ref{FIG:medbias_th} shows the amplitude of this bias for different log-normal flux dispersions, the highest dispersions producing the highest biases. \cite{white2007} argue that the median stacked flux is still a useful quantity, since it is actually a good measure of the \emph{mean} of the distribution, but this is only true in the limit of low signal to noise ratios. In their first example, a double normal distribution, the measured median reaches the true mean for ${\rm SNR} < 0.1$, but correctly measures the true median for ${\rm SNR} > 3$.

Of course these values depend on the distribution itself, as is shown in Fig.~\ref{FIG:medbias_th}. In particular, for a log-normal distribution with $0.3\,{\rm dex}$ scatter, the mean is reached for ${\rm SNR} < 0.4$, and the median for ${\rm SNR} > 3$. Theoretically, the difference between the mean and the median for a log-normal distribution is $\log(10)\ \sigma^2 / 2\,{\rm dex}$. In our simulations, the typical $100\,\mu{\rm m}$ flux dispersion within a stacking bin is $\sim0.45 \pm 0.1\,{\rm dex}$, which yields a factor $\sim1.7^{+0.5}_{-0.2}$, in agreement with the maximum observed bias.

\begin{table}[htdp]
\caption{Ratio of the $L_{\rm IR}$ values obtained from median and mean stacking using the same sample on the real {\it Herschel} images. \label{TAB:medmean}}
\begin{center}
\begin{tabular}{ccccccc}
    \hline
    \hline \\[-2.5mm]
 $\log_{10}(M_\ast/{\rm M}_\odot)$ & $z=0.5$ & $1.0$ & $1.5$ & $2.2$ & $3.0$ & $4.0$ \\
    \hline \\[-2.5mm]
    $11.2$ & $0.79$ & $0.95$ & $0.84$ & $0.88$ & $0.82$ & $0.86$ \\
    $10.8$ & $0.63$ & $0.90$ & $0.92$ & $0.94$ & $0.77$ &  ---   \\
    $10.2$ & $0.84$ & $0.98$ & $0.90$ & $0.97$ &  ---   &  ---   \\
    $9.8$  & $0.89$ & $0.91$ &  ---   &  ---   &  ---   &  ---   \\
    \hline
\end{tabular}
\end{center}
\end{table}

To see how this affects the measured $L_{\rm IR}$ in practice, we list in Table \ref{TAB:medmean} the ratio of the median to mean measured $L_{\rm IR}$ in each stacked bin, as measured on the real images. We showed in section \ref{SEC:sfrdisp} that the dispersion in $L_{\rm IR}$ is about $0.3\,{\rm dex}$. Therefore, assuming a log-normal distribution, we would theoretically expect the ratio of the median to mean $L_{\rm IR}$ to be close to $0.78$. It is, however, clear from Table \ref{TAB:medmean} that this is not the case in practice: the median is usually (but not always) much closer to the mean than expected for a noiseless situation. Therefore, the median stacked fluxes are often not measuring the median fluxes or the mean fluxes, but something in between. Since correcting for this bias requires assumptions on the flux distribution, we prefer (when possible) to use the more noisy but unbiased mean fluxes for this study.

\subsection{Clustering correction \label{APP:clustering}}

\begin{table}[htdp]
\setlength{\tabcolsep}{4pt}
\caption{Clustering bias in simulated {\it Herschel} images. \label{TAB:clustering}}
\begin{center}
\begin{tabular}{lccccc}
    \hline
    \hline \\[-2.5mm]
Method & $100\,\mu{\rm m}$ & $160\,\mu{\rm m}$ & $250\,\mu{\rm m}$ & $350\,\mu{\rm m}$ & $500\,\mu{\rm m}$ \\
    \hline \\[-2.5mm]
A & $0\%_{-7\%}^{+7\%}$ &  $3\%_{-8\%}^{+9\%}$ &  $8\%_{-8\%}^{+12\%}$ & $13\%_{-10\%}^{+12\%}$ &  $25\%_{-18\%}^{+19\%}$ \\[4pt]
B &  $0\%_{-12\%}^{+8\%}$ &  $3\%_{-12\%}^{+13\%}$ &  $19\%_{-11\%}^{+17\%}$ &
$33\%_{-19\%}^{+27\%}$ &  $58\%_{-31\%}^{+54\%}$ \\[4pt]
C & $0\%_{-7\%}^{+8\%}$ &  $7\%_{-9\%}^{+11\%}$ &  $14\%_{-9\%}^{+14\%}$ &
$22\%_{-14\%}^{+19\%}$ &  $39\%_{-23\%}^{+22\%}$ \\[2pt]
    \hline
\end{tabular}
\end{center}
These values were obtained by computing the ratio of measured mean stacked fluxes to the expected mean fluxes in simulated images using our flux extraction method (see section \ref{SEC:stackmethod}). Median stacked fluxes are affected the same way, after removing the noise bias described in Appendix \ref{APP:medmean}. We also show the $16$th and $84$th percentiles of the bias distribution. The methods are: {\bf A}, using our own flux extraction procedure section \ref{SEC:stackmethod}; {\bf B}, using the full PSF; and {\bf C}, using only the central pixel.
\end{table}

Among our $400$ random realizations, the measured mean fluxes do not show any systematic bias. However these simulations do not take the flux boosting caused by source physical clustering into account, because we assigned random positions to the sources in our catalog. To test the effect of clustering, we regenerate a new set of $200$ simulations, this time using the real optical positions of the sources and only varying the photometric noise and the ${\rm SFR}$s of the sources.

If galaxies are significantly clustered in the image, then the measured fluxes will be boosted by the amount of light from clustered galaxies that falls inside the beam. Since the beam size here is almost a linear function of the wavelength, we expect SPIRE bands to be more affected than PACS bands. Since the same beam at different redshifts corresponds to different proper distances, low redshift measurements ($z<0.5$) should be less affected. However, because of the flatness of the relation between redshift and proper distance for $z>0.5$, this should not have a strong impact for most of our sample. Indeed, we do not observe any significant trend with redshift in our simulations. No trend was found with stellar mass either, hence we averaged the clustering signal over all stacked bins for a given band, and report the average measured boost in Table \ref{TAB:clustering} (``method A'') along with the $16$th and $84$th percentiles. Although we limited this analysis to fluxes measured at better than $5\sigma$, the scatter in the measured bias is compatible with being only caused by uncertainties in flux extraction.

\begin{table}[htdp]
\caption{Ratio of the $L_{\rm IR}$s obtained after and before applying clustering corrections listed in Table \ref{TAB:clustering}. \label{TAB:lirdeclust}}
\begin{center}
\begin{tabular}{ccccccc}
    \hline
    \hline \\[-2.5mm]
 $\log_{10}(M_\ast/{\rm M}_\odot)$ & $z=0.5$ & $1.0$ & $1.5$ & $2.2$ & $3.0$ & $4.0$ \\
    \hline \\[-2.5mm]
    $11.2$ & $0.96$ & $1.01$ & $0.90$ & $0.93$ & $0.91$ & $0.75$ \\
    $10.8$ & $0.96$ & $1.02$ & $0.87$ & $0.97$ & $0.93$ &  ---   \\
    $10.2$ & $0.99$ & $0.98$ & $0.96$ & $0.99$ & $0.94$ &  ---   \\
    $9.8$  & $0.99$ & $0.95$ & $0.78$ &  ---   &  ---   &  ---   \\[1pt]
    \hline
    \end{tabular}
\end{center}
\end{table}

Although negligible in PACS, this effect can reach $30\%$ in SPIRE $500\,\mu{\rm m}$ data. Here we correct for this bias by simply deboosting the real measured fluxes by the factors listed in Table \ref{TAB:clustering}, band by band. The net effect on the total measured $L_{\rm IR}$ is reported in Table \ref{TAB:lirdeclust}.

By construction, these corrections are specific to our flux extraction method. By limiting the fitting area to pixels where the PSF relative amplitude is larger than $10\%$, we absorb part of the large scale clustering into the background level. If we were to use the full PSF to measure the fluxes, we would measure a larger clustering signal (see section \ref{SEC:stackmethod}). We have re-extracted all the fluxes by fitting the full PSF, and we indeed measure larger biases. These are tabulated in Table \ref{TAB:clustering} as ``method B''. An alternative to PSF fitting that is less affected by clustering consists of setting the mean of the flux map to zero before stacking and then only using the central pixel of the stacked cutout \citep{bethermin2012}. Because of clustering, the effective PSF of the stacked sources will be broadened, and using the real PSF to fit this effective PSF will result in some additional boosting. Therefore, by only using the central pixel, one can get rid of this effect. We show in Table \ref{TAB:clustering} as ``method C'' how the figures change using this alternative method. Indeed the measured boosting is smaller than when using the full PSF, and is consistent with that reported by B\'ethermin et al. (2014, submitted), but our method is even less affected thanks to the use of a local background.

\subsection{Error estimates \label{APP:errors}}

\begin{figure*}
    \centering
    \includegraphics*[width=8cm]{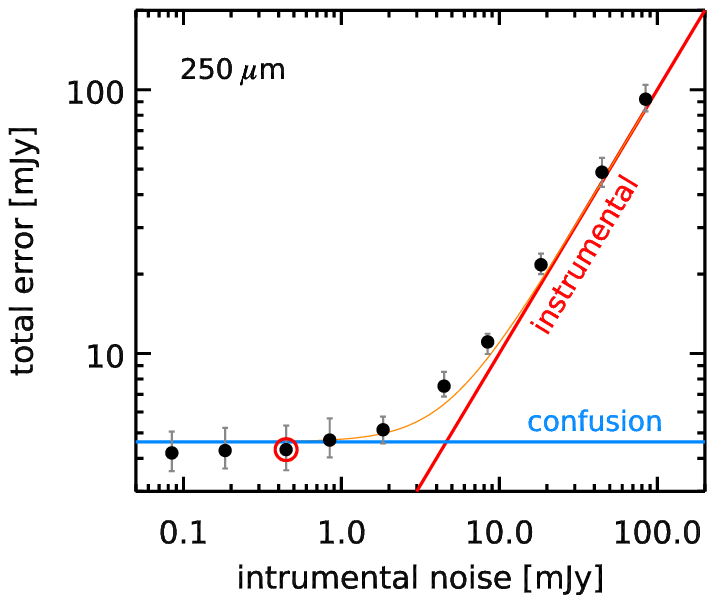}
    \includegraphics*[width=8cm]{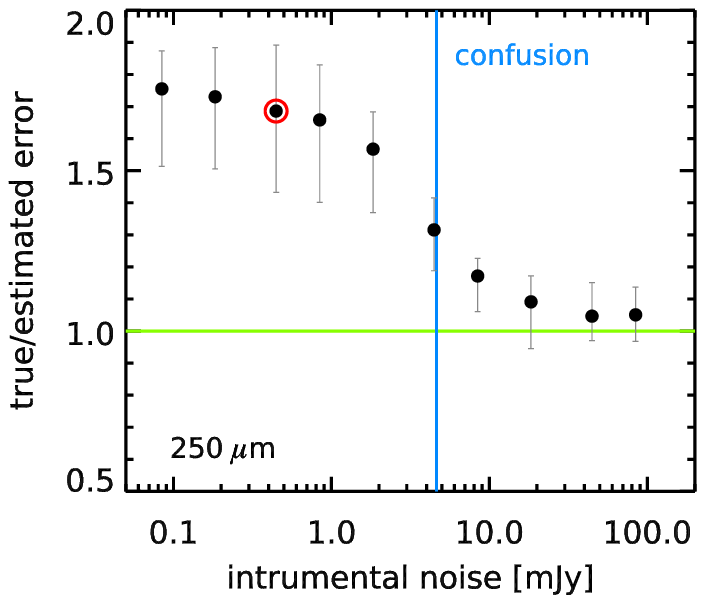}
    \caption{True error $\sigma$ on the stacked flux measurements as a function of the instrumental white noise level $\sigma_{\rm inst.}$ that is put on the image (here normalized to a ``PSF'' noise in ${\rm mJy}$, i.e., the error on the flux measurement of a point source in the absence of confusion). We generated multiple simulations of the $250\,\mu{\rm m}$ maps using varying levels of white noise, and compute $\sigma$ from the difference between the measured fluxes and their expected values. {\bf Left:} evolution of the average total noise per source $\sigma_{\rm tot.} = \sigma\times\sqrt{N_{\rm stack}}$ where $N_{\rm stack}$ is the number of stacked sources. This is the total error when extracting the flux of a single source on the map. When the instrumental noise (red line) is high, it dominates the error budget over the confusion noise. However, when reaching too low values, the measured total noise is dominated by the confusion noise $\sigma_{\rm conf.}$ (blue line). We fit this evolution as $\sigma^2_{\rm tot.} = \sigma^2_{\rm inst.} + \sigma^2_{\rm conf.}$ (orange line) to obtain $\sigma_{\rm conf.} = 4.6\,{\rm mJy}$. The red circle marks the instrumental noise level reached in the real maps. {\bf Right:} comparison between the estimated error from the stack residual $\sigma_{\rm IMG}$ and the true error $\sigma$. The points show the median of $\sigma/\sigma_{\rm IMG}$, and the error bars are showing the $16$th and $84$th percentiles of the distribution. The green horizontal line is the line of perfect agreement, and the blue vertical line marks the confusion noise at $250\,\mu{\rm m}$. The red circle marks the instrumental noise level reached in the real maps. }
    \label{FIG:errconv}
\end{figure*}

\begin{figure*}[ht]
    \centering
    \includegraphics*[width=18cm]{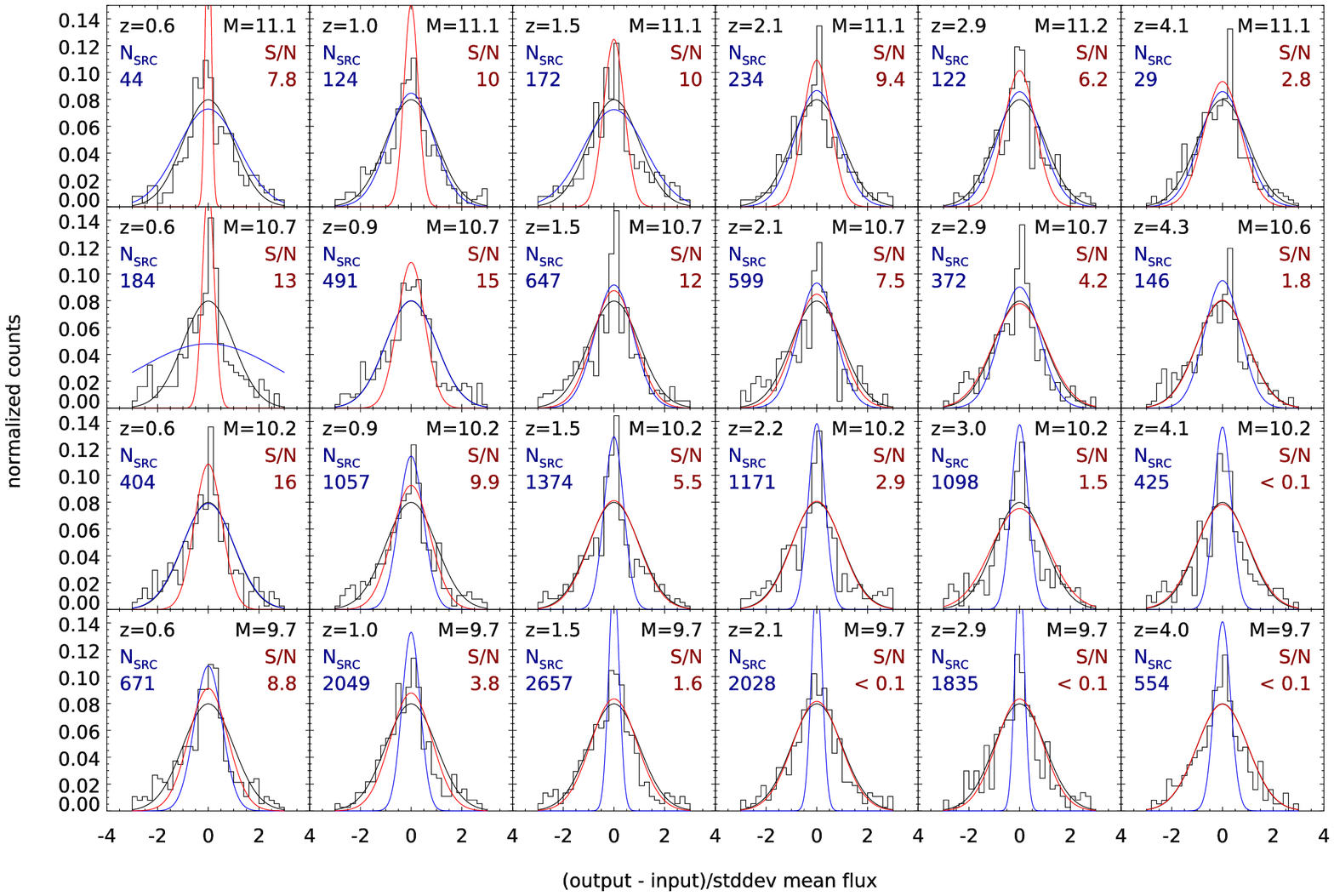}
    \includegraphics*[width=18cm]{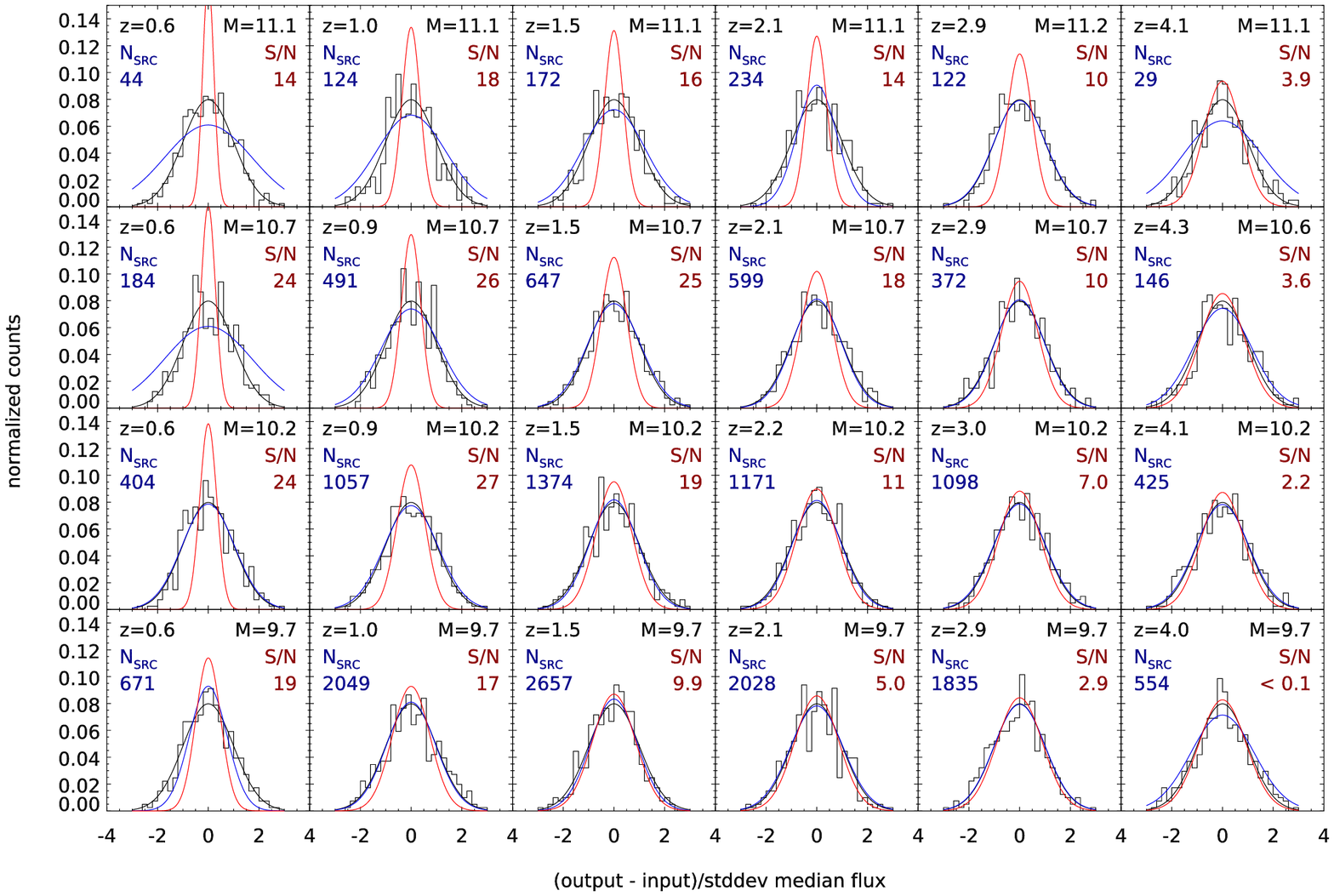}
    \caption{Normalized distribution of $(\Delta S - \left<\Delta S\right>)/\sigma$ of the mean (top) and median (bottom) stacked PACS $100\,\mu{\rm m}$ fluxes in each stacked bin. The black, blue, and red curves show Gaussians of width $1$, $\sigma_{\rm BS}/\sigma$ and $\sigma_{\rm IMG}/\sigma$, respectively. The estimation of the true signal to noise ratio of the flux measurement is displayed in dark red, while the average number of stacked sources is shown in dark blue.}
    \label{FIG:error_pacs}
\end{figure*}

We now study the reliability of our error estimates on the stacked fluxes. We compute the difference between the observed and input flux for each realization, $\Delta S$. We then compute the median $\left<\Delta S\right>$, which is essentially the value plotted in Fig.~\ref{FIG:medmean}, i.e., it is nonzero mostly for median stacked fluxes. We subtract this median difference from $\Delta S$, and compute the scatter $\sigma$ of the resulting quantity using median absolute deviation, i.e., $\sigma \equiv 1.48\times{\rm MAD}(\Delta S - \left<\Delta S\right>)$. We show in Fig.~\ref{FIG:error_pacs} the histograms of $(\Delta S - \left<\Delta S\right>)/\sigma$ for the mean and median stacked PACS $100\,\mu{\rm m}$ fluxes in each stacked bin. By construction, these distributions are well described by a Gaussian of width unity (black curve).

We have two error estimates at our disposal. The first, $\sigma_{\rm IMG}$, is obtained by measuring the RMS of the residual image (after the stacked fluxes have been fitted and subtracted), and multiplying this value by the PSF error factor (see Eq.~\ref{EQ:psferror}). The second, $\sigma_{\rm BS}$, is obtained by bootstrapping, i.e., repeatedly stacking half of the parent sample and measuring the standard deviation of the resulting flux distribution (again, see section \ref{SEC:stackmethod}). Each of these method provides a different estimation of the error on the flux measurement, and we want to test their accuracy.

In Fig.~\ref{FIG:error_pacs}, we show as red and blue lines the predicted error distribution according to $\sigma_{\rm IMG}$ and $\sigma_{\rm BS}$, respectively. When the predicted distribution is too narrow or too broad compared to the observed distribution (black curve), this means that the estimated error is respectively too low or too high.

For median stacked fluxes, it appears that $\sigma_{\rm BS}$ is accurate in all cases. It tends to slightly overestimate the true error on some occasions, but not by a large amount. On the other hand, $\sigma_{\rm IMG}$ dramatically underestimates the error when the measured $S/N$ of stacked sources is high (or the number of stacked sources is low).

The situation for mean stacked fluxes is quite different. The behavior of $\sigma_{\rm IMG}$ is the same, but $\sigma_{\rm BS}$ show the completely opposite trend, i.e., it underestimates the error at low signal to noise and high number of stacked sources. This may be caused by the fact that bootstrapping will almost always produce the same confusion noise, since it uses the same sources. The reason why this issue does not arise for median stacked fluxes might be because the median naturally filters out bright neighbors, hence reducing the impact of confusion noise.

The results are the same for the PACS $70$ and $160\,\mu{\rm m}$ band. Therefore, keeping the maximum error between $\sigma_{\rm IMG}$ and $\sigma_{\rm BS}$ ensures that one has an accurate error measurement in all cases for the PACS bands.

\begin{figure*}[ht]
    \centering
    \includegraphics*[width=18cm]{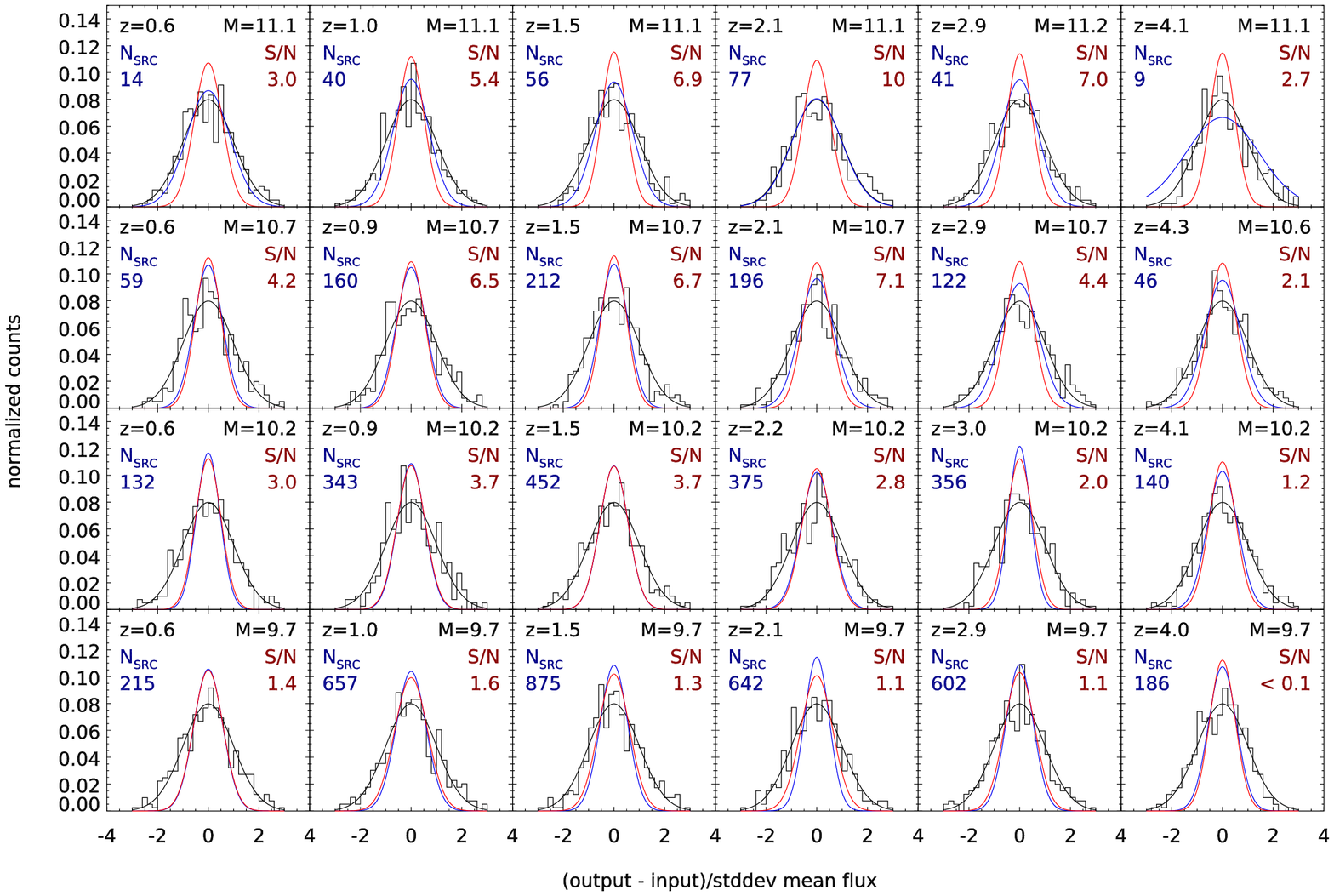}
    \includegraphics*[width=18cm]{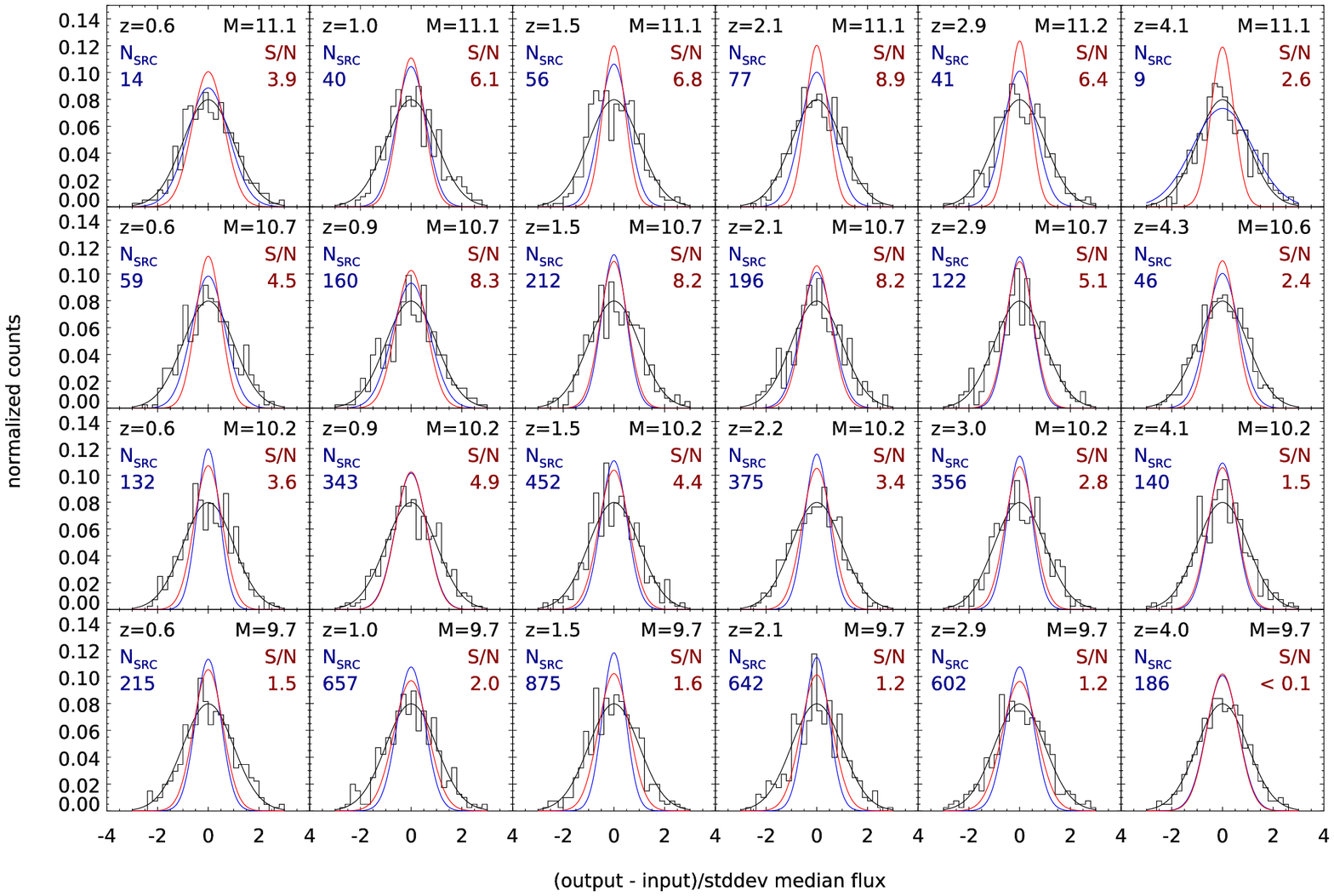}
    \caption{Same as Fig.~\ref{FIG:error_pacs} for SPIRE $350\,\mu{\rm m}$.}
    \label{FIG:error_spire}
\end{figure*}

The SPIRE fluxes on the other hand show a substantially different behavior. We reproduce the same figures in Fig.~\ref{FIG:error_spire}, this time for the SPIRE $350\,\mu{\rm m}$ band. Here, and except for the highest mass bin, the errors are systematically underestimated by a factor of $\sim 1.7$, regardless of the estimator used. We therefore use this factor to correct all our measured SPIRE errors in these bins.

We believe this underestimation of the error is an effect of confusion noise. Indeed, it is clear when looking at the stacked maps at these wavelengths (e.g., Fig.~\ref{FIG:stack}) that there is a substantial amount of large scale noise coming from the contribution of the neighboring bright sources. The main issue with this noise is that it is spatially correlated. This violates one of the assumptions that were made when deriving the error estimation of Eq.~\ref{EQ:psferror}, which may thus give wrong results. The reason why only the SPIRE bands are affected is because the noise budget here is (by design) completely dominated by confusion. This is clear from Fig.~\ref{FIG:errconv} (left): when putting little to no instrumental noise $\sigma_{\rm inst}$ on the simulated maps, the total error $\sigma_{\rm tot}$ on the flux measurements is completely dominated by the confusion noise $\sigma_{\rm conf}$ (blue line), and it is only by adding instrumental noise of at least $10\,{\rm mJy}$ (i.e., ten times more than what is present in the real maps) that the image becomes noise dominated. By fitting
\begin{equation}
    \sigma_{\rm tot} = \sqrt{\sigma^2_{\rm conf} + \sigma^2_{\rm inst}}\,,
\end{equation}
we obtain $\sigma_{\rm conf} = 4.6\,{\rm mJy}$. This value depends on the model we used to generate the simulated fluxes, but it is in relatively good agreement with already published estimates from the literature \citep[e.g.,][who predict $\sigma_{\rm conf} = 6\,{\rm mJy}$]{nguyen2010}.

We then show in Fig.~\ref{FIG:errconv} (right) that the error underestimation in the SPIRE bands, here quantified by the ratio $\sigma/\sigma_{\rm IMG}$, goes away when the image is clearly noise dominated, meaning that this issue is indeed caused by confusion and the properties of the noise that it generates.

%
%

\end{document}